\documentclass[fleqn,usenatbib]{mnras}

\usepackage{newtxtext,newtxmath}

\usepackage[T1]{fontenc}
\usepackage{ae,aecompl}

\usepackage{graphicx}	
\usepackage{amsmath}	
\usepackage{amssymb}	
\usepackage{xcolor}
\usepackage{soul}
\usepackage{subfigure}

\definecolor{bluehl}{rgb}{0.75,0.75,1}
\definecolor{greenhl}{rgb}{0.5,1.0,0.5}

\graphicspath{{./FiguresFeb2020/}}

\newcommand{\angstrom}{\mbox{\normalfont\AA}}

\title[Evaporative Transmission Spectra]{Alkaline Exospheres of Exoplanet Systems: Evaporative Transmission Spectra}

\author[Gebek \& Oza]{
Andrea Gebek$^{1}$\thanks{E-mail: agebek@student.ethz.ch}
and Apurva V. Oza$^{2}$
\\
$^{1}$Department of Physics, ETH Z\"{u}rich, Wolfgang-Pauli-Strasse 27, 8093 Z\"{u}rich, Switzerland\\
$^{2}$Physikalisches Institut, Universit\"{a}t Bern, Gesellschaftsstrasse 6, 3012 Bern, Switzerland\\
}

\date{Accepted XXX. Received YYY; in original form ZZZ}

\pubyear{2020}
\begin{document}
\label{firstpage}
\pagerange{\pageref{firstpage}--\pageref{lastpage}}
\maketitle

\begin{abstract}
Hydrostatic equilibrium is an excellent approximation for the dense layers of planetary atmospheres where it has been canonically used to interpret transmission spectra of exoplanets. Here we exploit the ability of high-resolution spectrographs to probe tenuous layers of sodium and potassium gas due to their formidable absorption cross-sections. We present an atmosphere-exosphere degeneracy between optically thick and optically thin mediums, raising the question of whether hydrostatic equilibrium is appropriate for Na I lines observed at exoplanets.
To this end we simulate three non-hydrostatic, \textit{evaporative}, density profiles: (i) escaping, (ii) exomoon, and (iii) torus to examine their imprint on an alkaline exosphere in transmission. By analyzing an \textit{evaporative} curve of growth we find that equivalent widths of $W_{\mathrm{Na D2}} \sim 1- 10\,\mathrm{m\angstrom}$ are naturally driven by evaporation rates $\sim 10^3 - 10^5$ kg/s of pure atomic Na. To break the degeneracy between atmospheric and exospheric absorption, we find that if the line ratio is $\mathrm{D2/D1} \gtrsim 1.2$ the gas is optically thin on average roughly indicating a non-hydrostatic structure of the atmosphere/exosphere. We show this is the case for Na I observations at hot Jupiters WASP-49b and HD189733b and also simulate their K I spectra. Lastly, motivated by the slew of metal detections at ultra-hot Jupiters, we suggest a toroidal atmosphere at WASP-76b and WASP-121b is consistent with the Na I data at present.   
\end{abstract}





\section{Introduction}

The alkali metals, sodium (Na I) and potassium (K I), have long been predicted to be observable spectroscopically during an exoplanet transit (\citealt{Seager2000}; \citealt{Hubbard2001}; \citealt{Brown2001}). While this launched the study of extrasolar \textit{atmospheres} we remark here that alkaline observations are also consistent with extrasolar \textit{exospheres} of diverse geometries in terms of the required source rates (\citealt{Johnson2006b}; \citealt{Oza2019}). This subtle degeneracy between collisional atmospheres and collisionless exospheres is due to the large absorption cross-sections of the Na \& K atoms, enabling small column densities of gas to illuminate optically thin gas. 
\\

While transmission spectra can be simulated analytically (\citealt{LecavelierdesEtangs2008}; \citealt{Wit2013}; \citealt{Betremieux2017}; \citealt{Heng2017}; \citealt{Jordan2018}; \citealt{Fisher2019}), these approaches rely on the assumption of an atmosphere being in local hydrostatic equilibrium, an assumption which generally holds in dense atmospheric layers where gravity and pressure gradients are the dominant forces. Hydrostatic equilibrium becomes less accurate in more tenuous atmospheric layers, and can break down in certain circumstances such as in an escaping atmospheric wind. In fact, soon after the first detection of Na I at HD209458b (interpreted as an exoplanet atmosphere by \citealt{Charbonneau2002}, and recently contested by \citealt{Casasayas-Barris2020} attributing the signal to the Rossiter-McLaughlin-effect and centre-to-limb variations), \citealt{Vidal-Madjar2003} detected neutral hydrogen beyond the Hill sphere and interpreted this as an evaporating component of the atmosphere. Close-in exoplanet atmospheres were hence observed to be nonhydrostatic, and shown to be hydrodynamically evaporating $\sim 10^{7}$ kg/s of gas due to XUV-driven escape (e.g. \citealt{Murray-Clay2009}). Considerable endogenic modeling of evaporative transmission spectra is indeed underway, largely aiming at exospheric signatures of atomic hydrogen (\citealt{BourrierLecavelier2013}; \citealt{Christie2016}; \citealt{Allan2019}; \citealt{Murray-Clay2019}; \citealt{Wyttenbach2020}), but also by atomic helium (\citealt{OklopvcicHirata2018}; \citealt{Lampon2020}), atomic magnesium (\citealt{Bourrier2015}) and ionized magnesium (\citealt{Dwivedi2019}). The problem of a hydrostatic assumption is further amplified and fundamentally different if the gas were to be detached from the planet. Such an exogenic\footnote{Exogenic refers to an external source whereas endogenic refers to a planetary source.} alkaline source, such as an outgassing satellite or a thermally desorbing torus, would not be in hydrostatic equilibrium. The impact of exogenic sources on transmission spectra, which is one of the key questions we seek to answer in this study, has not yet been investigated until present. In the following we shall use evaporative and non-hydrostatic interchangeably.

The retrieval of atmospheric parameters from observations using various techniques such as $\chi^2$-minimization (e.g. \citealt{Madhusudhan2009}), Bayesian analysis (e.g. \citealt{Madhusudhan2011}; \citealt{Lee2012}; \citealt{Benneke2013}), or advanced machine-learning methods (e.g. \citealt{Marquez-Neila2018}; \citealt{Hayes2020}) is generally carried out within a hydrostatic setup. Various hydrostatic retrieval codes exist in the community (e.g. \citealt{Madhudsudhan2009}; \texttt{TAU-REX I} \citealt{Waldmann2015}; \texttt{BART} \citealt{Blecic2017}; \texttt{Exo-Transmit} \citealt{Kempton2017}; \texttt{ATMO} \citealt{Goyal2018}; $^{\pi}\eta$ \citealt{Pino2018}\footnote{Based on the earlier $\eta$ code from \citet{Ehrenreich2006}.}; \texttt{Aura} \citealt{Pinhas2018}; \texttt{HELIOS-T} \citealt{Fisher2018}; \texttt{PLATON} \citealt{Zhang2019}; \citealt{Brogi2019}; \citealt{Fisher2019}), which mostly differ in terms of their treatment of atmospheric chemistry (ranging from constant abundances over chemical equilibrium to atmospheres in disequilibrium), assumptions of temperature profiles (using isothermal or parametric profiles, or invoking radiative-convective equilibrium modelling) and retrieval routines. For a comprehensive overview of different codes and retrieval techniques, see \citet{Madhusudhan2018}, here we briefly point out some of the codes which pose exceptions to the common assumption of an atmosphere in hydrostatic equilibrium. The \texttt{NEMESIS} (\citealt{Irwin2008}) and \texttt{CHIMERA} (\citealt{Line2013}) codes are flexible in terms of their atmospheric structure such that they can process arbitrary pressure profiles. The more recent \texttt{TAU-REX III} code (\citealt{Al-Refaie2019}) allows retrievals in chemical equilibrium or disequilibrium for custom (non-hydrostatic) pressure profiles (note that the default retrieval setup is still hydrostatic).

In the present study, we examine evaporative transmission spectra of the Na I doublet ($\lambda_{\mathrm{D2}}=5889.95\,\angstrom,\,\lambda_{\mathrm{D1}}=5895.92\,\angstrom$) and the K I doublet ($\lambda_{\mathrm{D2}}=7664.90\,\angstrom,\,\lambda_{\mathrm{D1}} =7698.96\, \angstrom$). The lines when viewed in absorption are extremely bright due to resonance scattering (\citealt{BrownYung1976}; \citealt{Draine2011}) off the sodium atoms. The resonant scattering cross section is large, producing considerable absorption in highly tenuous columns of gas at pressures where one expects large deviations from hydrostatic equilibrium. Fortunately astronomers and planetary scientists have had decades of fundamental understanding of the physics of the Na and K alkaline resonance lines by directly observing comets and moons in-situ from within our solar system (\citealt{Oza2019} Table 1 and references therein). The most spectrally conspicuous aforementioned \textit{evaporative} sources are Jupiter's and Saturn's moons Io \& Enceladus, motivating the mass loss model of an evaporating exomoon shown to be roughly consistent with several extrasolar gas giant planets today (\citealt{Oza2019}). Coupling a metallic evaporation model (termed \texttt{DISHOOM}) to a radiative transfer code capable of treating nonhydrostatic profiles (termed \texttt{Prometheus}) is the crux of our present study on \textit{evaporative transmission spectra}. This coupling permits a holistic approach to transit spectra, capable of breaking endogenic-exogenic degeneracies in alkaline exospheres today.

We use a custom-built radiative transfer code to simulate high-resolution transit spectra in the sodium doublet for four scenarios, seeking the precise imprint of an \textit{evaporative transmission spectrum} of an alkaline exosphere in regards to a canonical hydrostatic atmosphere. We present this code in Section \ref{Prometheus}. The mass loss model is laid out in Section \ref{dishoom}. Our four examined scenarios, corresponding to a particular geometry and spatial distribution of the sodium atoms, are presented in detail in Sections:

\begin{itemize}
\item \ref{hydrostatic} \underline{Hydrostatic}: A spherically-symmetric hydrostatic atmosphere.

\item \ref{escaping} \underline{Escaping}: A spherically-symmetric envelope undergoing atmospheric escape.

\item \ref{exomoon} \underline{Exomoon}: A spherically-symmetric cloud sourced by an outgassing satellite. 

\item \ref{torus} \underline{Torus}: An azimuthally-symmetric torus sourced by a satellite or debris.
\end{itemize}

We describe each of these scenarios with a particular number density profile $n(r)$ (since an exospheric collisionless gas cannot be described with a pressure profile) and two free parameters. These three components fully determine our simulated transit spectra. Of course, an exhaustive description of metals at a close-in gas giant system would involve a careful treatment of atmospheric collisional processes (e.g. \citealt{Huang2017}), exospheric physical processes (\citealt{Leblanc2017}), an atmospheric escape treatment in 3-D (e.g. \citealt{Debrecht2019}) and also the ability to track ions in the presence of a magnetic field (e.g. \citealt{Carnielli2020}). Since we reduce every scenario to a number density profile with two free parameters, our model is heuristic at present. Hence, this study isn't targeted at providing an exhaustive model of hot Jupiter exospheres with the corresponding transit spectra, but rather encourages a novel approach by elucidating fundamental differences between hydrostatic and non-hydrostatic assumptions.

We use the synergy of our two distinct approaches (radiative transfer \& mass loss) to gain a physical intuition on the four density scenarios described above, starting in Section \ref{physics}. We find that evaporative sodium profiles naturally allow vastly more extended yet tenuous distributions of the absorbing atoms than hydrostatic profiles. This property of evaporative sodium can lead to absorption in a primarily optically thin regime, resulting in high-resolution transit spectra differing from spectra computed within a hydrostatic framework. In Section \ref{d2d1method} we provide a simple diagnostic gathered from physics of the interstellar medium (e.g. \citealt{Draine2011}), to determine if an alkaline gas is optically thin or thick at a transiting exoplanet. Given that several hot Jupiters observed in high-resolution appear to reveal an optically thin regime, we simulate the spectral imprint for each evaporative scenario at HD189733b in Section \ref{Forward Modeling}. The forward model in this section serves to demonstrate the differences between hydrostatic and evaporative transmission spectra by coupling mass loss \textit{a priori}. 

For those more acquainted with inverse modeling, we use the observed transit spectra of the hot Jupiters WASP-49b and HD189733b to portray the behavior of evaporating alkalis in parameter space (Section \ref{Inverse Modeling}). In our observational analysis, best-fit parameters are found within the radiative transfer code using a bounded $\chi^2$-minimization. We then check the plausibility of these retrieved parameters by comparing to the calculated values from the alkali mass loss model. We choose to comparatively analyze these two planets as they were observed by the same spectrograph (HARPS) and reduced by the same authors (\citealt{Wyttenbach2015}; \citealt{Wyttenbach2017}) enabling a consistent comparison to the data of both hot Jupiters. These planets represent the first high-resolution Na I detections and therefore have been extensively studied by independent groups (\citealt{Louden2015}; \citealt{Cubillos2017}; \citealt{Huang2017}; \citealt{Pino2018}; \citealt{Fisher2019}) compared in Section \ref{comparisons}. Furthermore, these planets were shown to be able to host exogenic sources of alkali metals (i.e. satellites) in terms of tidal stability and average column densities (\citealt{Oza2019}). For our study of alkaline exospheres, we focus on the physics of evaporating Na I acknowledging the behavior of evaporating K I is nearly identical, simulated in Section \ref{potassium}. We discuss the imminence of metal detections at ultra-hot Jupiters (WASP-76b \& WASP-121b) in the context of toroidal atmospheres in Section \ref{ultrahot} and conclude in Section \ref{Conclusions}.




\section{Methods}\label{Methods}

\subsection{\texttt{Prometheus}: Alkaline Transmission Spectra of Atmospheres and Exospheres}\label{Prometheus}

\begin{figure}
    \centering
    \includegraphics[width=\columnwidth]{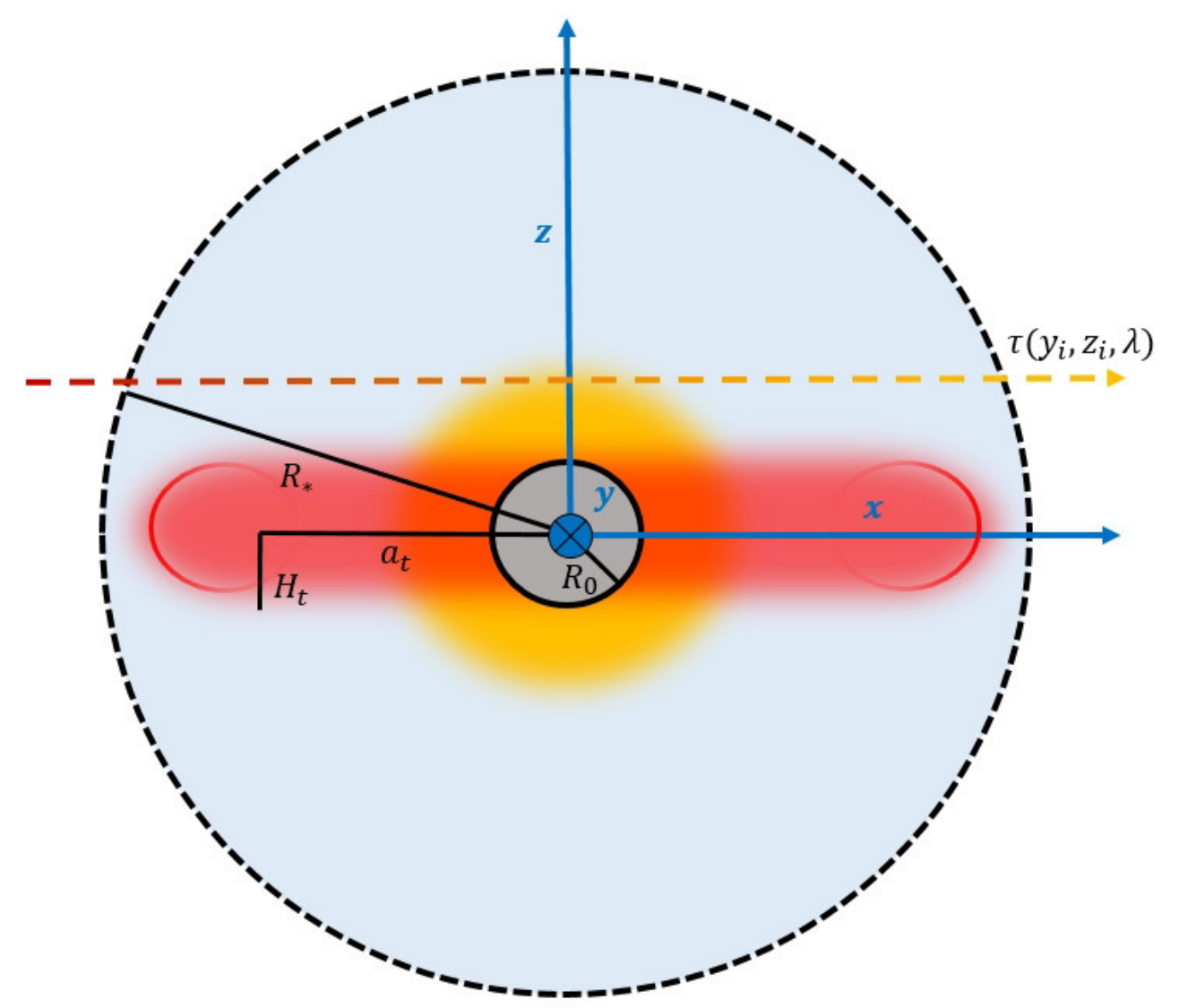}
    \caption{Sketch of an exoplanet in transit geometry. In this scheme, the three coordinate axes build a regular right-handed coordinate system and the host star is located at $(-a_P,0,0)$, with $a_P$ being the orbital radius of the exoplanet. Our four scenarios exhibit two different geometries: Spherical symmetry (the orange cloud) and azimuthal symmetry (the red torus). For a detailed scheme of the architecture of a sodium exosphere with exogenic sources illustrated see Figure 4 of \citet{Oza2019}.}
    \label{Geometry}
\end{figure}

We simulate transit spectra using a simple and flexible custom-built \texttt{Python} code: \texttt{Prometheus}. Our code computes absorption either in the Na I or K I doublet for an exoplanet \textit{system} in transit geometry. For a spherically-symmetric system (our first three scenarios), we divide the atmosphere using a linear grid in $x$- and a logarithmic one in $z$-direction with adjustable spatial resolutions to adapt to each scenario (see Figure \ref{Geometry}). For the torus scenario exhibiting azimuthal geometry we also linearly discretize the $y$-axis which affects the computation of the transit spectra (Eqns. \ref{tau} and \ref{flux}), see Section \ref{torus} for details. The transit spectrum is computed in the following two steps. First, we calculate the optical depth along the chord (x-axis, Figure \ref{Geometry}) at a certain altitude $z$ and wavelength $\lambda$:

\begin{equation}\label{tau}
    \tau (z,\lambda)=2\cdot\int_0^{\infty}\! n(r)\cdot \sigma\bigl(\lambda,T(r)\bigr)\cdot\chi_{\mathrm{i}}(r) \,\mathrm{d}x,
\end{equation}

given that $r=\sqrt{x^2+z^2}$ and where $\chi_{\mathrm{i}}(r)$ is the volume mixing ratio\footnote{Relative abundance by number. We denote mass mixing ratios by $x_{\mathrm{i}}$.} of the \textit{neutral} absorber (either Na I or K I), $\sigma(\lambda,T(r))$ the absorption cross section and $n(r)$ the number density profile which depends on the scenario (Eqns. \ref{n_hyd}, \ref{n_esc}, \ref{n_out}, \ref{n_tor}). $n(r)$ is generally calculated assuming hydrostatic equilibrium, in the present study we investigate how different spatial distributions of the absorbing atoms - i.e. different number density profiles - compare to the canonical hydrostatic atmosphere in terms of transmission spectra.

In addition, our code can be coupled to a chemistry code such as \texttt{FastCHEM} (\citealt{Stock2018}) to calculate these mixing ratio profiles, but we only examine constant mixing ratio profiles in the scope of this paper. Given that the mixing ratio $\chi_{\mathrm{i}}$ for a species i can be described as: $\chi_{\mathrm{i}}(r)=\chi_{\mathrm{i}}(r)\cdot (1-f_{\mathrm{ion}}(r))$, the assumption of a constant mixing ratio of the absorber implies a constant ionization fraction $f_{\mathrm{ion}}(r)$ for the absorbing species, given that the total (neutral plus ionized) mixing ratio of the absorber isn't expected to vary significantly. At present we focus solely on the absorption of alkali metals as they dominate the absorption in the narrow wavelength regions around the doublets. We also include Rayleigh scattering from a background $\mathrm{H}_2$ atmosphere, but find this to be negligible and therefore do not include it for the present application. 

We compute the absorption cross section $\sigma(\lambda,T)$ as the sum of the individual D2 and D1 absorption lines. These lines are modeled as Voigt profiles using \texttt{scipy.special.wofz}. Line broadening is included in our model due to the gas temperature $T(r)$ (Doppler broadening). Since the absorbing atoms reach velocities significantly above thermal speeds for our three evaporative, non-hydrostatic scenarios, we must also consider this non-thermal Doppler broadening. We incorporate this broadening by treating the average velocity $\bar{v}_{\mathrm{i}}$ of the absorbing atoms as an effective line temperature:

\begin{equation}\label{lineTeqn}
    T(r)=\frac{\pi m_{\mathrm{i}}}{8 k_B}\bar{v}_{\mathrm{i}}(r)^2,
\end{equation}

where $m_{\mathrm{i}}$ is the atomic mass of the absorber. The line is then broadened using Voigt profiles, with a Doppler broadening parameter given by this line temperature for the three evaporative scenarios. We remark that the alkali D lines have strong sub-Lorentzian wings due to pressure broadening (e.g. \citealt{Heng2015}; \citealt{Allard2019}), which is an observable effect in low-resolution transmission spectroscopy (e.g. \citealt{Nikolov2018}). Since we investigate high-resolution spectra in this analysis, the probed pressures are comparatively lower. Upon verifying the influence of pressure broadening in our models, we find the effect negligible for our simulated spectra. In line with \citet{Heng2015}, we neglect pressure broadening throughout this work.

Second, we average all chords over the stellar disk to obtain the flux decrease due to a canonical atmosphere and more comprehensively, any exospheric source of absorption above the reference radius $R_0$, where $F_{out}$ denotes the flux out of transit and $F_{in}$ the flux during transit of the hot Jupiter:

\begin{equation}\label{flux}
    \Re:=\frac{F_{in, \lambda}}{F_{out, \lambda}}=\frac{2}{R_{\ast}^2-R_0^2}\int_{R_0}^{R_{\ast}}\!z\cdot e^{-\tau(z,\lambda)}\,\mathrm{d}z.
\end{equation}

This scheme is a general prescription for the computation of transmission spectra. To compare our calculations to observations WASP-49b (\citealt{Wyttenbach2017}) and HD189733b (\citealt{Wyttenbach2015}) we use a convolution, binning and normalization routine for all simulated transit spectra as in \citet{Pino2018} (see Section \ref{observation} for details). While our code is comparatively simple in terms of atmospheric chemistry, wavelength coverage, line broadening and absorbing species, it has the stark advantage that the number density profile can be arbitrarily defined without relying on the assumption of the atmosphere being in local hydrostatic equilibrium. This allows the simulation of transit spectra in two endogenic (hydrostatic \& escaping) and two exogenic scenarios (exomoon \& torus).

\begin{table}
	\centering
	\caption{System parameters of WASP-49b (\citealt{Wyttenbach2017}) and HD189733b (\citealt{Wyttenbach2015}), lifetimes of neutral sodium ($\tau_{\mathrm{Na}}$) are estimated from \citet{Huebner2015}. $\gamma$ denotes the systemic velocity of the exoplanetary systems, which is important for the comparison of the observational data to our simulated spectra.}
	\begin{tabular}{lcr} 
\hline
 Parameter& WASP-49b & HD189733b \\
\hline
$R_{\ast}$ [$R_{\odot}$]& 1.038&0.756\\ 

$R_0$ [$R_J$]& 1.198&1.138\\

$M_p$ [$M_J$]&0.399&1.138\\

$T_{\mathrm{eq}}$ [K]& 1400&1140\\

$\gamma$ [km/s]&  41.7&-2.28\\

$t_{\mathrm{Na}}$ [s] & 241&1010
	\end{tabular}
	\label{systemparameters}
\end{table}

\subsection{\texttt{DISHOOM}: Alkaline Mass Loss to Atmospheres and Exospheres}\label{dishoom}
We couple a semi-analytic metal mass loss model \texttt{DISHOOM} (Desorbing Interiors via Satellite Heating to Observing Outgassing Model) described comprehensively in \citet{Oza2019} Sections 3 (Jupiter's Atmospheric Sodium) and 4 (Jupiter's Exospheric Sodium) to \texttt{Prometheus} in order to simulate evaporative transmission spectra for the Na I and K I lines.

\subsubsection{Observing Outgassing During a Planetary Transit}\label{observingoutgassing}

An outgassing source undergoing mass loss can produce a prodigious quantity of foreground gas capable of generating a spectral signature during transit. Our escape model generalizes outgassing of planetary surfaces to include a wide-range of phenomena capable of producing spectral signatures as observed in the solar system. Escape of the so-called supervolatiles N$_2$, CH$_4$, and CO is included (see \citealt{Johnson2015}), as well as water-products H$_2$O, O$_2$ (see \citealt{Oza2019A}) where the mechanisms leading to outgassing and eventual escape include solar heating, magnetospheric ion sputtering, and general space weathering. For close-in systems, like the ones studied here, nearly molten temperatures lead to volcanic and magma-products due to tidal heating or direct sublimation of grains (see Section \ref{metallicmassloss} below). For a volcanically-active system Na \& K are products of the parent molecules NaCl and KCl as observed at Io. The close-in systems are jeopardous in that the atomic lifetimes are strongly limited by photoionization. Following \citet{Huebner2015} we can estimate the atomic lifetime for a species i as $t_{\mathrm{i}} \sim \frac{1}{k_{\mathrm{i}}}$ where $k_{\mathrm{i}}$ is the photoionization rate coefficient in s$^{-1}$:

\begin{equation}\label{ratecoeff}
  k_{\mathrm{i}, \lambda} = \int_{\lambda}^{\lambda + \Delta \lambda } \!\sigma_{\gamma}(\lambda) \Phi (\lambda)\, \mathrm{d}\lambda,
\end{equation}

in the wavelength of interest $\lambda + \Delta \lambda$, a photoionization cross section $\sigma_{\gamma}(\lambda)$ and the spectral photon flux:
\begin{equation}\label{ratecoeff2}
\Phi (\lambda) = \frac{2 \pi c}{4 \lambda^4 [\exp(hc/ \lambda k_B T) -1 ]},
\end{equation}

for a Blackbody of temperature $T$ (e.g. \citealt{RybickiLightman1979}). As we describe in Section \ref{metallicmassloss}, photoionization timescales provide a critical lower limit to the atomic lifetime, as recombination processes due to the ambient plasma could boost these lifetimes. Independent of the mass loss mechanism, we can write the general mass loss rate for a source venting a species i depending on the number of atoms $\mathcal{N}_{\mathrm{i}}$: 

\begin{equation}\label{Mdot}
   \dot{M}_{\mathrm{i}}\approx   m_{\mathrm{i}} k_{\mathrm{i}}  \mathcal{N}_{\mathrm{i}} \approx  m_{\mathrm{i}} t_{\mathrm{i}}^{-1}  \mathcal{N}_{\mathrm{i}} .
\end{equation}

The number of evaporating atoms $\mathcal{N}_{\mathrm{i}}$ is then fed into the number density profiles for each evaporative mechanism, enabling a straightforward computation of an evaporative transmission spectrum.

\subsubsection{Metallic Mass Loss}\label{metallicmassloss}

In addition to the evaporation of Na\,I and K\,I, the code is equipped to estimate the destruction of rocky bodies of arbitary composition (e.g. MgSiO$_3$; Fe$_2$SiO$_4$). This description of atmospheric loss holds observational relevance given the recent explosion of heavy metal detections at gas giants (\citealt{Hoeijmakers2018}, \citealt{Hoeijmakers2019}; \citealt{Sing2019}; \citealt{Cabot2020}; \citealt{Gibson2020}; \citealt{Hoeijmakers2020}). In our study of the alkali metals, we shall use chondritic ratios based off of \citet{FegleyZolotov2000} for Na and K, and rely on observations by \citet{Lellouch2003} and \citet{Postberg2009} to constrain the Na I abundance when estimating mass loss rates due to thermal evaporation of silicate grains (i.e. Section \ref{torus}).

In practice, the metal evaporation code computes different regimes of escape (e.g. thermal, nonthermal) due to several heating mechanisms (e.g. XUV, tidal heating) for a close-in irradiated body. The dominant mass loss rate is then either supplied to \texttt{Prometheus} to generate a forward-model transit spectrum, or compared to retrieved mass loss rates and velocities in the inverse modeling as a plausibility check if the different evaporative scenarios can indeed provide the required absorber source rates. The dominant mass loss mechanism varies between the different evaporative scenarios. We show the equations we use to estimate $\dot{M}_{\mathrm{i}}$ in the respective subsection of the scenario (Sections \ref{escaping} to \ref{torus}). Generally, as data regarding the chemistry and plasma conditions are largely unknown at an exoplanetary system at present, we will find it useful to use scaling relations to solar system bodies observed in-situ (see \citealt{Gronoff2020} and references therein for a full review on escape from solar system and exoplanet bodies).

In our heuristic model, we provide upper limits to the required mass loss rate based on the assumption that photoionization is the dominant process regulating the alkali lifetime. While to first order this is valid, two of our evaporative scenarios (Exomoon: \ref{exomoon} \& Torus: \ref{torus} described below) are directly analogous to the radiation environment of a gas giant magnetosphere. As observed and simulated in the Jupiter-Io plasma torus system, recombination and charge exchange could considerably extend the net lifetime of the alkali atoms (see \citealt{Wilson2002} and references therein). Radiative recombination is additionally important for a close-in magnetosphere as ions sourced by photo or electron-impact ionization of the evaporating atoms have the ability to accumulate in the magnetosphere so long as advection is small. We find, similar to \citet{Vidal-Madjar2013} for Mg at HD209458b, that $n_e \sim 10^8$ cm $^{-3}$ are required for electronic recombination of Na. In a toroidal B-field, we remark that ion-recombination and charge exchange can be effective to source the extended alkali clouds we describe here. In the absence of a toroidal B-field, a conservative ion density assuming charge neutrality based on the simulations by \citealt{Dwivedi2019}, ion densities of $\sim 10^7 - 10^8\,\mathrm{cm}^{-3}$ were simulated up to $\sim 3 R_0$ validating a viable recombination mechanism at distances corresponding to our exogenic evaporating scenarios (Sections \ref{exomoon} \& \ref{torus}). Furthermore, we find that for the alkali atoms studied here, loss rates due to radiation pressure are small mostly due to the domineering ionization rates discussed above. Balancing the radiation pressure force with that of gravitation (e.g. \citealt{Vidal-Madjar2003}; \citealt{Murray-Clay2009}), upper limits to the loss rate due to radiation pressure  $\dot{M}_{\mathrm{rad.press.}} \sim 10 - 10^4\,$kg/s are found for the range of sodium column densities studied in Section \ref{physics}. In Tables \ref{plasmaprocesses} and \ref{velocities} we qualitatively present the plasma processes regulating our evaporative scenarios and the average speeds of the alkali atoms $\bar{v}_{\mathrm{i}}$. 

The simple coupling via mass loss rate and Equation \ref{Mdot} we perform is to demonstrate the dramatic influence of mass loss and number densities on an alkaline transmission spectrum. We remark that more robust hydrodynamic codes focusing on individual planets are especially suited for this problem, as has been shown in Ly $\alpha$ at HD189733b (\citealt{BourrierLecavelier2013}; \citealt{Christie2016}), H$\alpha$, H$\beta$, and H$\gamma$ at KELT-9b (\citealt{Wyttenbach2020}), Mg I at HD209458b (\citealt{Bourrier2015}), Mg II at WASP-12b (\citealt{Dwivedi2019}) and He I at HD209458b (\citealt{OklopvcicHirata2018}; \citealt{Lampon2020}). 


\begin{table}
	\centering
	\caption{Sodium plasma processes.}
	\label{plasmaprocesses}
	\begin{tabular}{lr}
\hline
Plasma Process & Reaction \\
\hline
Photoionization & Na + $\gamma \rightarrow$ Na$^{+} + e^{-}$ \\ 
Recombination & Na$^{+} + e^{-} \rightarrow$ Na + $\gamma$ \\ 

Charge Exchange & Na$^{+}$ + Na $\rightarrow $Na$^{*}$ + Na$^{+}$  \\ 

Dissociative Recombination & NaX$^{+} + e^{-} \rightarrow$ Na$^{*}$ + X$^{*}$ \\ 

	\end{tabular}
\end{table}

\begin{table}
	\centering
	\caption{Sodium velocity distributions due to plasma-driven escape of Na at an exo-Io or exo-torus. The nominal velocity distribution gives the range of velocities based on thermal and nonthermal flux distributions used to model the analogous physical process in the solar system. If an example is available the reference is provided: 
	$^a$ \citet{Smyth1988}; \citet{Smyth1992};
	$^b$ \citet{Wilson2002};
	$^c$ \citet{Cassidy2009}.}
	\label{velocities}
	\begin{tabular}{lr}
\hline
Physical Process & Nominal velocity distribution \\
(Analog) & $\bar{v}$ \\
\hline
Atmospheric Sputtering$^{a,b}$ & 1 - 30 km/s \\ 
(Atmospheric Escape at Io) & 10 km/s\\ \hline
Charge Exchange$^{b}$ & 30 - 100 km/s \\ 
(Sodium Exosphere at Jupiter) & 60 km/s\\ \hline
Pickup Ions$^{b}$& 10 - 60 km/s \\ 
(Io plasma torus) & 74 km/s\\ \hline
Resonant Orbit$^{c}$ & 10-42 km/s \\
(Io motion) &  17 km/s \\ \hline
Thermal & 0.6 - 2 km/s \\
(Volcano temperature)  & 1.4 km/s \\ \hline 
	\end{tabular}
\end{table}


\subsection{Hydrostatic scenario}\label{hydrostatic}

\begin{figure}
    \centering
    \includegraphics[width=\columnwidth]{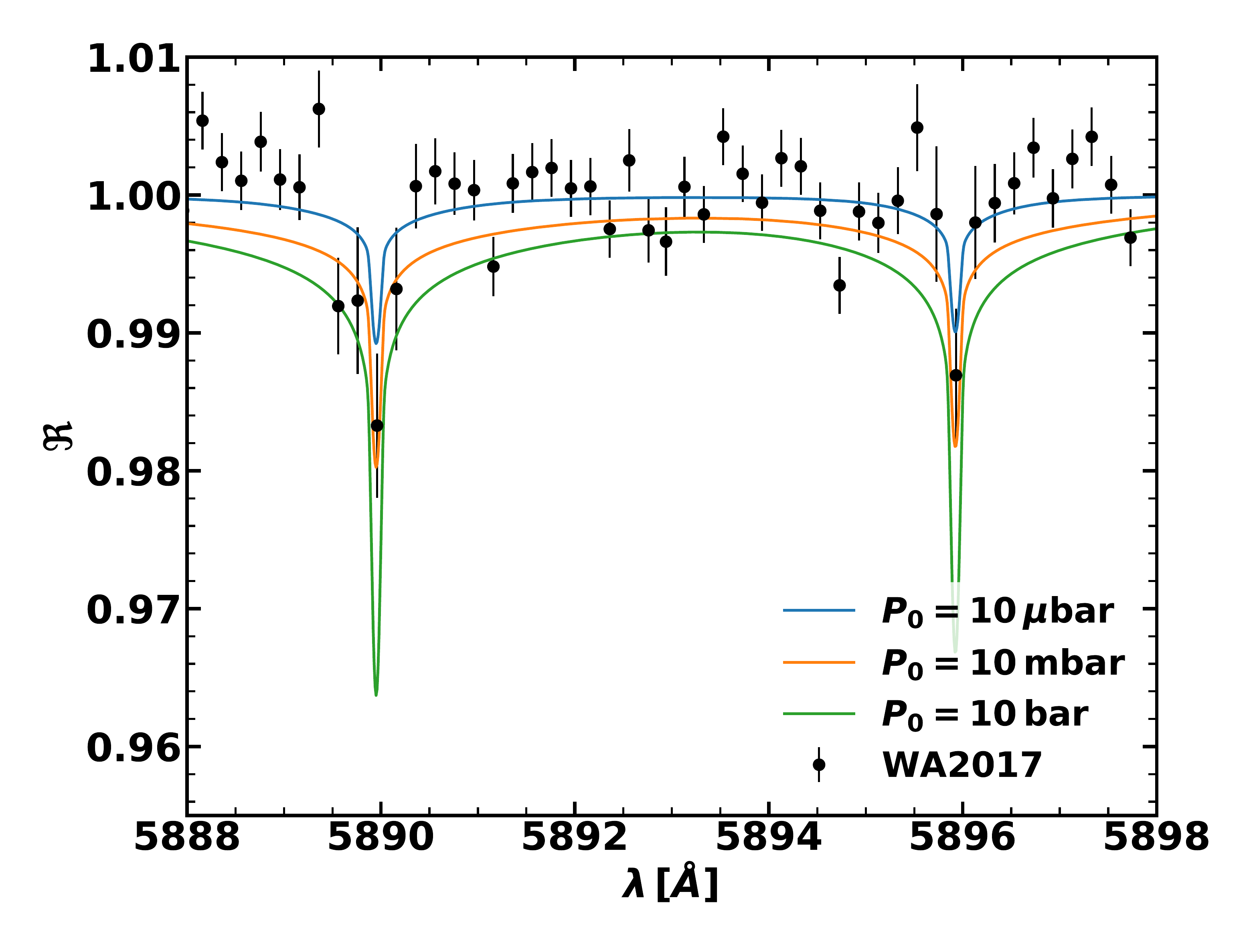}
    \caption{Simulation of high-resolution transit spectra in the sodium doublet, for different reference pressures in the hydrostatic scenario. The shape of the transit spectra isn't determined by $P_0$ but by the partial pressure of sodium at the reference radius, $P_{0,\mathrm{Na}}$. For better readability we fix $\chi_{\mathrm{Na}}$ and vary $P_0$. We set $\chi_{\mathrm{Na}}=1.7\,$ppm (the solar mixing ratio) and $T=3000\,$K.  The planetary parameters and data points are for WASP-49b, taken from \citet{Wyttenbach2017}.}
    \label{hydVARP}
\end{figure}

\begin{figure}
    \centering
    \includegraphics[width=\columnwidth]{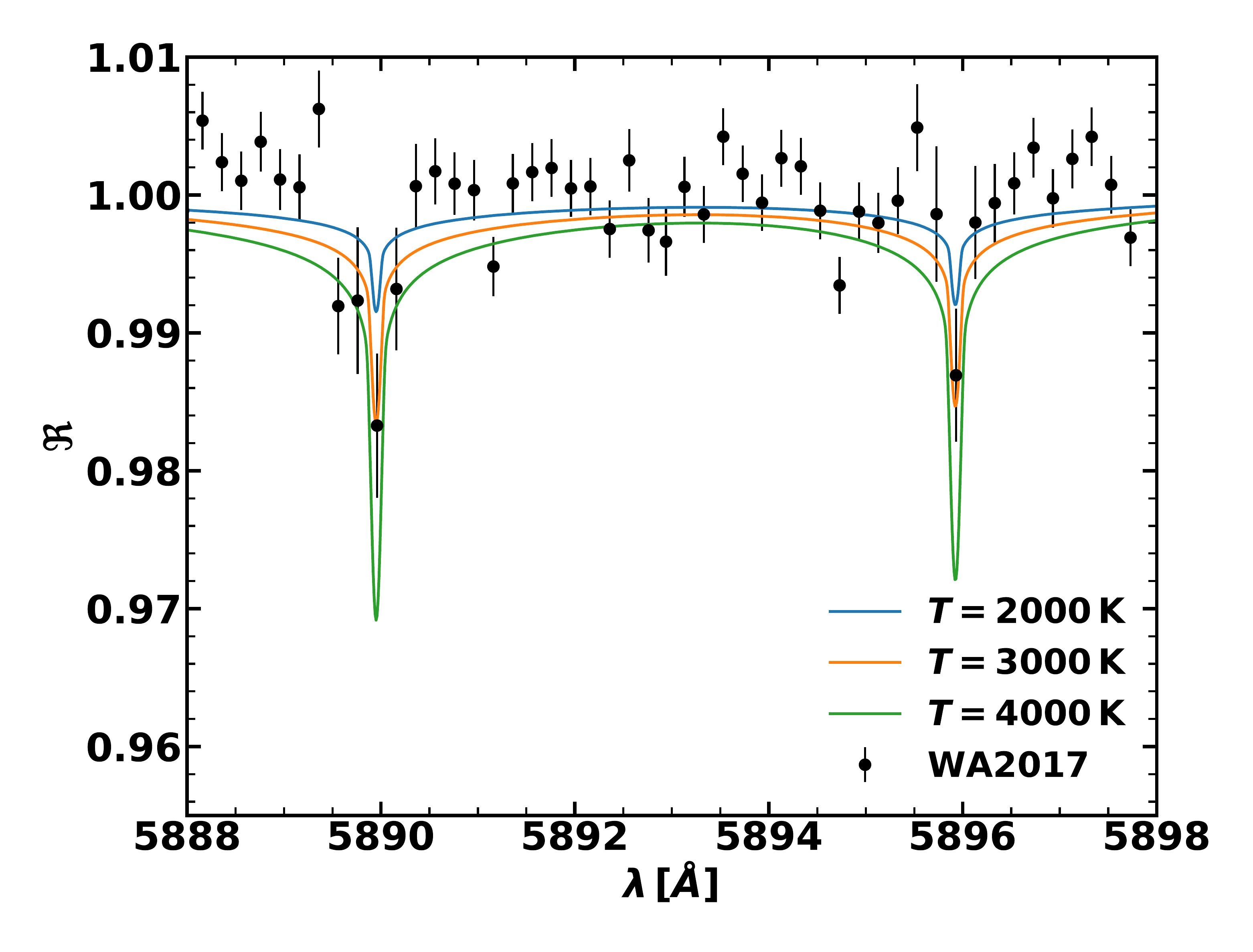}
    \caption{Simulation of high-resolution transit spectra in the sodium doublet, for different temperatures in the hydrostatic scenario. We set $\chi_{\mathrm{Na}}=1.7\,$ppm (the solar mixing ratio) and $P_0=1\,$mbar. The planetary parameters and data points are for WASP-49b, taken from \citet{Wyttenbach2017}.}
    \label{hydVART}
\end{figure}

This is the canonical scenario for transmission spectra. The number density profile is straightforward to derive using the definition of the pressure scale height: $H(r) = \frac{k_B T(r)}{\mu(r) g(r)}$, where $\mu(r)$ is the mean molecular weight and $g(r)$ the local gravitational acceleration. Solving the equation of hydrostatic equilibrium $\mathrm{d}P(r)/\mathrm{d}r = -n(r) \mu(r) g(r)$ for number density, $n(r)$, yields the hydrostatic number density profile:

\begin{equation}\label{n_hyd_general}
n_{\mathrm{hyd}} (r) = n_0\cdot\frac{T_0}{T(r)}\cdot\exp\Bigl[-\!\int_{R_0}^r\!\frac{\mathrm{d}r'}{H(r')}\Bigr],
\end{equation}

where $R_0$ is a fixed reference radius, and quantities with subscript zero are evaluated at this reference radius. Although our code can process arbitrary temperature and mixing ratio profiles, we restrict ourselves to isothermal and vertically-mixed (i.e. with constant mixing ratios of the neutral absorber throughout the atmosphere) models, as we want to isolate the effect of varying the number density profiles within our different scenarios. Therefore, $T(r)$ and $\mu(r)$ are constant and the number density profile in Equation \ref{n_hyd_general} can be integrated:

\begin{equation}\label{n_hyd}
n_{\mathrm{hyd}} (r) = n_0\cdot\exp\bigl(\lambda(r)-\lambda_0\bigr),
\end{equation}

where $\lambda(r)$ is the Jeans parameter given by $\lambda(r)=\frac{GM_p \mu }{r k_BT}$. We use the canonical value of $\mu=2.3\,\mathrm{amu}$ (corresponding to a \textit{mass} mixing ratio $x_{\mathrm{He}} = 0.25$) for our hydrostatic model. Using the number density profile of Equation \ref{n_hyd} with constant temperature and mixing ratios throughout the atmosphere leads to three free parameters in this scenario: temperature $T$, pressure at reference radius $P_0$ and absorber mixing ratio $\chi_{\mathrm{i}}$. However, $P_0$ and $\chi_{\mathrm{i}}$ are mutually degenerate in our model since we neglect pressure broadening, collision-induced absorption and other absorbers (see \citealt{Heng2017} and \citealt{Welbanks2019} for a detailed discussion on this degeneracy). Therefore, we have the partial pressure of the absorber at the reference radius ($P_{0,\mathrm{i}}=P_0\cdot\chi_{\mathrm{i}}$) as a second free parameter. We show how these two parameters, $T$ and $P_{0,\mathrm{i}}$, affect the transmission spectrum for a hot Jupiter with planetary parameters of WASP-49b in Figures \ref{hydVARP} and \ref{hydVART}. We will find it useful to define auxiliary parameters which are easier to interpret than the free parameters, but don't fundamentally affect the transmission spectrum. For the hydrostatic scenario we use $P_0$ as an auxiliary parameter (e.g. in Figure \ref{hydVARP}), calculated from $P_{0,\mathrm{i}}$ under the assumption of a specific absorber mixing ratio. A summary of free, auxiliary and fixed parameters for all scenarios can be found in Table \ref{retrievalparameters}.

\subsection{Escaping scenario}\label{escaping}

Hot Jupiters are observed to have escaping planetary winds (e.g. \citealt{Vidal-Madjar2003}). Currently, there are significant modeling efforts to simulate escaping atmospheres by either solving the Navier-Stokes equations or more computationally expensive yet robust, by directly solving the Boltzmann equation on a particle to particle basis (see \citealt{Gronoff2020} for both approaches to atmospheric escape). While atmospheric escape is not the prime focus of the current study, we emphasize that the latter kinetic model is sorely needed at present for a full description of an exoplanet exosphere. Nevertheless, the hydrodynamical approach of escaping atmospheres has been applied to different hot Jupiters, computing pressure, temperature and velocity profiles of the wind. \citet{Murray-Clay2009} found that at pressures $P>1\,\mathrm{nbar}$, the atmosphere can still be treated as hydrostatic, while at lower pressures the atmospheric structure strongly deviates from a hydrostatic profile. As a first-order approximation for the number density profile we use a power law with $q_{\mathrm{esc}} = 6$ for our analysis, which is roughly in line with the profiles found for WASP-49b (\citealt{Cubillos2017}) and HD209458b (\citealt{Murray-Clay2009}).

\begin{equation}\label{n_esc}
    n(r)_{\mathrm{esc}} = n_0 \Bigl ( \frac{R_0}{r}\Bigr )^{q_{\mathrm{esc}}}.
\end{equation}

 The escape index $q_{\mathrm{esc}} = 6$ is indicative of ion-neutral scattering for an escaping neutral gas interacting with a plasma (\citealt{Johnson1990}; \citealt{Johnson2006O2}). The reference radius $R_0$ corresponds to the base of the wind in this scenario. As we want to isolate the different scenarios we neglect the hydrostatic layer below the base of the wind for the computation of the transit spectra. Since the hydrodynamical simulations generally treat only hydrogen and helium we don't know the mixing ratio profile of the absorber, $\chi_{\mathrm{i}}(r)$. As in the hydrostatic scenario we set this profile to a constant value, meaning that we encounter a similar degeneracy between $n_0$ and $\chi_{\mathrm{i}}$ as in the hydrostatic scenario\footnote{We encountered this degeneracy in the hydrostatic setting between $P_0$ and $\chi_{\mathrm{i}}$, as our knowledge of the temperature allowed for a direct conversion between pressures and number densities. As we don't know the temperature in the escaping wind we formulate the degeneracy in this scenario as one between $n_0$ and $\chi_{\mathrm{i}}$.}. Hence we have the number density of the absorber at the base of the wind, $n_{0,\mathrm{i}}$, as a free parameter. We can convert this quantity into the total number of absorbing atoms as the volume integral of Equation \ref{n_esc}: 

\begin{equation}\label{N_esc} 
\begin{aligned}
    \mathcal{N_{\mathrm{i}}}= \iiint_V \!\chi_{\mathrm{i}}\cdot n_{\mathrm{esc}}(r)\, \mathrm{d}V  &=4\pi\chi_{\mathrm{i}}\cdot\int_{R_0}^{\infty}\!r^2\cdot n_{\mathrm{esc}}(r)\, \mathrm{d}r \\ &=\left ( \frac{4\pi}{q_{\mathrm{esc}}-3}\right ) n_{0,\mathrm{Na}}R_0^3.
\end{aligned}
\end{equation}

\begin{figure}
    \centering
    \includegraphics[width=\columnwidth]{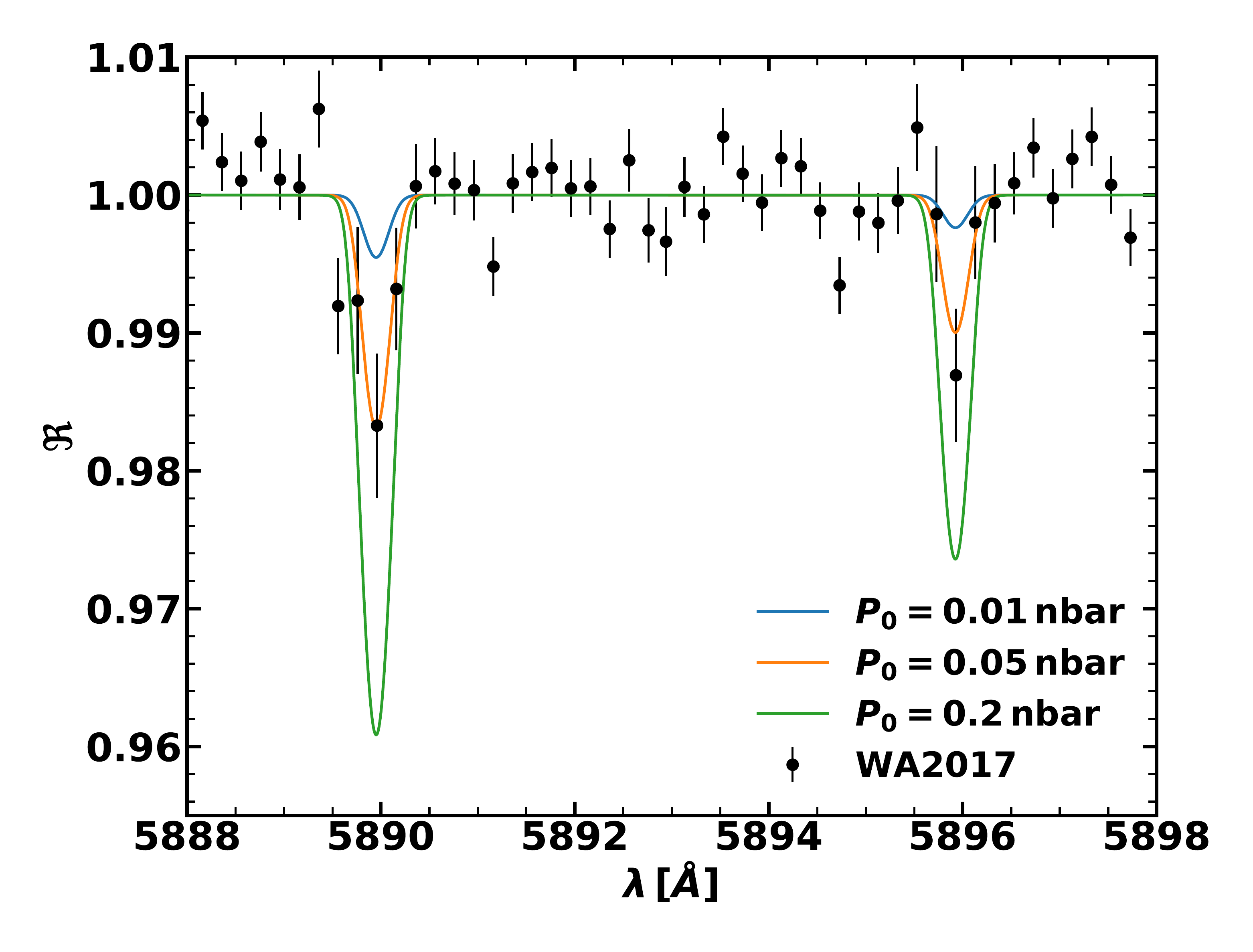}
    \caption{Simulation of high-resolution transit spectra in the sodium doublet, for different reference pressures in the escaping scenario. The shape of the transit spectra is determined by $\mathcal{N}_{\mathrm{Na}}$, but for better readability we show the auxiliary parameter $P_0$. We use the planetary equilibrium temperature (Table \ref{systemparameters}) and a solar mixing ratio of $\chi_{\mathrm{Na}}=1.7\,$ ppm to calculate the reference pressures. We fix $\bar{v}_{\mathrm{Na}}=10\,$km/s in this Figure. The planetary parameters and data points are for WASP-49b, taken from \citet{Wyttenbach2017}.}
    \label{escVARP}
\end{figure}

Given that $R_0$ and $q_{\mathrm{esc}}$ are fixed we can use $\mathcal{N}_{\mathrm{i}}$ and $n_{0,\mathrm{i}}$ interchangeably as free parameters. Since we use $\mathcal{N}_{\mathrm{i}}$ to compute the required source rate, comprising the coupling between our codes (Equation \ref{Mdot}), we choose the total number of absorbing atoms in the systems as a free parameter. Since this quantity is difficult to interpret physically, we use the pressure at the base of the wind as auxiliary parameter. Assuming a certain absorbing mixing ratio and temperature we calculate this pressure as $P_0=n_{0,\mathrm{i}}/\chi_{\mathrm{i}}\cdot k_BT$. We expect photoionization of the alkali metals to be significant in the escaping scenario. We remark here that while photoionization probably affects $\chi_{\mathrm{i}}(r)$, it doesn't change the validity of setting $\chi_{\mathrm{i}}(r)$=const. as long as the ionized fraction is constant in the escaping wind. However, the retrieved value of $\mathcal{N}_{\mathrm{i}}$ can't be accurately converted into $P_0$ as we don't know the mixing ratio of the neutral absorber, $\chi_{\mathrm{i}}$. We show how $P_0$ affects the transmission spectrum for a hot Jupiter with planetary parameters of WASP-49b in Figure \ref{escVARP}. Apparently, the escaping transit spectrum depends strongly on the value of $P_{0}$ (compared to the hydrostatic transit spectrum, Figure \ref{hydVARP}). We elucidate this behavior in Section \ref{Forward Modeling}.

\begin{figure}
    \centering
    \includegraphics[width=\columnwidth]{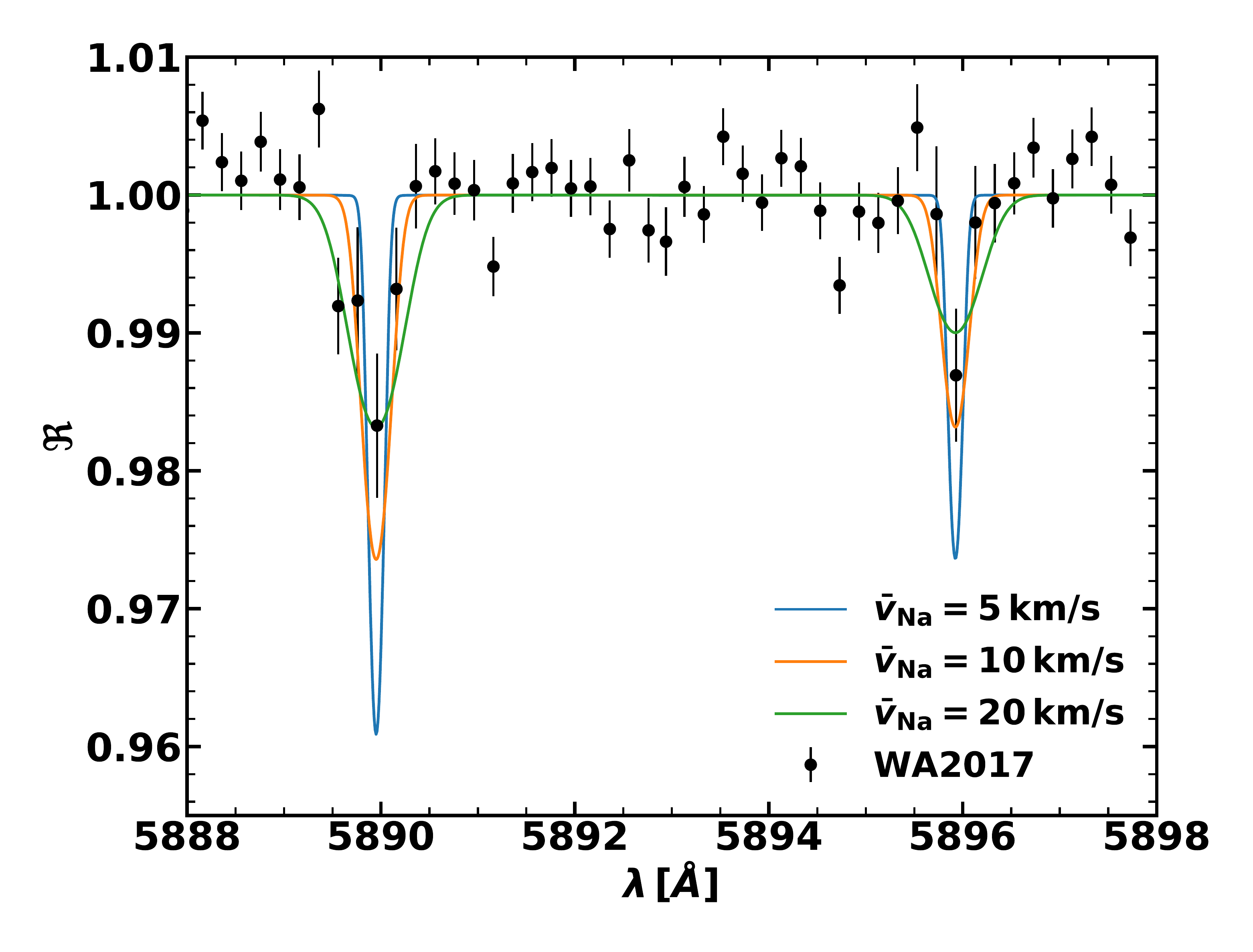}
    \caption{Simulation of high-resolution transit spectra in the sodium doublet, for different $\bar{v}_{\mathrm{Na}}$ in the escaping scenario. We set $\mathcal{N}_{\mathrm{Na}}=2.16\cdot10^{33}$, corresponding to a pressure at the base of the wind of $P_0=0.1\,$nbar (using the planetary equilibrium temperature from Table \ref{systemparameters} and $\chi_{\mathrm{Na}}=1.7\,$ppm). The planetary parameters and data points are for WASP-49b, taken from \citet{Wyttenbach2017}.}
    \label{escVARv}
\end{figure}

The speeds of the absorbing atoms in escaping winds greatly exceed the thermal speeds. As described in Section \ref{Prometheus}, we use the wind speed to calculate a line temperature, which is then treated as thermal Doppler broadening of the line. We simplify the velocity profile of the escaping wind to a constant value (as in the hydrostatic model with vertical winds of \citealt{Seidel2020}). We therefore employ the mean speed of the absorbing atoms in the escaping wind, $\bar{v}_{\mathrm{i}}$, as second free parameter. The impact of this parameter on a transmission spectrum for a hot Jupiter with planetary parameters of WASP-49b is shown in Figure \ref{escVARv}. We note that line broadening in an escaping wind consists of both the broadening due to the stochastic, thermal motions and of the wavelength shifts according to the line-of-sight wind velocities (\citealt{Seidel2020}). Our simplified treatment of line broadening consists only of the latter effect (but treating the wind velocity as thermal speeds), since we expect this to be the dominant source of line broadening. Nevertheless, the inclusion of the broadening due to the inherent thermal motions of the escaping gas would alter the Doppler core of the Voigt profile and thereby also the transmission spectrum, which could become important for lower wind speeds.


We estimate the mass loss rate in this evaporative scenario using \texttt{DISHOOM}. Hydrodynamically escaping atmospheres have two escape regimes: radiation-recombination limited (\citealt{Murray-Clay2009}) and energy-limited escape (\citealt{Watson1981}), where the latter regime is found to be accurate within a factor of 1.1 for a $1\,M_J$ planet and within a factor of $\sim 1/4$ for a $0.3\,M_J$ planet (\citealt{Allan2019}). Here we choose to crudely estimate the mass loss rate of the escaping absorber using the energy-limited approximation to hydrodynamic escape (e.g. \citealt{Johnson2013}; described as $\dot{M}_U$ in \citealt{Oza2019}):

\begin{equation}\label{mdotELescape1}
    \dot{M}_{\mathrm{esc,i}} \sim x_{\mathrm{i}}m_{\mathrm{i}} \frac{Q}{U},
\end{equation}
where $Q$ is the XUV-heating rate of the upper atmosphere given a heating efficiency between 0.1 - 0.4, $U$ is the binding energy of the atmosphere, $x_{\mathrm{i}}$ the mass mixing ratio for the atom with mass $m_{\mathrm{i}}$. For sodium, we use $x_{\mathrm{Na}}=1.7\times10^{-5}$, which corresponds to the solar volumetric mixing ratio of sodium of 1.7\,ppm, multiplied by $m_{\mathrm{Na}}/\mu$ (with $\mu=2.3\,$amu).
This mass loss rate, together with the required source rate estimated from equation \ref{Mdot}, comprises the coupling between \texttt{DISHOOM} and \texttt{Prometheus} for the escaping scenario. We note that the upper limits we use here are to examine the extremeties of the evaporative wind scenario. As the escaping atmosphere passes the exobase, the energy-limited approximation used here will overestimate the source rate (\citealt{Johnson2013}). Furthermore, as hot Jupiters are expected to have strong magnetic fields (\citealt{YadavThorngren2017}), the mass loss rates may be far less as pointed out by \citet{Christie2016} based on the MHD simulations of \citet{Trammell2011}; \citet{Trammell2014ApJ}; \citet{OwenAdams2014} as well as charge-exchange and hydrodynamical simulations by \citet{TremblinChiang2013}.

\subsection{Exomoon scenario}\label{exomoon}

As observed in the Jupiter-Io system, a tidally-heated satellite can outgas significant amounts of particles, especially neutral sodium, via volcanism. For an exo-Io orbiting a hot Jupiter, the outgassing is further enhanced by sublimating the silicate surface (see Eqn. \ref{mdotsub}). At Io the observed volcanic escape is not due to outgassing however, but sublimation \textit{and} volcanic outgassing (see Eqn. 5 in \citealt{Oza2019} and surrounding text)  coupled with plasma-driven escape to space. We shall therefore refer to the observed gas near the satellite as atmospheric sputtering (\citealt{haff81}; \citealt{Johnson2004}), although outgassing is often used interchangeably to describe the satellite itself. To first order, we approximate the number density profile of the absorber by scaling $n_{\mathrm{i}}(r)=n(r)\cdot\chi_{\mathrm{i}}(r)$ to the sputtered Na I number density profile observed at Io (\citealt{Burger2001}) with a power law exponent of $q_{\mathrm{moon}}=3.34$: 

\begin{equation}\label{n_out}
    n_{\mathrm{moon,i}}(r)=n_{0,\mathrm{i}}\cdot \bigl(\frac{R_s}{r}\bigr)^{q_{\mathrm{moon}}},
\end{equation}

where $R_s$, the radius of the satellite, is set to Io's radius. Since we directly calculate the number density of the absorber in this scenario we can drop the mixing ratio profile $\chi_{\mathrm{i}}(r)$ for the computation of the optical depth along the chord in Equation \ref{tau}. As in the escaping scenario we use the total number of the absorbing atoms in the system as a free parameter. We compute this quantity analogously to Equation \ref{n_esc} in the escaping scenario:

\begin{equation}
\mathcal{N_{\mathrm{i}}}=\iiint_V \,n_{\mathrm{i}}(r) \mathrm{d}V=\left ( \frac{4\pi}{q_{\mathrm{moon}}-3}\right ) n_{0,\mathrm{i}}R_s^3 .
\end{equation}

\begin{figure}
    \centering
    \includegraphics[width=\columnwidth]{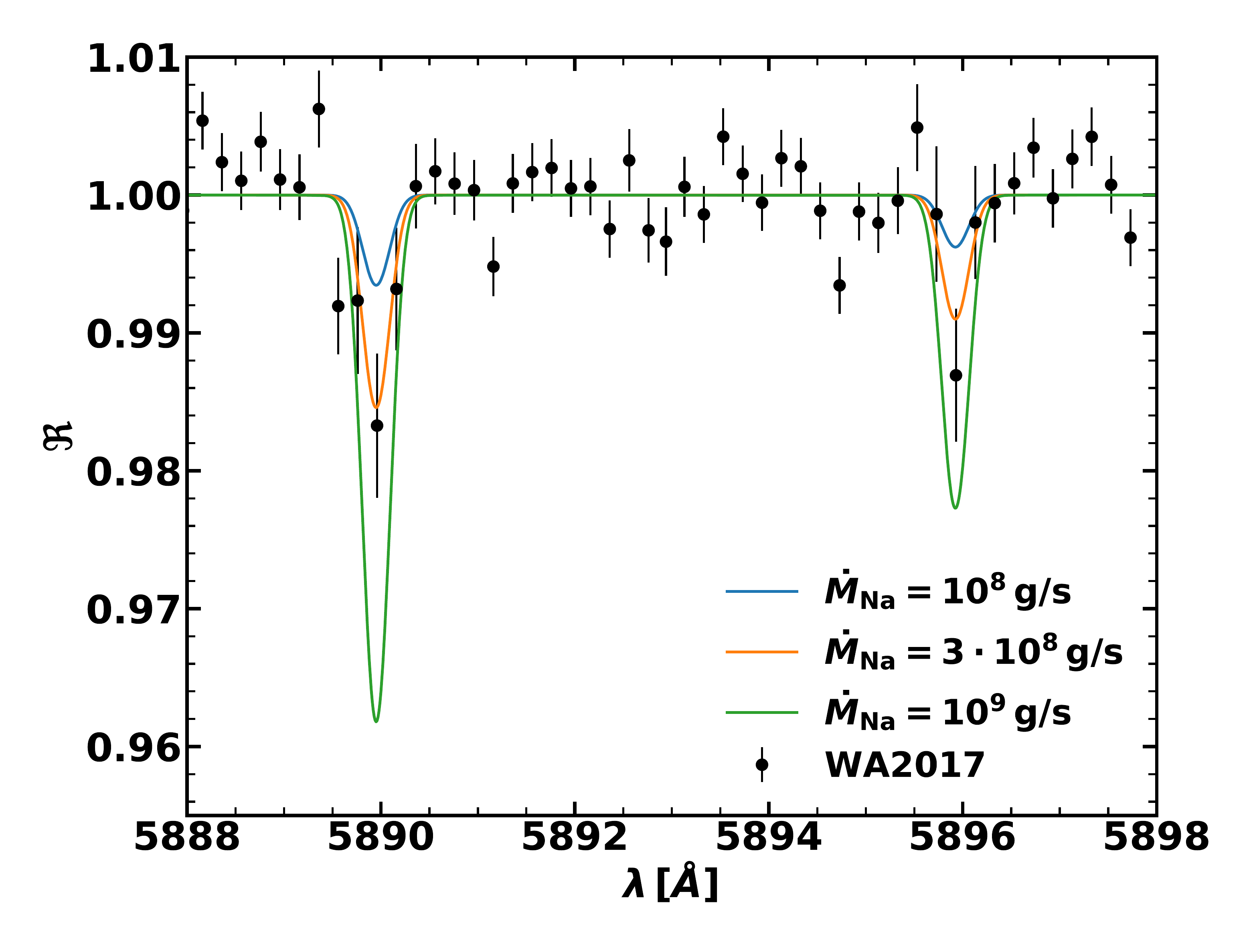}
    \caption{Simulation of high-resolution transit spectra in the sodium doublet, for different mass loss rates of the sodium atoms in the exomoon scenario. We set $\bar{v}_{\mathrm{Na}}=10\,$km/s. The planetary parameters and data points are for WASP-49b, taken from \citet{Wyttenbach2017}.}
    \label{outgassing_mdots}
\end{figure}

Note that the gas in such a sputtered cloud isn't in local thermodynamic equilibrium, therefore we don't use pressure as an auxiliary parameter. Instead, we convert $\mathcal{N_{\mathrm{i}}}$ into a source rate using Equation \ref{Mdot}. Transit spectra for different values of this mass loss rate are shown in Figure \ref{outgassing_mdots}. As we want to isolate the different scenarios we neglect the planetary atmosphere in this scenario, and consider only the sputtered cloud from the satellite for the computation of the transit spectrum. For computational convenience, we place the satellite in the center of our coordinate system (Figure \ref{Geometry}) to exploit the spherical symmetry of the system. The reference radius in this scenario corresponds to the surface of the satellite, $R_0=R_s=R_{\mathrm{Io}}$.

We will again refer to the speeds of atoms instead of temperatures in this scenario as $T$ isn't well-defined, moreover irrelevant for an exosphere/collisionless gas. Similar to the escaping scenario, we describe the particles with a line temperature calculated from the mean velocity of the absorbing atoms (which we assume to be constant throughout the system). Therefore, we have the mean velocity of the absorbing atoms in the system, $\bar{v}_{\mathrm{i}}$, as our second free parameter.\\

We calculate the plasma-driven mass loss rate of the absorber within \texttt{DISHOOM} using the scaling parameters $\mathcal{P} = \frac{P_s}{P_{\mathrm{Io}}}$ = $f_{\mathrm{mag}}$ + $f_{\mathrm{ram}}$, $\mathcal{V} = \frac{v_s}{v_{\mathrm{Io, torus}}}$, $\mathcal{U} = \frac{U_s}{U_{\mathrm{Io}}}$, $\mathcal{R}_x = \frac{\mathcal{R}_{x,s}}{\mathcal{R}_{x,\mathrm{Io}}}$ describing the total plasma pressure, relative ion velocities, binding energies, and exobase altitudes respectively. We can therefore express $\dot{M}_P$ (Eqn. 7 of \citealt{Oza2019}) more tractably in terms of a scaled plasma-heating rate $\propto \mathcal{P} \mathcal{V} \mathcal{R}_x^2$ as: 

\begin{equation}\label{mdotplasma}
\begin{aligned}
  \dot{M}_{\mathrm{moon,i}} &\sim x_{\mathrm{i}} \frac{\mathcal{P} \cdot \mathcal{V}}{\mathcal{U}} \mathcal{R}_x^2  \dot{M}_{\mathrm{Io}} \\
  &\sim \frac{x_{\mathrm{i}} } {\mathcal{U}} \left ( \frac{B_r^2}{2 \mu_0} P_{\mathrm{mag,Io}}^{-1} + n_{\mathrm{i}} m_{\mathrm{i}} u_{\mathrm{i}}^2 P_{\mathrm{ram,Io}}^{-1} \right ) \mathcal{V} \mathcal{R}_x^2   \dot{M}_{\mathrm{Io}}.
\end{aligned}
\end{equation}

The dominant components of the total plasma pressure are the magnetic pressure, where $P_{\mathrm{mag,Io}} = 1.5 \times 10^{-5}\,\mathrm{dynes\,cm}^{-2}$, dependent on the magnetic field strength at the satellite radius ($B_r$) given $\mu_0$ the permeability of free space . The ram pressure is also critical, where $P_{\mathrm{ram,Io}} = 2.6 \times 10^{-6}\,\mathrm{dynes\,cm}^{-2}$, dependent on $n_{\mathrm{i}}$, $m_{\mathrm{i}}$, and $u_{\mathrm{i}}$ the ion number density, mass, and velocity respectively. We adopt a mass mixing ratio with respect to SO$_2$, $x_{\mathrm{Na}} = 0.1$, corresponding to the that used for an exo-Io; for a full description on extrasolar atmospheric sputtering see Section 4.2.1 in \citet{Oza2019}. As \texttt{DISHOOM} does not explicitly track the ions that drive escape we use the velocity distributions modeled explicitly for Io by  \citet{Smyth1988}; \citet{Smyth1992} to constrain the velocities of the sodium gas sputtered from our satellite. The velocities range between 2-30 km/s, whereas speeds approaching 100 km/s are possible due to charge exchange. A summary of the plasma processes and mean velocities associated with various processes are presented in Tables \ref{plasmaprocesses} and \ref{velocities} respectively. 


\begin{figure}
    \centering
    \includegraphics[width=\columnwidth]{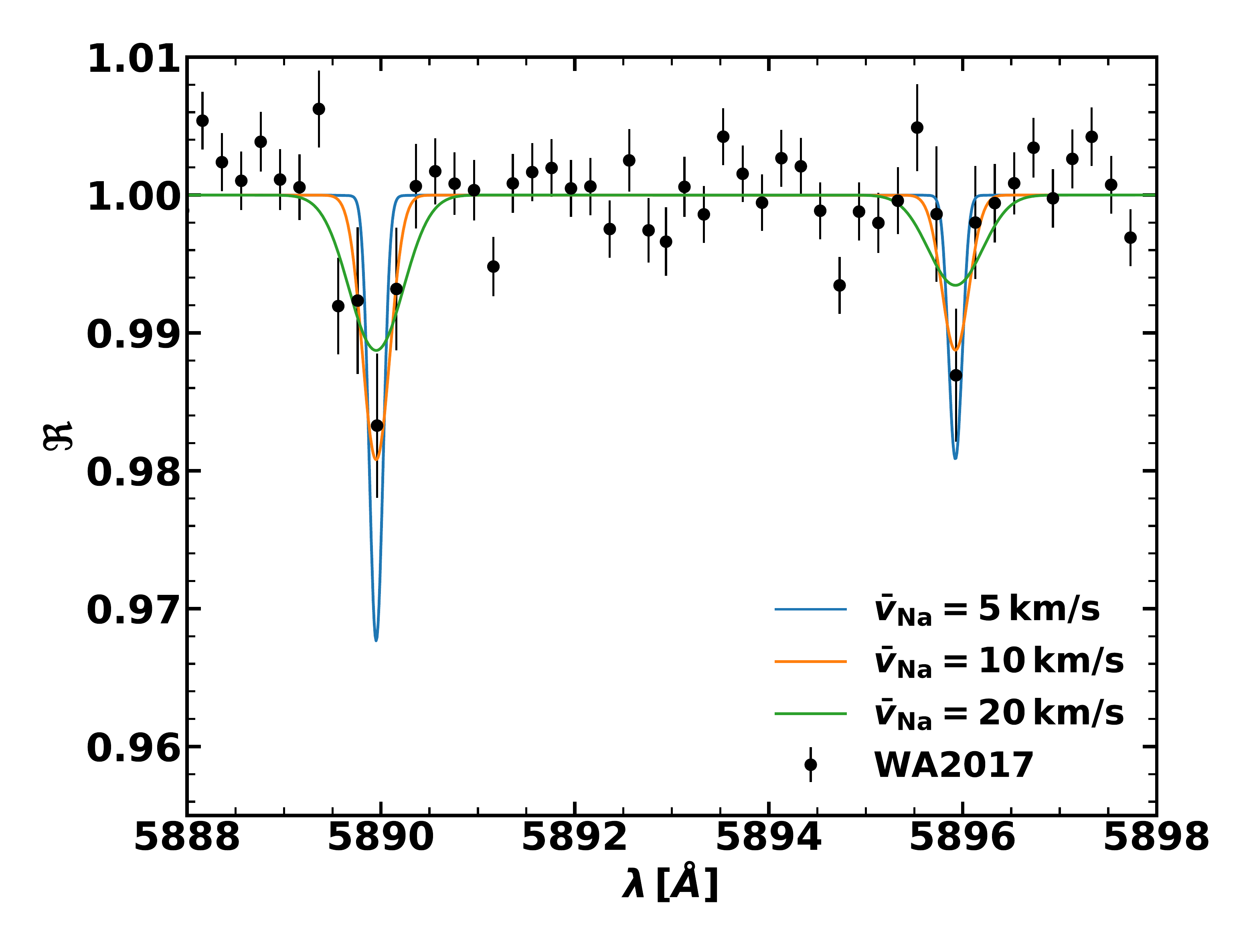}
    \caption{Simulation of high-resolution transit spectra in the sodium doublet, for different mean velocities of the sodium atoms (corresponding to different effective temperatures) in the exomoon scenario. We set $\dot{M}_{\mathrm{Na}}=4\cdot10^8\,$g/s, which corresponds to $\mathcal{N}_{\mathrm{Na}}=2.5\cdot10^{33}$ neutral sodium atoms in the system. The planetary parameters and data points are for WASP-49b, taken from \citet{Wyttenbach2017}.}
    \label{outgassing_velocities}
\end{figure}

\subsection{Torus scenario}\label{torus}

As observed at Saturn, a moon or debris around a planet can source a circumplanetary torus with neutrals (in our study Na \& K) in two ways. Case (1) Direct outgassing from an active satellite: Enceladus outgasses water-products generating an OH torus along its orbital path (\citealt{Johnson2006_OH}). Case (2) Desorption of grains from a \textit{toroidal atmosphere} \cite{Johnson2006b}: UV photoionization of ice grains generates a close-in O$_2$ torus at Saturn (\citealt{Johnson2006O2}). We can model both sources of an exoplanet torus via the number density of the absorber as:

\begin{equation}\label{n_tor}
    n_{\mathrm{tor,i}}(a,z)=n_{0,\mathrm{i}}\cdot\exp\bigl[- \left (\frac{z}{H_t} \right )^2\bigr]\cdot\exp\bigl[-\left (\frac{a-a_t}{4H_t}\right )^2\bigr],
\end{equation}

which depends on the torus scale height $H_t$ and the satellite semimajor axis $a_t$, which approximates the distance between the planet and the circumplanetary torus. $a$ denotes the radial distance in the orbital $x-y$-plane (see Figure \ref{Geometry}), $a=\sqrt{x^2+y^2}$. We set $a_t=2 R_0$ for our analysis as in \citet{Oza2019} based on \citet{Domingos2006} and \citet{Cassidy2009}, although in principle this number is allowed to decrease until the Roche radius for a silicate grain (\citet{Roche1849}). The torus scale height can be expressed as $H_t=a_t\cdot v_{\mathrm{ej}}/v_{\mathrm{o}}$, where $v_{\mathrm{o}}=\sqrt{GM_p/a_t}$ is the orbital velocity of the debris or venting moon and $v_{\mathrm{ej}}$ denotes the ejection velocity of the atoms, which we fix to 2\,km/s based on \citet{Johnson2006b}. As in the exomoon scenario we can drop the mixing ratio profile of the absorber $\chi_{\mathrm{i}}(r)$ for the computation as it is already incorporated into the number density profile.

The geometry for the computation of the transit spectra is more complex in this scenario due to the number density profile being azimuthally symmetric instead of spherically symmetric. Hence we also need to (linearly) discretize our coordinate grid along the $y$-axis. Equation \ref{tau} then needs to be adjusted to the new geometry:
 
\begin{equation}
    \tau (y,z,\lambda)=2 \int_0^{\infty}\! n_{\mathrm{tor,i}}\Bigl(a,z\Bigr) \cdot \sigma \bigl(\lambda,T(a,z)\bigr) \mathrm{d}x\,.
\end{equation}

To obtain the flux decrease at a certain wavelength, the optical depth has to be averaged both over the $y$- and $z$-coordinate. Equation \ref{flux} changes to

\begin{equation}
    \Re=\frac{4}{\pi(R_{\ast}^2-R_0^2)}\int_0^{R_{\ast}}\!\mathrm{d}z\int_{\operatorname{Re}\bigl(\sqrt{R_0^2-z^2}\bigr)}^{\sqrt{R_{\ast}^2-z^2}}\!\mathrm{d}y\,\Bigl[1-e^{-\tau(y,z,\lambda)}\Bigr].
\end{equation}

\begin{figure}
    \centering
    \includegraphics[width=\columnwidth]{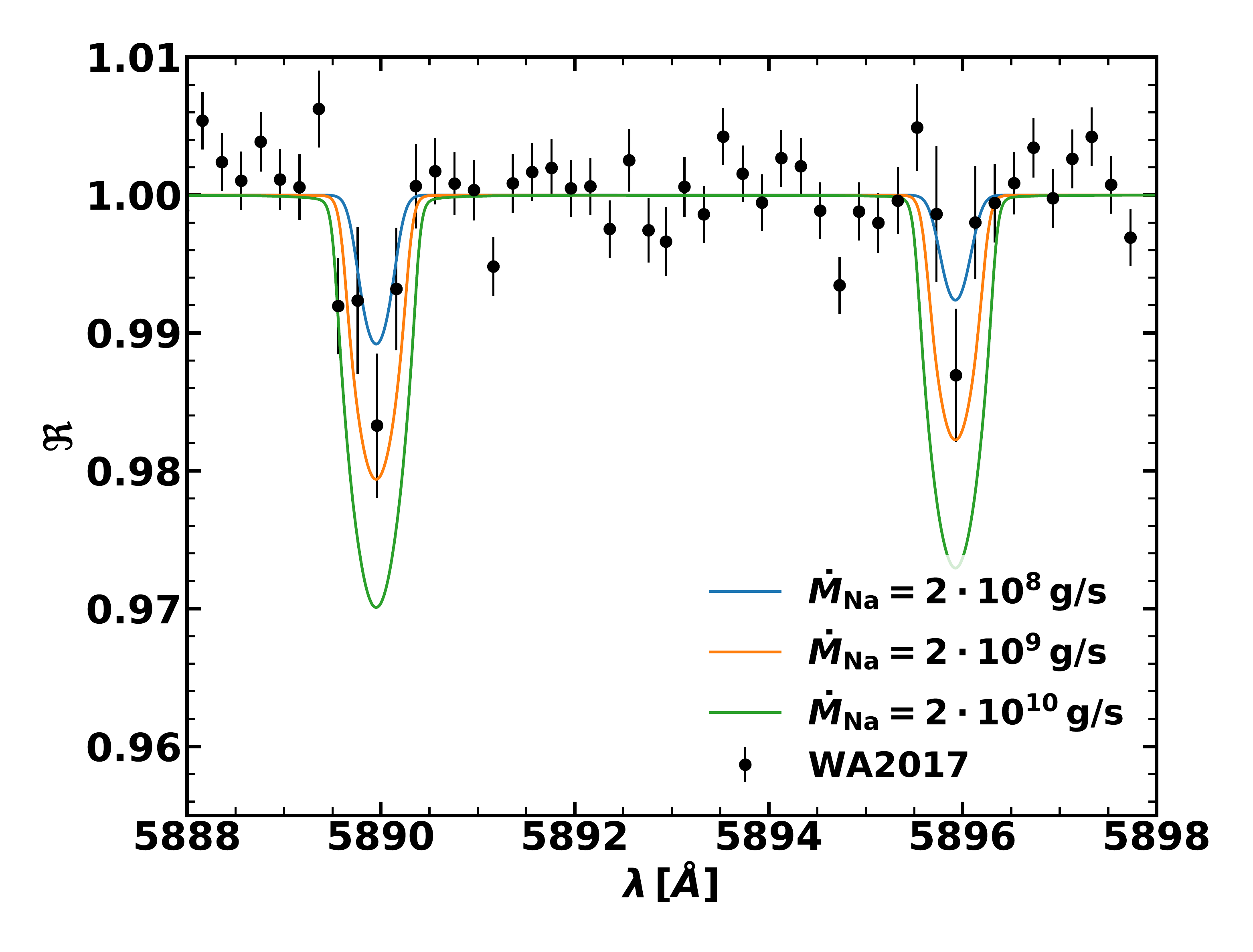}
    \caption{Simulation of high-resolution transit spectra in the sodium doublet, for different mass loss rates of the sodium atoms in the torus scenario. We set $\bar{v}_{\mathrm{Na}}=10\,$km/s. The planetary parameters and data points are for WASP-49b, taken from \citet{Wyttenbach2017}.}
    \label{torus_mdots}
    
\end{figure}

\begin{figure}
    \centering
    \includegraphics[width=\columnwidth]{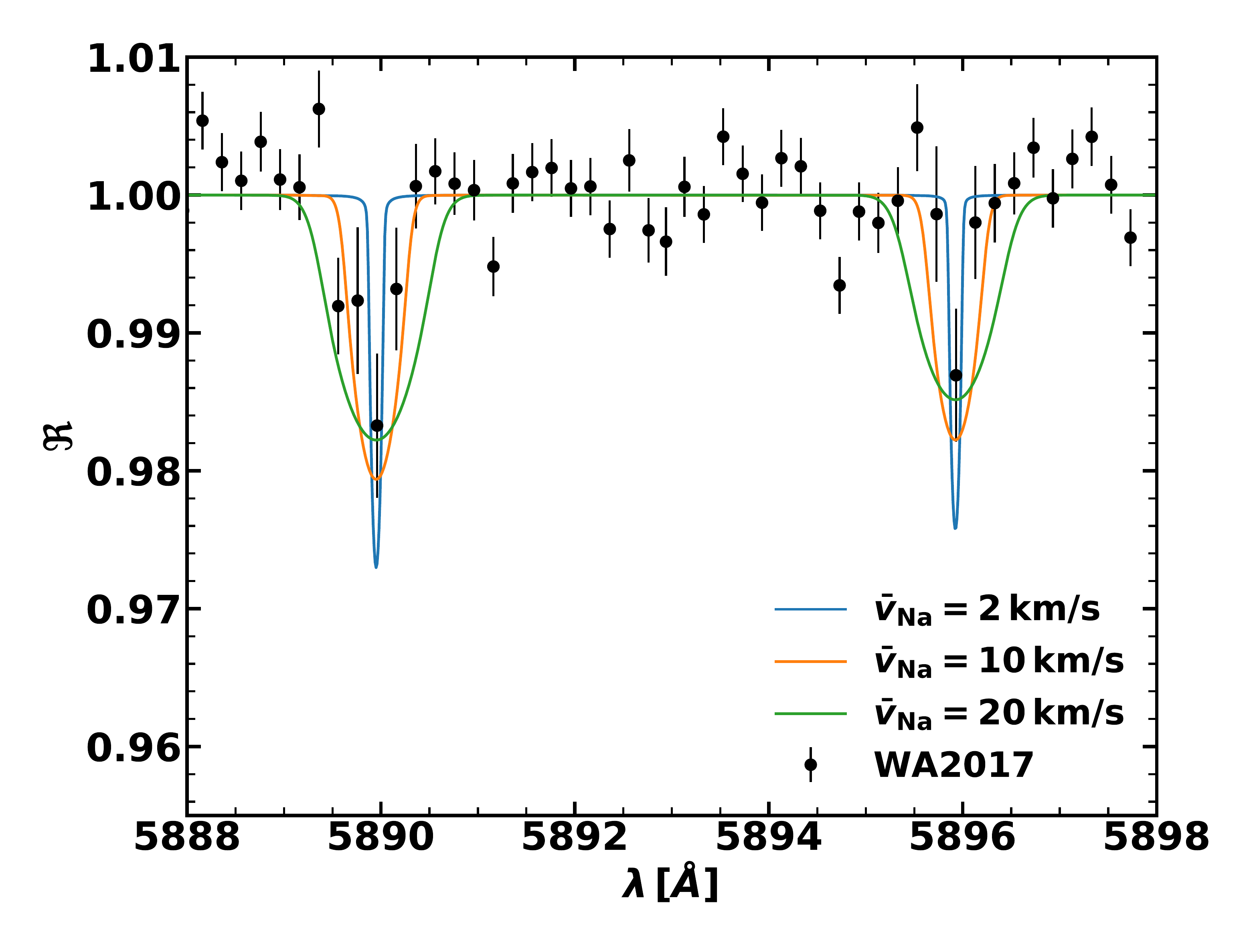}
    \caption{Simulation of high-resolution transit spectra in the sodium doublet, for different mean velocities of the sodium atoms in the torus scenario. We set $\dot{M}_{\mathrm{Na}}=2\cdot10^9\,$g/s, which corresponds to $\mathcal{N}_{\mathrm{Na}}=1.3\cdot10^{34}$ neutral sodium atoms in the system. The planetary parameters and data points are for WASP-49b, taken from \citet{Wyttenbach2017}.}
    \label{torus_velocities}
\end{figure}

As in the exomoon scenario we have $\bar{v}_{\mathrm{i}}$ and $\mathcal{N_{\mathrm{i}}}$ as our free parameters. Transit spectra for some choices of these parameters are shown in Figures \ref{torus_mdots} and \ref{torus_velocities}. There is again the following conversion between $\mathcal{N_{\mathrm{i}}}$ and $n_{0,\mathrm{i}}$ which has to be solved numerically: $\mathcal{N_{\mathrm{i}}}=\iiint_V \!n_{\mathrm{tor,i}}(a,z)\,\mathrm{d}V $. Since the circumplanetary torus isn't in thermodynamical equilibrium, we again use the absorber mass loss rate $\dot{M}_{\mathrm{i}}$ (instead of pressures) as an auxiliary parameter for an easier interpretation of $\mathcal{N_{\mathrm{i}}}$ and to couple the radiative transfer code to calculations within \texttt{DISHOOM}. 
 

\underline{Case 1: Outgassed Torus}: While the dominant escape mechanism which sources a cloud uniquely due to a satellite (exomoon scenario) is atmospheric sputtering, a plasma torus can be fueled by several of the plasma processes described in Tables \ref{plasmaprocesses} and \ref{velocities}. We first focus on a \textit{directly} outgassed torus: a tenuous Na I torus fueled by a \textit{small} outgassing satellite similar to Enceladus (e.g. \citealt{Porco2006}; \citealt{Johnson2006_OH};  \citealt{Postberg2009}) where the probability of escape, $\mathcal{R} (1+ \lambda_0) \exp(-\lambda_0) \approx 1$. Here $\lambda_0$ is the Jeans parameter, and $\mathcal{R}$ a parameter to account for surface heating (see \citealt{Johnson2015}). This case of an \textit{exo-Enceladus} fueling a torus is therefore in contrast with the exo-Ios described in \citet{Oza2019} in that the thermal desorption $\dot{M}_{0, d}$ (see Equations 10 -12 in \citealt{Oza2019} and Table 5) approaches the thermal evaporation rate of silicate grains so that:

\begin{equation}\label{mdotsub}
\begin{aligned}
  \dot{M}_{\mathrm{tor1,i}} 
   &\sim \dot{M}_{0, \mathrm{evap}}  \\
   &\sim x_{\mathrm{i}} 4 \pi R_s^2 P_{\mathrm{vap}} (T_0) \left (\frac{m_{\mathrm{i}}}{2 \pi k_B T_0} \right )^{1/2},
\end{aligned}
\end{equation}

The toroidal mass rate for Case 1 is then limited by the grain temperature T$_0$ and the surface area of the evaporating body, $4\pi R_s^2$. A fundamental parameter driving the evaporation rate is the vapor pressure of the rocky mineral in question, $P_{\mathrm{vap}}$. In principle, this formalism can compute the sublimation of arbitrary grain compositions, given that experiments have constrained the necessary coefficients to estimate the vapor pressure (c.f. Table 3, Eqn. 13 \citealt{vanlieshout2014} where we use: MgSiO$_3$ and Fe$_2$SiO$_4$). $x_{\mathrm{i}}$ is the mass fraction of the absorbing atom outgassing off of grains of exo-Io composition (\citealt{Oza2019}), and $m_{\mathrm{i}}$ the mass. As little is known regarding the volatile composition of silicate grains at hot Jupiters, we use a conservative estimate of $x_{\mathrm{Na}} = 0.003$ assuming the grain has experienced substantial volatile loss, like at Io. The value is consistent with direct obserations of NaCl by \citet{Lellouch2003}, where outgassed values of NaCl/SO$_2$ = 0.3 - 1.3 \% are consistent with CI chondritic compositions of Na/S = 0.13 and Cl/S = 0.01. The value is also roughly consistent with grains ejected from Enceladus (NaCl/H$_2$O = 0.005 - 0.02) observed by the Cosmic Dust Analyser aboard Cassini  (\citealt{Postberg2009}).  We comment that due to the magmatic nature of these grains, more recent geophysical modeling of high-temperature rocky bodies is warranted to better predict $x_{\mathrm{Na}}$ (e.g. \citealt{Noack17}; \citealt{Bower19}; \citealt{sossi2019}) and its origin.

\underline{Case 2: Desorbing Torus}: In the case of a directly outgassed torus, the surface area is quite small yielding a lower limit to the toroidal mass rate. However for a toroidal atmosphere, the desorption of Na from a silicate torus (similar to Saturn's O$_2$ ring atmosphere from H$_2$O grains, \citealt{Johnson2006O2}) can considerably enhance the toroidal mass rate. The enhancement factor $\xi$ is simply the ratio of the surface areas:

\begin{equation}\label{Enhancement torus}
\begin{aligned}
    \dot{M}_{\mathrm{tor2,i}}=\xi\cdot\dot{M}_{\mathrm{tor1,i}} &\sim \frac{4 \pi^2 a_t H_t}{4 \pi R_s^2} \dot{M}_{\mathrm{tor1,i}} \\&=4\pi \left (\frac{R_p}{R_s} \right )^2 \left ( \frac{v_{\mathrm{ej}}}{v_{\mathrm{o}}} \right ) \dot{M}_{\mathrm{tor1,i}}.
\end{aligned}
\end{equation}

For a range of semimajor axes $a_t$ we find this enhancement ranges from $\xi \sim  10^{3 \pm 0.5}$. In this way, distinguishing between a directly outgassed toroidal exosphere and a desorbing toroidal atmosphere may be achieved by high-resolution datasets (see Section \ref{toroidalatmospheres}).

\begin{table}
	\centering
	\caption{Summary of free, auxiliary and fixed parameters. The free parameters fundamentally define the transmission spectrum in our setting. We derive auxiliary parameters from them to improve the physical interpretability. For instance, we calculate the reference pressure $P_0$ in the hydrostatic scenario as auxiliary parameter from the free parameter $P_{0,\mathrm{i}}$ assuming a specific mixing ratio of the absorber.}
	\begin{tabular}{lccr}
\hline
Scenario& Free & Auxiliary & Fixed \\
\hline
Hydrostatic&$P_{0,\mathrm{i}},\,T$&$P_0$&$R_0,\,\mu$\\ \hline
Escaping&$\mathcal{N_{\mathrm{i}}},\,\bar{v}_{\mathrm{i}}$&$P_0,\,\dot{M}_{\mathrm{i}}$&$R_0,\,q_{\mathrm{esc}}$\\ \hline
Exomoon&$\mathcal{N_{\mathrm{i}}},\,\bar{v}_{\mathrm{i}}$&$\dot{M}_{\mathrm{i}}$&$R_s,\,q_{\mathrm{moon}}$ \\ \hline
Torus&$\mathcal{N_{\mathrm{i}}},\,\bar{v}_{\mathrm{i}}$&$\dot{M}_{\mathrm{i}}$&$R_0,\,a_t,\,v_{\mathrm{ej}}$\\ \hline
	\end{tabular}
	\label{retrievalparameters}
\end{table}

\subsection{Comparison to observations}\label{observation}

We compare simulated spectra for all four scenarios to high-resolution observations of the sodium D doublet at WASP-49b and HD189733b. We restrict our retrieval analysis to the data points within the wavlength interval $[5888,5898]\,\angstrom$. We note that the Na I detections at these two planets were reduced and analyzed by the same algorithm, using the same instrument (HARPS) thereby validating a 1 to 1 comparison. We retrieve the two free parameters in every scenario using the reduced chi-squared statistic (e.g. \citealt{Ocvirk2006}; \citealt{Andrae2010a}):

\begin{equation}
    \chi^2_r=\frac{\chi^2}{\nu}=\frac{1}{\nu}\sum_i\Bigl(\frac{O_i-C_i}{\sigma_i}\Bigr)^2,
\end{equation}

where $O_i$ are observations with the corresponding errors $\sigma_i$, $C_i$ are computed values and $\nu$ are the degrees of freedom of the model given by $\nu=N_O-N_P$, the difference between the number of data points and the number of free parameters (in our analysis, $N_P=2$ for all scenarios). The standard deviation of this distribution is given by $\sigma=\sqrt{2/\nu}$. Lower values of $\chi^2_r$ indicate that the model is a better fit to data. If $\chi^2_r<1$ the model overfits the data.

To compare our simulated transit spectra to the observations we apply a convolution with the instrumental line-spread function (LSF), normalization and binning routine to the raw simulated spectra. For the convolution of the raw spectrum with the LSF of the instrument we use a Gaussian with FWHM of 0.048\,$\angstrom$. We then bin the convolved spectrum to 0.2\,$\angstrom$ wide bins centered on the D2 and D1 absorption lines, which maximizes the signal-to-noise ratio according to \citet{Wyttenbach2017}. Finally, we normalize the spectrum by the average transit spectrum $\Re$ in two reference bands, $B=[5874.94;5886.94]\,\angstrom$ and $R=[5898.94;5910.94]\,\angstrom$.

\section{Optically Thin Gas in High-Resolution Transmission Spectra of Exoplanets}\label{physics}

Consider a foreground gas of atomic species i, illuminated by a background radiation field $F_{out, \lambda}$. If we treat the gas as a slab the observed intensity can be written according to Beer's law or Lambert's law (e.g. \citealt{RybickiLightman1979}):

\begin{equation}\label{fluxslab} 
F_{in, \lambda} = F_{out, \lambda} e^{-\tau(\lambda)},
\end{equation}

appropriate for a single line-of-sight. Using Equation \ref{tau} and assuming that the absorption cross section is solely a function of wavelength (which holds for our scenarios as we don't vary the Doppler broadening parameter and neglect pressure broadening), the slant optical depth for the slab can be written as:

\begin{equation}\label{tauslab}
\tau(\lambda) = N_{\mathrm{i}} \sigma(\lambda),
\end{equation}

with the line-of-sight column density of species i:

\begin{equation}\label{columndef}
    N_{\mathrm{i}}=\int_{-\infty}^{+\infty}\!n_{\mathrm{i}}(x)\,\mathrm{d}x.
\end{equation}

For transmission spectroscopy geometry, Eqn. \ref{fluxslab} needs to be averaged over infinitely many line-of-sights. In the context of a hydrostatic atmosphere, this can be done analytically. We briefly review this formalism in the next section.

\subsection{Canonical Hydrostatic Gas: an Effective Column Density at $\tau \sim 1$}\label{hydrostaticgas}

For a hydrostatic planetary atmosphere with a constant pressure scale height $H$, the line-of-sight column is given by (\citealt{Fortney2005}): 

\begin{equation}\label{chapman}
N_{\mathrm{i}}(z) = N_{v,\mathrm{i}}(z)\sqrt{\frac{2 \pi R_0}{ H}}= n_{\mathrm{i}}(z)\sqrt{2 \pi R_0 H},
\end{equation}

where $N_{v,\mathrm{i}}(z)=n_{\mathrm{i}}(z) H$ denotes the \textit{vertical} column density above altitude $z$ of species i and $\sqrt{(2 \pi R_0)/ H}$ is the Chapman enhancement factor. We have assumed that the mixing ratio of atomic species i doesn't change throughout the line-of-sight. Combining Eqns. \ref{tauslab} and \ref{chapman}, one can define a reference optical depth:

\begin{equation}
   \tau(R_0,\lambda)\equiv \tau_0(\lambda) = n_{0,\mathrm{i}}\sigma(\lambda) \sqrt{2 \pi R_0H},
\end{equation}

which can also be written in terms of a reference pressure. Continuing with a hydrostatic profile decaying over an altitude $z$: 

\begin{equation}\label{tau_hyd}
\tau(z,\lambda) = \tau_0(\lambda) e^{- z/H}.
\end{equation}

Equation \ref{tau_hyd} is an analytical expression for the general optical depth profile (Eqn. \ref{tau}) needed for the  calculation of transmission spectra. Integrating over all lines of sight (Eqn. \ref{flux}) using an identity (\citealt{Chandrasekhar1960}), a closed form expression for the atmospheric transit radius $R_{\lambda}$ can be obtained (\citealt{Wit2013}; \citealt{Betremieux2017}; \citealt{Heng2017}; \citealt{Jordan2018}) given reasonable assumptions for an isothermal atmosphere and $\tau_0(\lambda) \rightarrow \infty$. The transit radius and transit depth are related via (c.f. Eqn. \ref{flux})

\begin{equation}\label{Transit Radius}
 1 - \Re(\lambda) = \frac{R_{\lambda}^2 - R_{0}^2 }{R_{*}^2}.
\end{equation}

Remarkably, by assuming an optically thick reference pressure ($\tau_0(\lambda) \rightarrow \infty$), it turns out that the optical depth at the transit radius $\tau(R_{\lambda},\lambda)\equiv\tau_{\mathrm{eff}}(\lambda)$ (termed \textit{effective optical depth}) is $\approx0.56$, \textit{independent} of wavelength. This analytical result is elegant in that the notion of an atmospheric transit radius can be interpreted as an approximate boundary between opaque and transparent layers. In this sense, it is the (wavelength-dependent) location $z$ at which $\tau(\lambda,z) \approx0.56$, which determines the transit radius and simultaneously the transit depth. This formalism implies that the effective atmospheric column density, at the atmospheric transit radius $R_{\lambda}$, converges for all wavelengths as $N (R_{\lambda}) \approx \frac{0.56}{\sigma(\lambda)}$. At $T = 10^3 - 10^5$ K, the atmospheric column density probed at Na D2 line center is $N(R_{\mathrm{NaD2}}) \sim 5 \times 10^{11} - 1.5 \times 10^{12}\,\mathrm{cm}^{-2}$.


\subsection{Non-hydrostatic Gas: Evaporative Column Densities at $\tau \ll 1$}\label{nonhydrostaticgas}


In solving the same problem for a collisionless exosphere of arbitrary number density profile $n(r)$, one \textit{cannot} pinpoint the location of such a boundary, rendering the concept of an atmospheric transit radius inappropriate. On the other hand, as $\tau \ll 1$, the transit depth reveals the \textit{total number of absorbing atoms} $\mathcal{N}_{\mathrm{i}}$ spread over the stellar disk. The remarkable ability to constrain the total number of absorbing atoms $\mathcal{N_{\mathrm{i}}}$ is a fundamental property of foreground gas well known from studies of the interstellar medium (\citealt{Spitzer78}; \citealt{Draine2011}) using measurements of equivalent widths $W_{\lambda}$ (in units of $\angstrom$):

\begin{equation}\label{w}
W_{\lambda} = \int_{\lambda - \Delta \lambda}^{\lambda + \Delta \lambda} \!\Bigl ( 1 - \Re(\lambda) \Bigr)\, \mathrm{d} \lambda.
\end{equation}

We can then give a lower bound to the column density of the absorbing species i (Eqn. \ref{columndef}) using the approximation of the curve of growth in an optically thin regime (\citealt{Draine2011}; $\langle N \rangle_{obs}$ in \citealt{Oza2019}) :

\begin{equation}\label{Nobs2}
N _{\mathrm{min,i}}= 1.13\cdot10^{-12}\,\mathrm{cm}^{-1} \frac{W_{\lambda}}   {f_{ik} \lambda_{ik}^2}, 
\end{equation}

where $f_{ik}$ is the oscillator strength of the line and $\lambda_{ik}$ the wavelength of the transition. Unlike the interstellar medium where a single line-of-sight is considered, the background radiation field (star) is far larger than the foreground gas (the planetary atmosphere/exosphere). Therefore, a transmission spectrum requires averaging an arbitrary number density distribution $n(r)$ over infinitely many line-of-sights (or chords) since the gas is not necessarily homogeneous\footnote{For an analogous derivation of optically-thin gas please see Appendix B of \citet{Hoeijmakers2020} for a derivation of a homogenous, optically thin slab of gas. Here we present a heterogenous, dynamic gas.}. In order to apply Equation \ref{Nobs2}, we reduce the various line-of-sight column densities in an exosphere to a \textit{disk-averaged} column:

\begin{equation}\label{Nobs1}
\langle N_{\mathrm{i}} \rangle \equiv \frac{\mathcal{N}_{\mathrm{i}}}{\pi R_*^2}.
\end{equation}

Provided that that an outgassing or evaporative source is present, $\mathcal{N}_{\mathrm{i}}$, the number of evaporating atoms, can be described by an arbitrary mass loss rate (e.g. Eqns. \ref{mdotELescape1}, \ref{mdotplasma}, \ref{mdotsub}). The mass loss rates then directly supply the above disk-averaged column density. In other words, an evaporative column density $\langle N_{\mathrm{i}}\rangle$ is equivalent to the observed column density $N_{\mathrm{min,i}}$ (\citealt{Johnson2006b}; \citealt{Oza2019}), given that the absorption occurs in an optically thin regime. As we show in the remainder of this paper, the evaporative scenarios do indeed absorb in a primarily optically thin regime. In Table \ref{DISHOOM results} we show calculations from \texttt{DISHOOM} demonstrating that predicted mass loss rates for hot Jupiters roughly align with the required equivalent widths (converted into minimal mass loss rates using Eqns. \ref{Mdot}, \ref{Nobs2} and \ref{Nobs1}) found by high-resolution spectroscopy. 

\begin{table}
    \centering
	\caption{Sodium source rates computed within \texttt{DISHOOM}. The enhanced sodium source rate $\dot{M}_{\mathrm{tor2,Na}}$ corresponds to a desorbing torus (Eqn. \ref{Enhancement torus}). The minimally required source rate $\dot{M}_{\mathrm{min,Na}}$ is calculated using Equation \ref{Nobs2} with the observed equivalent widths at the Na D2 line (\citealt{Wyttenbach2015}; \citealt{Wyttenbach2017}) for the respective planet.}
	\label{DISHOOM results}
	\begin{tabular}{lcr} 
\hline
Source rate &WASP-49b&HD189733b\\

\hline
$\dot{M}_{\mathrm{esc,Na}}$ [kg/s] &$10^{2.4\pm0.3}$&$10^{2.3\pm0.3}$\\ 
$\dot{M}_{\mathrm{moon,Na}}$ [kg/s] &$10^{4.3\pm1.5}$&$10^{4\pm1.2}$\\
$\dot{M}_{\mathrm{tor1,Na}}$ [kg/s] &$10^{3.1\pm1.3}$&$10^{1.0\pm1.5}$\\
$\dot{M}_{\mathrm{tor2,Na}}$ [kg/s] &$10^{6.4\pm1.3}$&$10^{4.1\pm1.5}$\\
$\dot{M}_{\mathrm{min,Na}}$ [kg/s] &$10^{4.7\pm0.5}$&$10^{3.9\pm0.1}$\\ \hline

	\end{tabular}
\end{table}

\subsection{Evaporative Curve of Growth for Sodium at an Exoplanet}

\begin{figure}
    \centering
    \includegraphics[width=\columnwidth]{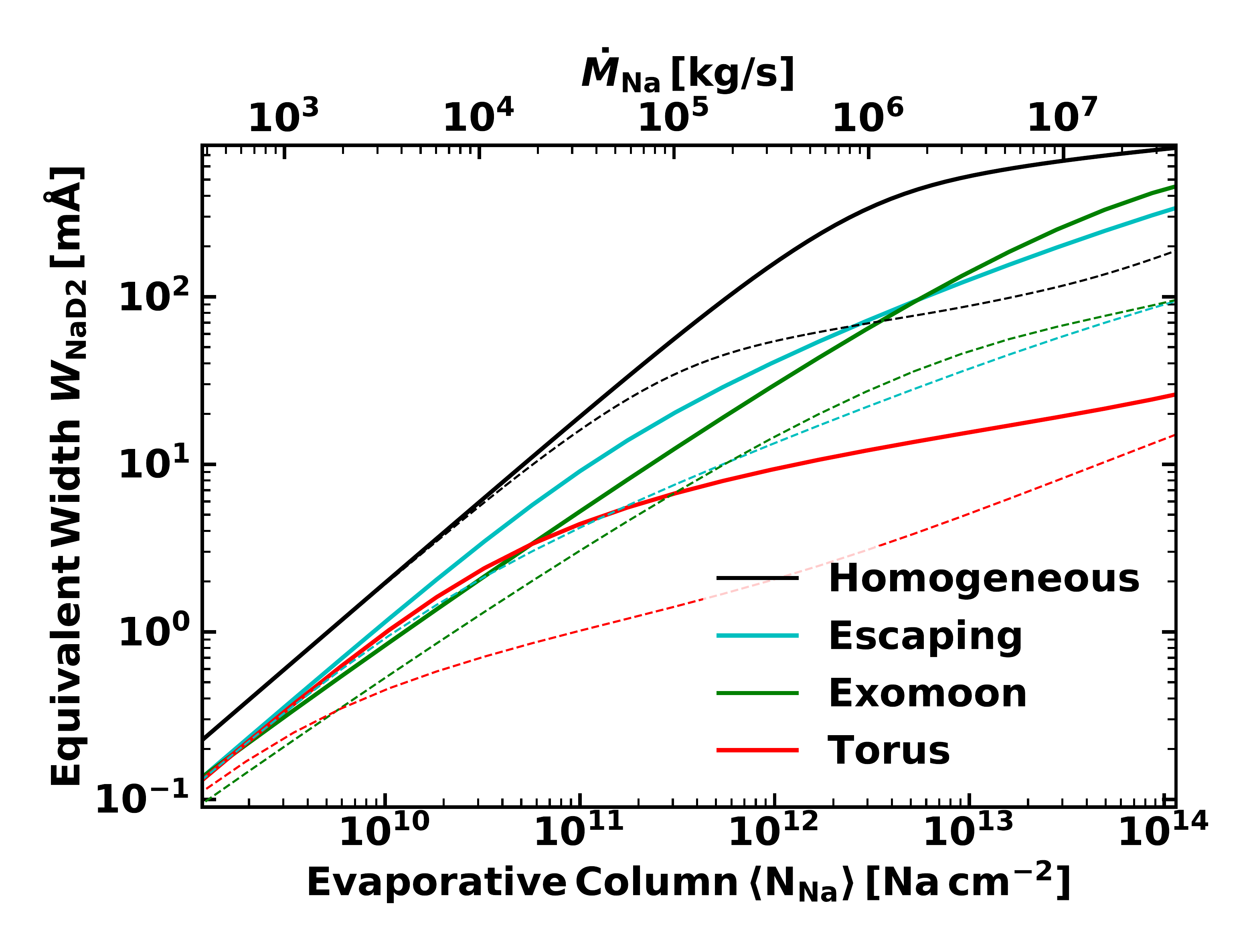}
    \caption{Equivalent width at the Na D2 line versus evaporating atoms during transit, for planetary parameters of HD189733b (Table \ref{systemparameters}). The evaporative column density is essentially a disk-averaged column (Eqn. \ref{Nobs1}). We also show the classical curve of growth ('homogeneous'), which is the equivalent width of a single line-of-sight as a function of column density. We convert $\langle N_{\mathrm{i}} \rangle$ into sodium source rates using Eqns. \ref{Mdot}, \ref{Nobs2} and \ref{Nobs1}. Bold lines are with $\bar{v}_{\mathrm{Na}}=10\,$km/s, dashed lines have $\bar{v}_{\mathrm{Na}}=1\,$km/s.}
    \label{ecog2}
\end{figure}

Using our radiative transfer code, we calculate equivalent widths at Na D2 line center for our three evaporative scenarios (Figure \ref{ecog2}). The equivalent width depends on the free parameters $\bar{v}_{\mathrm{Na}}$ and $\mathcal{N}_{\mathrm{Na}}$ (equivalently: $\mathcal{N}_{\mathrm{Na}}$, $\langle N_{\mathrm{Na}}\rangle$ or $\dot{M}_{\mathrm{Na}}$). The dependence of the equivalent width on a (disk-averaged) column density is reminiscent of the classical curve of growth for a single line-of-sight. If the equivalent width is known to high precision, one can constrain an evaporative column density of occulting atoms depending on the scenario in Figure \ref{ecog2}. These values coincide with the values from \citet{Oza2019} (their Table 5) and indicate how effective each evaporative scenario is in generating the observed absorption. 

To generate typical observed equivalent widths of $W_{\mathrm{NaD2}}\sim1-10\,\mathrm{m}\angstrom$ at hot Jupiters, the required evaporative column densities range from $\sim 10^{10}$ to $\sim10^{12}\,\mathrm{Na\,cm}^{-2}$. For a torus, whose number density is a Gaussian distribution (Eqn. \ref{n_tor}), the column densities can be far larger. We note that absorption along a single line-of-sight (or, equivalently, in a homogeneous cloud as in \citealt{Hoeijmakers2020}) is always absorbing more efficiently than our evaporative sodium distributions.

Independent of the source driving mass loss, we confirm that extreme mass loss rates are required for observation of extrasolar evaporating sodium. To observe an equivalent width of 10 m$\angstrom$, for instance, roughly $\sim 10^{5}$ kg/s of sodium gas at 10 km/s is required for an escaping atmosphere or exo-Io as estimated by \citet{Oza2019}. In comparison, Io's volcanism outgasses $\sim 7 \times 10^6$ kg/s of SO$_2$ (\citealt{Lellouch2015}).  A thermally desorbing torus, however, requires $\dot{M}_{\mathrm{Na}}\sim 5 \times 10^{5}$ kg/s based on the scenario described. We note, based on the discussion in sections \ref{torus} (Eqn. \ref{Enhancement torus}) and \ref{ultrahot}, that these rates can be achieved if the grains are thermally desorbing over a large surface area, while being trapped in a toroidal magnetic field for instance.  This analysis provides an \textit{evaporative} curve of growth for atomic sodium viewed at an exoplanet. The quantity of evaporating gas $\sim 10^{31} - 10^{34}$ Na atoms, is able to govern the detection and non-detection of optically-thin absorption during transit. The evaporative curve of growth is drastically influenced by the spatial distribution of the Na atoms across the star during transit.

\subsection{The D2-to-D1 line ratio}\label{d2d1method}

At the Jupiter-Io system, the D2/D1 ratio is able to provide information on the velocity of the Na I atoms, which is observed to be variable over decades of observations. Therefore in our application to extrasolar systems, the ratio may be able to inform predictions on the average velocity distributions of the Na atoms and their ongoing behavior. However, the ratio first and foremost provides a very simple result. 

Fortuitously, the oscillator strength of the Na D2 transition is twice as large as the D1 transition, meaning that the absorption cross section and optical depth at D2 line center is double the respective values at D1 line center (\citealt{Draine2011}). In this way the optical depth at sodium D2 line center is $\tau(\lambda_{\mathrm{NaD2}})\equiv\tau_{\mathrm{NaD2}}=\tau(\lambda_{\mathrm{NaD1}})/2$. This reveals an easy diagnostic:

\begin{equation}\label{f_D2D1}
    f_{\mathrm{D2/D1}}=\frac{1-e^{-\tau_{\mathrm{NaD2}}}}{1-e^{-\tau_{\mathrm{NaD2}}/2}}.
\end{equation}

\begin{figure}
    \centering
    \includegraphics[width=\columnwidth]{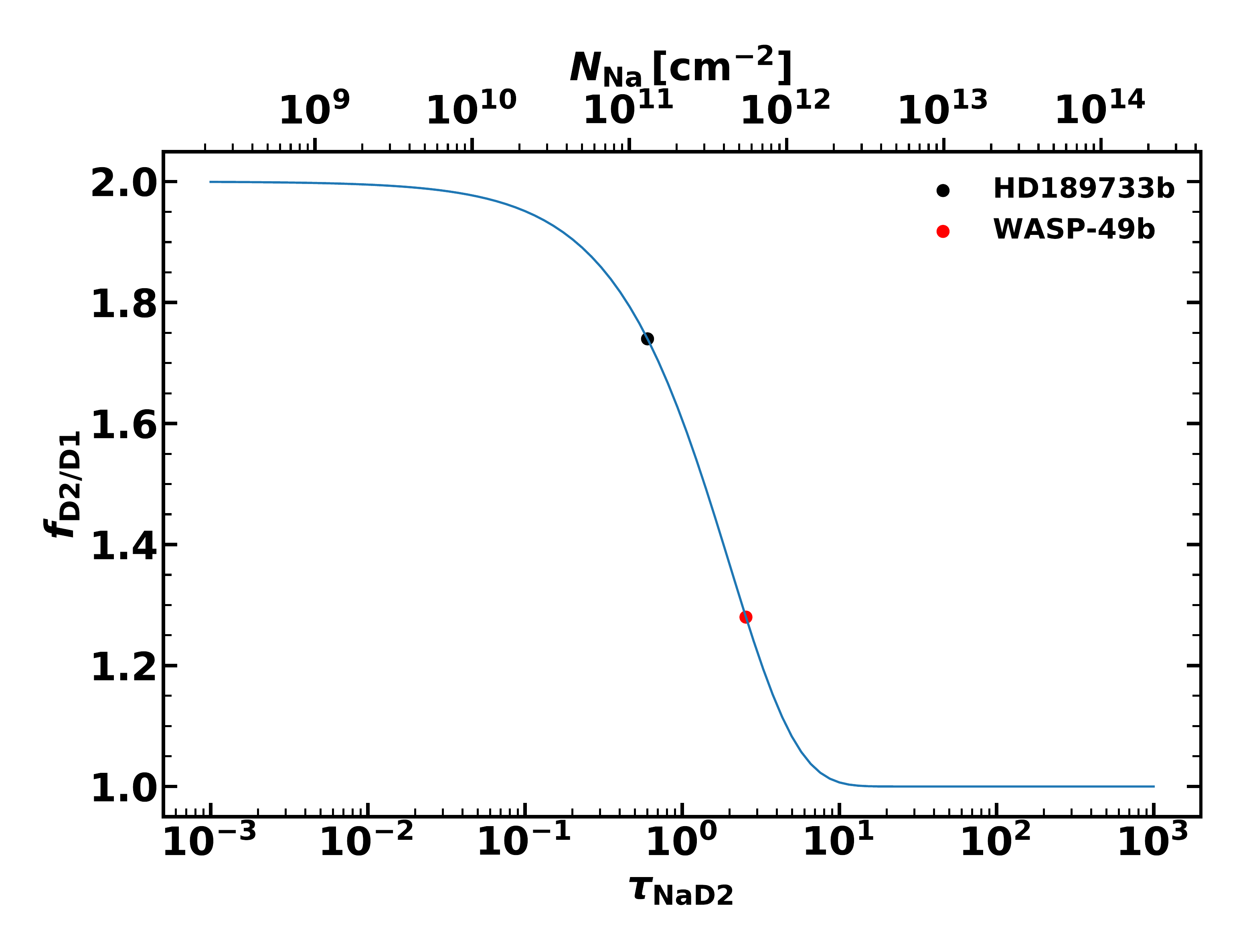}
    \caption{Computation of the D2/D1-ratio $f_{\mathrm{D2/D1}}$ as function of the optical depth at sodium D2 line center (Equation \ref{f_D2D1}). In the optically thin regime this ratio approaches two, while in the optically thick regime $f_{\mathrm{D2/D1}}$ goes to one. For the conversion of $\tau_{\mathrm{NaD2}}$ into a column, we assume a fixed absorption cross section of $\sigma_{\mathrm{NaD2}} =  4.62 \times 10^{-12}$ corresponding to a gas at $T = 2000$\,K (Eqn. 6.39 in \citealt{Draine2011}).}
    \label{d2d1ratio}
\end{figure}

This relation, together with the measured line ratios for HD189733b and WASP-49b, is shown in Figure \ref{d2d1ratio}. Throughout this paper, we calculate line ratios by using the transit depths averaged over bandwidths of $0.2\,\angstrom$, centered on the absorption lines. We note that the exact value of the line ratio depends on the choice of the bandwidth, a negligible effect for modeled transmission spectra but not for the observations. For small bandwidths, the measurement error associated with the binned transit depth is large, while for large bandwidths, the observations at wavelengths which are more than $\sim0.5\,\angstrom$ away from the line center mostly contain noise. Hence, we choose an intermediate bandwidth of $0.2\,\angstrom$ (which maximizes the signal-to-noise ratio according to \citealt{Wyttenbach2017}) for the calculation of the line ratios. While optically-thick chords with $\tau_{\mathrm{NaD2}}>10$ lead to a line ratio of one, $f_{\mathrm{D2/D1}}$ transitions over two orders of magnitude in $\tau_{\mathrm{NaD2}}$ to $f_{\mathrm{D2/D1}}=2$ for optically-thin chords, $\tau_{\mathrm{NaD2}}<0.1$. 

Unfortunately, for the setting of transmission spectroscopy, this relation isn't applicable in a straightforward way since one observes a flux decrease averaged over infinitely many chords. For example, a line ratio of 1.5 can be achieved in different ways: By having a constant line-of-sight column density throughout the stellar disk with $\tau_{\mathrm{NaD2}}\approx1$, or by having ten percent of the area of the stellar disk blocked with $\tau_{\mathrm{NaD2}}\approx10$ and the remaining ninety percent having $\tau_{\mathrm{NaD2}}\approx0.1$. Both of these models (and infinitely many other spatial distributions of the absorber) would lead to $f_{\mathrm{D2/D1}}\approx1.5$. For $f_{\mathrm{D2/D1}}=1.5$, given that the column density is a smooth function of impact parameter, we can therefore only state that the majority of the absorption occurs along chords with $\tau_{\mathrm{NaD2}}$ roughly around 0.5, but chords with vastly different values of $\tau_{\mathrm{NaD2}}$ are probably also present in the system. Hence, the D2-to-D1 ratio tells us in which regime (optically thin/thick) the \textit{majority} of the absorption occurs. 


In the following, we shall test how the predicted quantities of evaporating gas fare against canonical hydrostatic assumptions for high-resolution Na I spectra at hot Jupiters. 






\section{Forward modeling and scenario comparison}\label{Forward Modeling}

\begin{figure}
    \centering
    \includegraphics[width=\columnwidth]{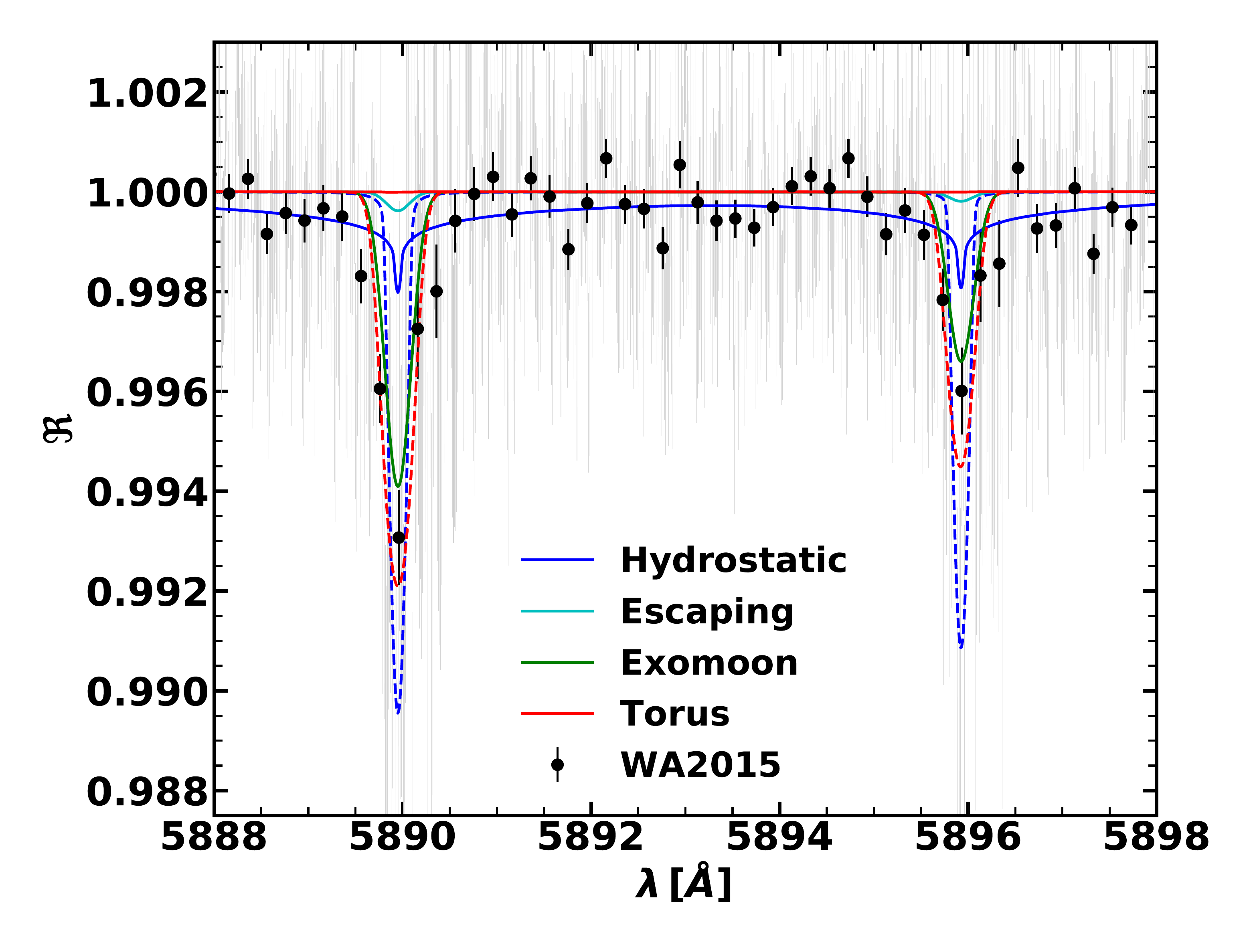}
    \caption{Forward model transmission spectra of each scenario studied, for the hot Jupiter HD189733b. The dashed blue line represents an extremely heated hydrostatic atmosphere with $T=10\,T_{\mathrm{eq}}$. The dashed red line has an enhanced sodium source rate corresponding to a desorbing torus (Eqn. \ref{Enhancement torus}). Planetary parameters and data points are taken from \citet{Wyttenbach2015}.}
    \label{FORWARDMODEL}
\end{figure}

We elucidate the general features of evaporative transmission spectra in this section by constructing a forward model for the hot Jupiter HD189733b, using \texttt{DISHOOM} to calculate sodium source rates for the evaporative scenarios (listed in Table \ref{DISHOOM results}). These rates are converted into $\mathcal{N}_{\mathrm{Na}}$ (using Eqn. \ref{Mdot}), a parameter which is then fed into \texttt{Prometheus} to compute the transmission spectra. We emphasize that these values of $\mathcal{N}_{\mathrm{Na}}$ should be regarded as lower limits, since we use a lower limit to the lifetime of neutral sodium in Equation \ref{Mdot}. We uniformly set $\bar{v}_{\mathrm{Na}}=10\,$km/s for the three evaporative scenarios in this forward model, a simplification which doesn't affect the following analysis as the dominant effect of $\bar{v}_{\mathrm{Na}}$ is only the broadening of the lines. We also examine two hydrostatic scenarios: Our first hydrostatic model has $T=T_{\mathrm{eq}}=1'140\,$K and $P_0=1\,$bar, the second model\footnote{The purpose of this model is to demonstrate the spectral imprint of extreme thermospheric heating.} has $T=10 T_{\mathrm{eq}}=11'400\,$K and $P_0=0.1\,\mu$bar (for the conversion of the reference pressures into $P_{0,\mathrm{Na}}$ we assume $\chi_{\mathrm{Na}}=1.7\,$ppm for both scenarios). This choice of parameters for the hydrostatic scenario is arbitrary, and intentionally spans a large range within the free parameters. The forward-model transmission spectra are shown in Figure \ref{FORWARDMODEL}.

The four evaporative transmission spectra in figure \ref{FORWARDMODEL} share some features such as negligible absorption between the lines and a D2-to-D1 line ratio significantly larger than one. The transit depth on the line cores is much larger for the exomoon scenario and for the desorbing torus (Eqn. \ref{Enhancement torus}) compared to the two other evaporative transmission spectra, stemming from the almost three orders of magnitude larger sodium source rates (Table \ref{DISHOOM results}). The two hydrostatic scenarios have, in contrast to the evaporative scenarios, a D2-to-D1 line ratio only slightly larger than one. Furthermore, the hydrostatic scenario with $T=T_{\mathrm{eq}}$ exhibits significant absorption between the line cores, due to the reference pressure of $P_0=1\,$bar leading to a very thick atmosphere.

\begin{figure}
    \centering
    \includegraphics[width=\columnwidth]{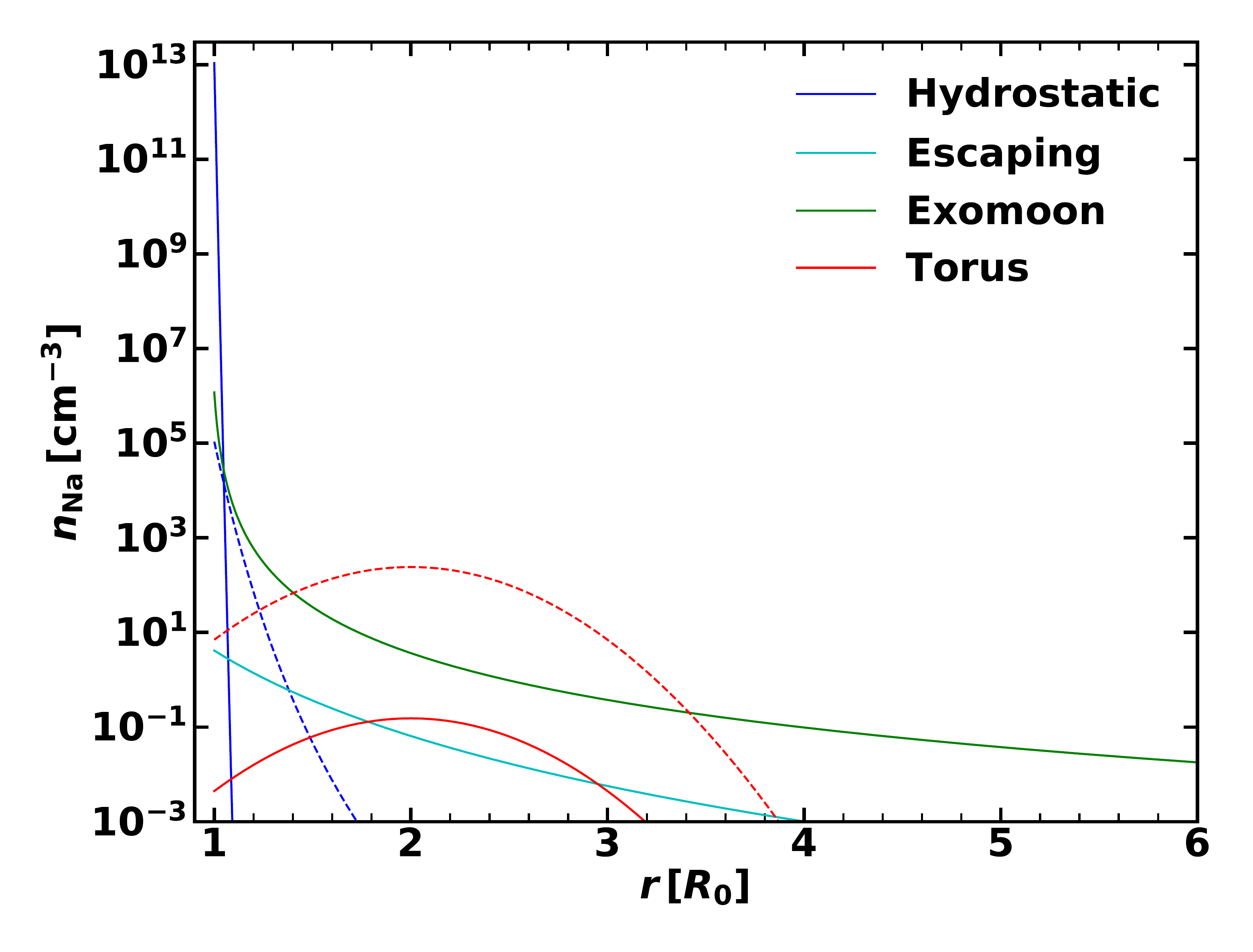}
    \caption{Sodium number density profiles of each scenario studied. The number density profiles correspond to our forward models for the hot Jupiter HD189733b. The dashed blue line represents an extremely heated hydrostatic atmosphere with $T=10\,T_{\mathrm{eq}}$, the dashed red line has an enhanced sodium source rate corresponding to a desorbing torus (Eqn. \ref{Enhancement torus}). The exomoon profile starts in our model already at $r=R_{\mathrm{Io}}$, we shift the curve to the planetary radius in this plot to enhance readability. For the torus scenario we show the number density profiles through the orbital plane, $n_{\mathrm{tor,Na}}(r,0)$.}
    \label{n_profiles}
\end{figure}

\begin{figure}
    \centering
    \includegraphics[width=\columnwidth]{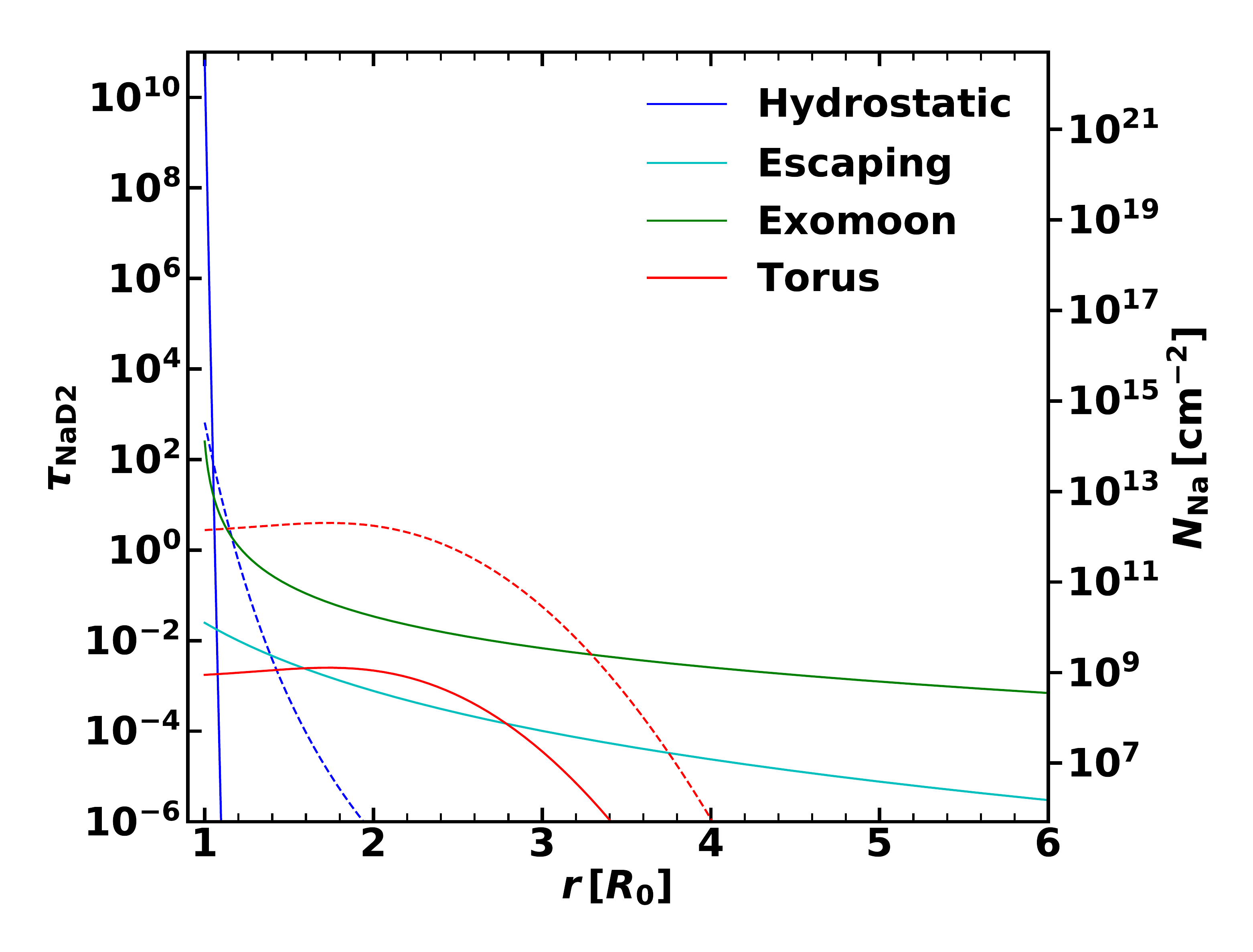}
    \caption{Optical depth (at Na\,D2 line center) profiles of each scenario studied. The optical depth profiles correspond to our forward models for the hot Jupiter HD189733b. The dashed blue line represents an extremely heated hydrostatic atmosphere with $T=10\,T_{\mathrm{eq}}$, the dashed red line has an enhanced sodium source rate corresponding to a desorbing torus (Eqn. \ref{Enhancement torus}). The exomoon profile starts in our model already at $r=R_{\mathrm{Io}}$, we shift the curve to the planetary radius in this plot to enhance readability. For the torus scenario we show the optical depth profiles through the orbital plane, $\tau(r,0,\lambda_{\mathrm{Na\,D2}})$. The line-of-sight column density is shown for a Doppler broadening parameter corresponding to $T=10T_{\mathrm{eq}}$, allowing a conversion from $\tau_{\mathrm{Na\,D2}}(r)$ to $N_{\mathrm{Na}}(r)$ using Eqn. \ref{tauslab}.}
    \label{tau_profiles}
\end{figure}

The origin of the variety of transit spectra becomes apparent when examining the spatial structures of the different forward models. We show the corresponding sodium number density profiles in Figure \ref{n_profiles}. While the sodium number density in the hydrostatic scenarios drops fast (especially in the lower-temperature case), the three evaporative scenarios lead to more extended and tenuous exospheres. The sodium number density of the torus scenario peaks at $r=2\cdot R_0$ due to the fixed orbital separation of the torus of $a_t=2\cdot R_0$ (see Section \ref{torus}). These number density profiles can be converted into optical depth profiles at a fixed wavelength using Equation \ref{tau}. We fix the wavelength to the Na\,D2 line center and show $\tau_{\mathrm{Na\,D2}}(r)$ in Figure \ref{tau_profiles}. Since the optical depth profiles\footnote{If not specified otherwise, by optical depth we mean the optical depth at sodium D2 line center for the following discussion.} are equivalent to the line-of-sight integrals of the sodium number density profiles times a constant depending on the Doppler broadening (Eqns. \ref{tauslab} and \ref{columndef}), the curves are similar in Figures \ref{n_profiles} and \ref{tau_profiles}, but decay slower in the latter plot (note that both plots cover approximately sixteen orders of magnitude). From the optical depth profiles we can derive the properties of the different transmission spectra in Figure \ref{FORWARDMODEL}.
\begin{itemize}
    \item \underline{Hydrostatic, $T=T_{\mathrm{eq}}$}: The hydrostatic scenario with $T=T_{\mathrm{eq}}$ exhibits an optical depth which drops very fast with increasing altitude. This behaviour leads to a small and very optically thick layer, which absorbs all incoming starlight nearly uniformly up to a certain altitude, followed by negligible absorption. This leads to $f_{\mathrm{D2/D1}}$ being very close to one in this model. We have a large reference pressure of $P_0=1\,$bar in this scenario, which leads to the slant optical depth at the reference radius $\tau_{0,\mathrm{Na\,D2}}$ being of the order $10^{10}$ for the Na\,D2 line center. Since the absorption cross section does not drop by ten orders of magnitude between the line centers, we have significant absorption also between the lines.
    \item \underline{Hydrostatic, $T=10\,T_{\mathrm{eq}}$}: At a larger temperature the hydrostatic scenario has a more puffed up atmosphere. However, the optical depth still drops very fast, leading to $f_{\mathrm{D2/D1}}\approx1.1$. Since the slant optical depth at the reference radius is only of the order $10^3$, the atmosphere does not produce significant absorption between the lines.
    \item \underline{Evaporative}: On the other hand, the three evaporative scenarios have extended and mostly optically thin exospheres. Both the escaping and the exomoon scenario have long tails in their optical depth profiles. Since $\tau_{\mathrm{Na\,D2}}$ is lower than unity in these two scenarios for the largest part of the exosphere, the resulting transmission spectrum is optically thin. For optically thin chords, the proportion of absorbed light is directly proportional to the optical depth, leading to $f_{\mathrm{D2/D1}}\approx2$. Since the optical depth profile in the escaping scenario is offset by two orders of magnitude due to the lower sodium source rate, the flux decrease in this scenario is accordingly lower. The torus scenario sourced by direct outgassing (Eqn. \ref{mdotsub}) has a plateau with $\tau_{\mathrm{Na\,D2}}$ slightly lower than one percent, extending over multiple radii, before the optical depth drops. Hence, the exosphere of the torus in this forward model doesn't generate significant absorption as seen in Figure \ref{FORWARDMODEL}. On the other hand, the desorbing torus (Eqn. \ref{Enhancement torus}) with a source rate three orders of magnitudes larger has its optical depth profile shifted up by the same factor, leading to significant absorption.
\end{itemize}

We conclude from this analysis that the three evaporative scenarios generally produce optically thin, extended exospheres (especially the escaping \& exomoon scenarios), while hydrostatic models lead to a small, optically thick atmosphere. We find here that despite an atmosphere with a temperature ten times larger than the planetary equilibrium temperature, effectively enhancing the atmospheric scale height by a factor of ten, the hydrostatic atmosphere still drops very fast in number density and optical depth, such that the optically thin layer is negligibly small. This confirms our analysis in Section \ref{nonhydrostaticgas}, in that the evaporative scenarios absorb mostly in an optically thin regime, unlike the hydrostatic scenario.

\section{Observational Analysis of Evaporative Sodium Transit Spectra}\label{Inverse Modeling}

In the following we will perform a simple retrieval on two hot Jupiters (WASP-49b and HD189733b) applying the reduced $\chi^2$-statistic to determine best fits for all four scenarios. Using high-resolution observations of the sodium doublet, we determine the two free parameters in all scenarios. Next, we scrutinize the physical validity of the retrieved parameters based on our sodium source rate calculations from \texttt{DISHOOM} (Table \ref{DISHOOM results}), which comprises the coupling between our codes in this inverse modeling. The retrieval results for both planets are summarized in Table \ref{RetrievalResults}. Error bars for the retrieved parameters correspond to the 1-$\sigma$-interval around the best-fitting model, where $\sigma$ is the standard deviation of the $\chi_r^2$-statistic (see Section \ref{observation}). We have $\sigma=\sqrt{2/\nu}\approx0.2$, where $\nu$ are the degrees of freedom of the model. We also evaluate $f_{\mathrm{D2/D1}}$ for our best-fitting models. The error bars for the line ratios are calculated such that $f_{\mathrm{D2/D1}}$ lies within the error bars with a likelihood\footnote{We use that the probability of a model is proportional to $\exp(-\chi^2/2)$.} of $\approx68\%$.

\subsection{Inverse Modeling of WASP-49b}\label{Inverse Modeling W49}
\begin{figure}
    \centering
    \includegraphics[width=\columnwidth]{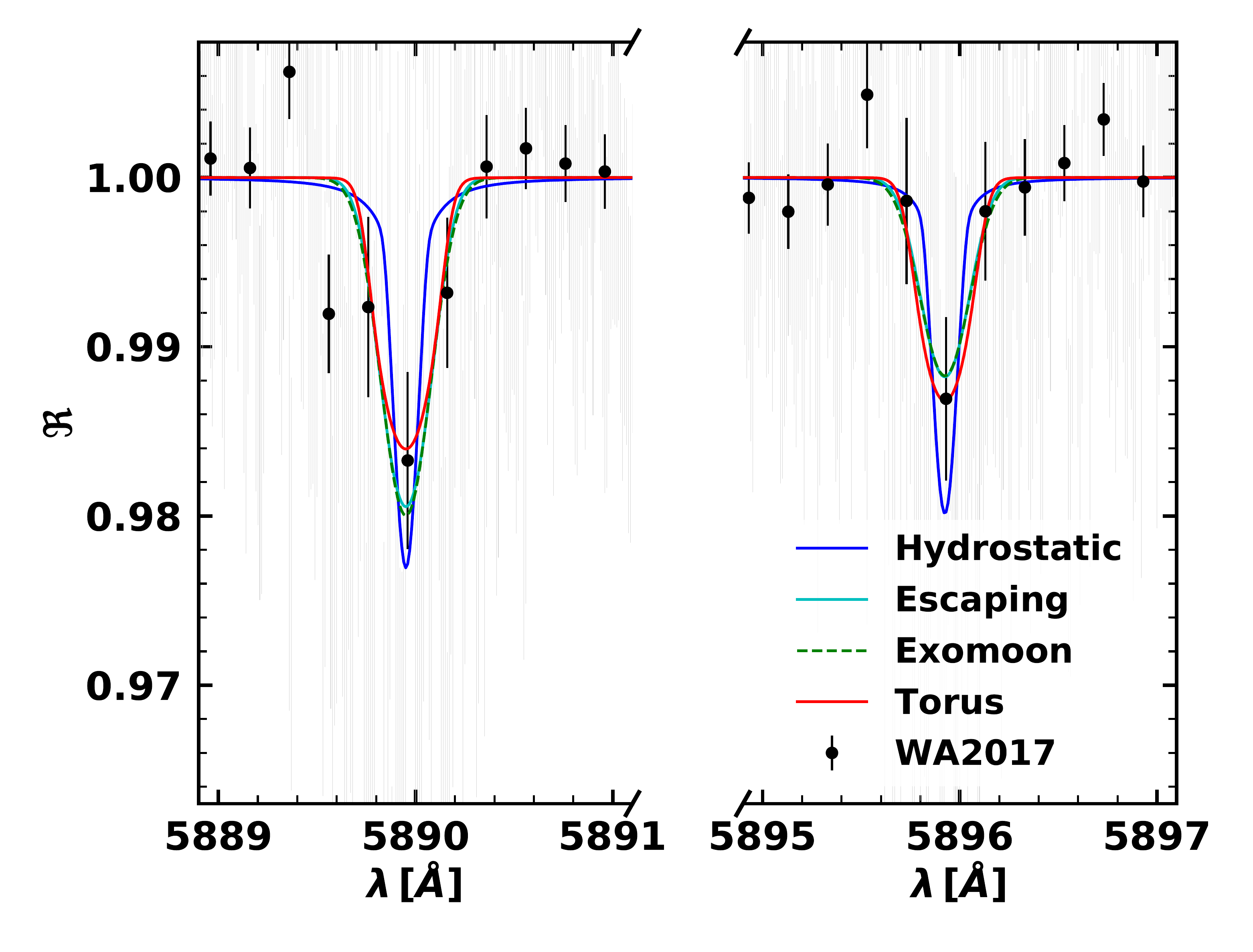}
    \caption{Transit spectrum for WASP-49b with our best-fitting models for each scenario. Planetary parameters and data points are taken from \citet{Wyttenbach2017}. Note the break in the x-axis.}
    \label{w49bestfits}
\end{figure}

\citet{Wyttenbach2017} obtained  a  high-resolution  spectrum  of  the  hot Jupiter WASP-49b, observing significant absorption in the sodium doublet (more than two percent of the D2 line center). The spectrum shows negligible absorption between the line cores and a D2-to-D1-line ratio which is $f_{\mathrm{D2/D1}}=1.28\pm0.62$ (calculated in bands of $0.2\,\angstrom$ centered on the sodium D lines). The observation and the best-fitting models for each scenario are shown in Figure \ref{w49bestfits}, we summarize the corresponding parameters in Table \ref{RetrievalResults}. Given the large uncertainty in $f_{\mathrm{D2/D1}}$, all four scenarios have line ratios within the observational error bars. We remark that for the case of WASP-49b, this ratio converges to two for larger bandwidths since the D2 line is broader than the D1 line (\citealt{Wyttenbach2017}). If this line ratio was confirmed in a more precise measurement, the hydrostatic and the torus scenario would significantly underestimate $f_{\mathrm{D2/D1}}$. The achieved goodness-of-fit is very similar in all four scenarios $1.44<\chi_r^2<1.49$, with slightly better fits for the evaporative scenarios. From the $\chi_r^2$-statistic, a model which correctly describes data with proper error bars should achieve $\chi_r^2=1$. The fact that our best-fitting models have values of $\chi_r^2$ which are $\gtrsim2\sigma$ away from $\chi_r^2=1$ implies that our models either poorly describe the data or that the observational error bars are underestimated. One can rescale the $\chi_r^2$-values such that the best-fitting model achieves $\chi_r^2=1$, which is essentially a rescaling of the error bars. However, this procedure is statistically incorrect (\citealt{Andrae2010b}), hence we refrain from such a rescaling and keep our measured values of $\chi_r^2$. We remark that over the spectral range investigated here, our models seem to fit the data well when roughly checked by eye, which leads us to suggest that the oddly large values of $\chi_r^2$ can primarily be attributed to an underestimation of the observational error bars.

\begin{table*}
    \centering
	\caption{Retrieval results for WASP-49b and HD189733b. Note that due to the irregularity of the $\chi_r^2$-maps of WASP-49b (Figure \ref{RetrievalPlotsW49}), the error bars for the retrieved parameters of this planet sometimes extend in only one direction. We also evaluate the D2-D1 line ratio $f_{\mathrm{D2/D1}}$ for our models. From the measurements, we estimate $f_{\mathrm{D2/D1, W49}} =  1.28 \pm 0.62$ (\citealt{Wyttenbach2017}) and $f_{\mathrm{D2/D1, HD189}} =  1.74 \pm 0.45$ (\citealt{Wyttenbach2015}).}
	\label{RetrievalResults}
	\begin{tabular}{lccccccr}
\hline
&&&WASP-49b&&&HD189733b&\\
Scenario& Parameter& Retrieved & $f_{\mathrm{D2/D1}}$ & $\chi_r^2$&Retrieved & $f_{\mathrm{D2/D1}}$ & $\chi_r^2$\\

\hline
 Hydrostatic&$P_{0,\mathrm{Na}}$ [bar]&$10^{-12.2^{+1.4}}$&$1.17^{+0.12}_{-0.02}$&1.49&$10^{-10.7\pm0.5}$&$1.086^{+0.002}_{-0.002}$& 1.49\\
&$T$ [K]&$6100_{-3400}$&&&$5200\pm1500$&&\\ \hline

 Escaping&$\mathcal{N}_{\mathrm{Na}}$ [Na atoms]&$10^{33.1\pm1}$&$1.68^{+0.04}_{-0.06}$&1.44&$10^{32.5\pm0.2}$&$1.930^{+0.005}_{-0.005}$&1.51\\
&$\bar{v}_{\mathrm{Na}}$ [km/s]&$9^{+13}$&&&$19\pm9$&&\\ \hline

 Exomoon&$\mathcal{N}_{\mathrm{Na}}$ [Na atoms]&$10^{33.4\pm0.6}$&$1.71^{+0.01}_{-0.02}$&1.45&$10^{32.7\pm0.2}$&$1.737^{+0.002}_{-0.001}$& 1.50\\
&$\bar{v}_{\mathrm{Na}}$ [km/s]&$9^{+13}$&&&$18\pm8$&&\\ \hline

 Torus&$\mathcal{N}_{\mathrm{Na}}$ [Na atoms]&$10^{33.5^{+2.3}_{-0.8}}$&$1.24^{+0.07}_{-0.04}$&1.45&$10^{32.6\pm0.2}$&$1.64^{+0.02}_{-0.04}$& 1.53\\
&$\bar{v}_{\mathrm{Na}}$ [km/s]&$7^{+14}$&&&$18\pm9$&&\\ \hline
	\end{tabular}
\end{table*}

\begin{figure*}
    \centering
    \subfigure[Hydrostatic]{\includegraphics[width=\columnwidth]{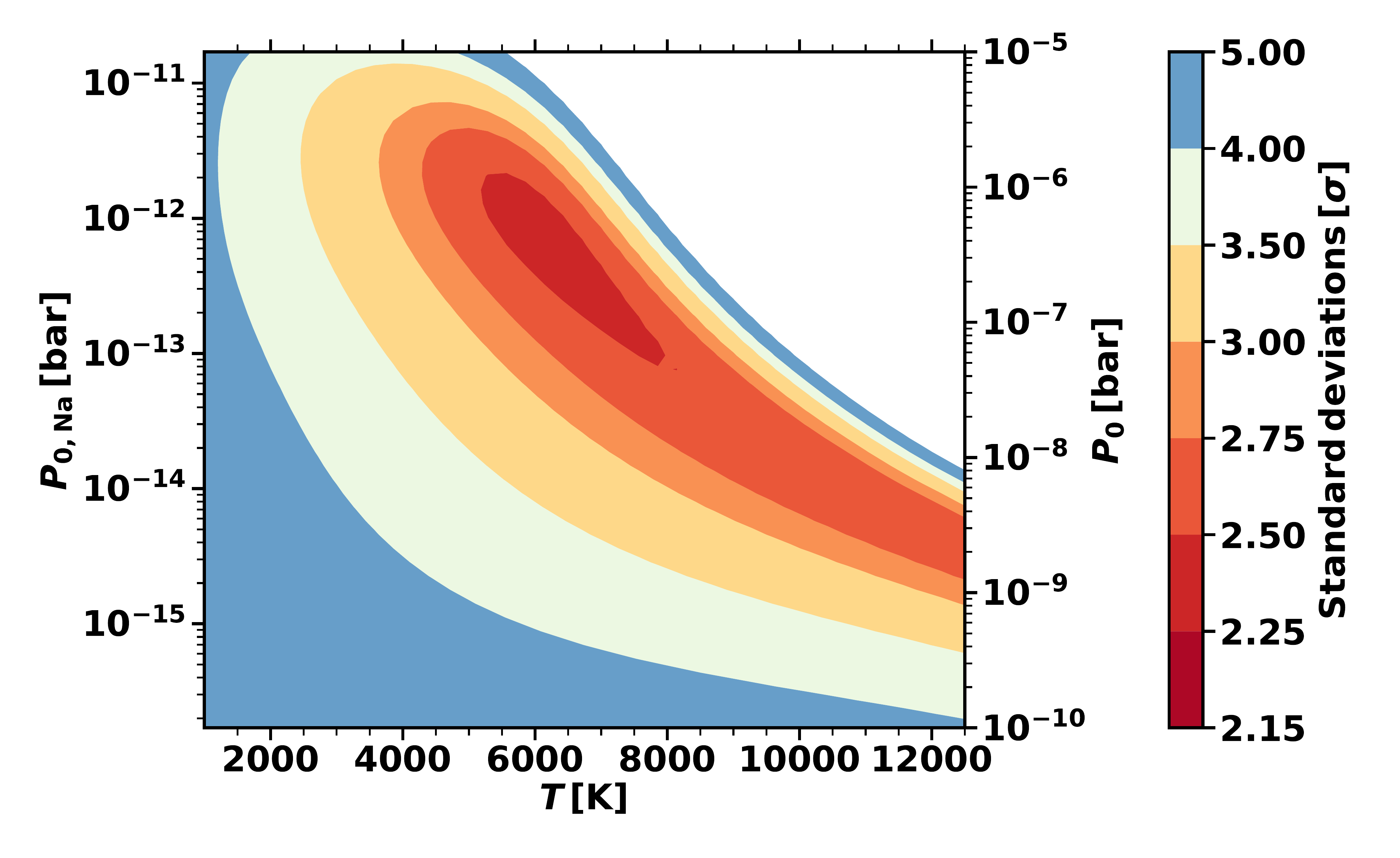}}\quad
    \subfigure[Escaping]{\includegraphics[width=\columnwidth]{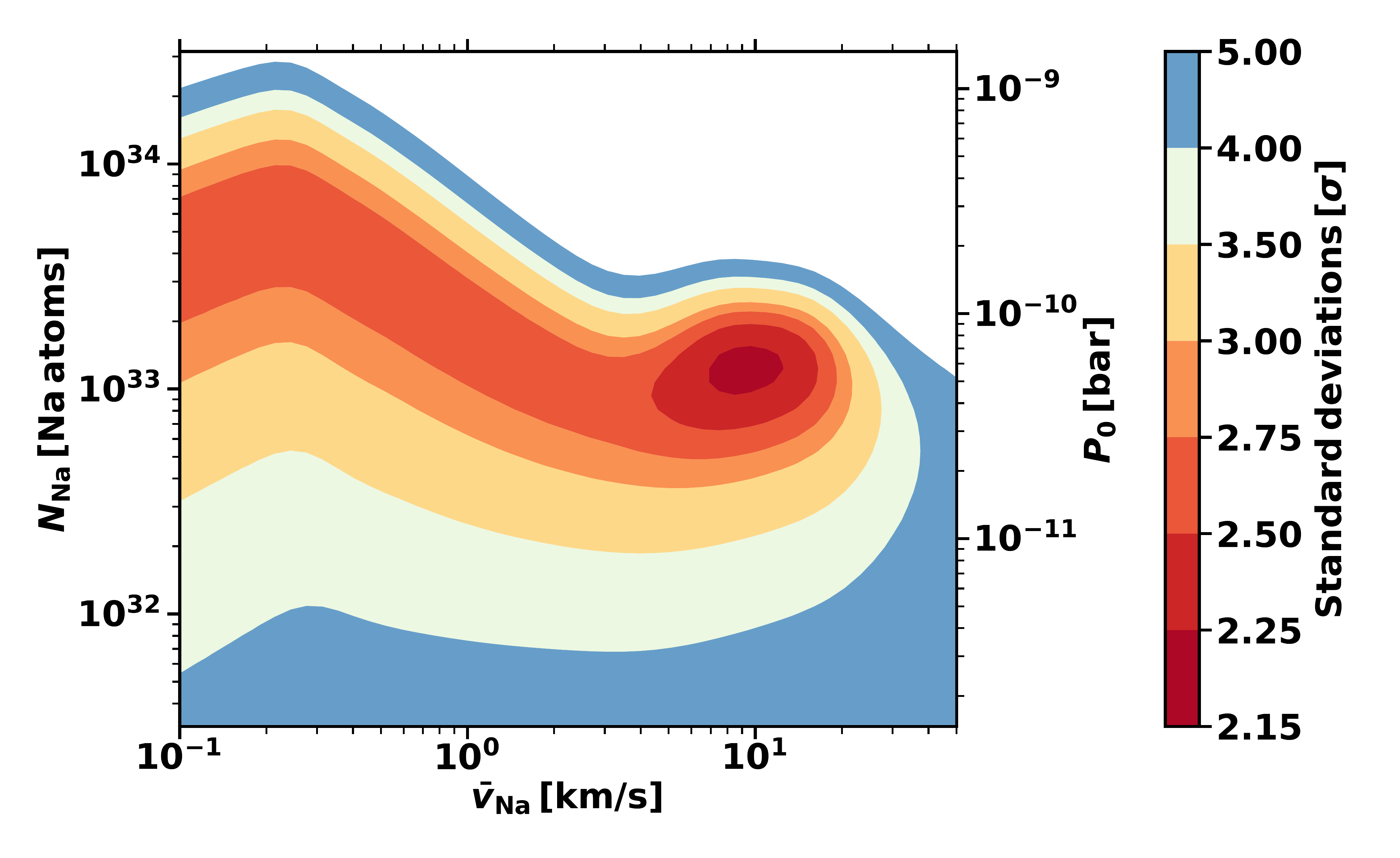}}
    \subfigure[Exomoon]{\includegraphics[width=\columnwidth]{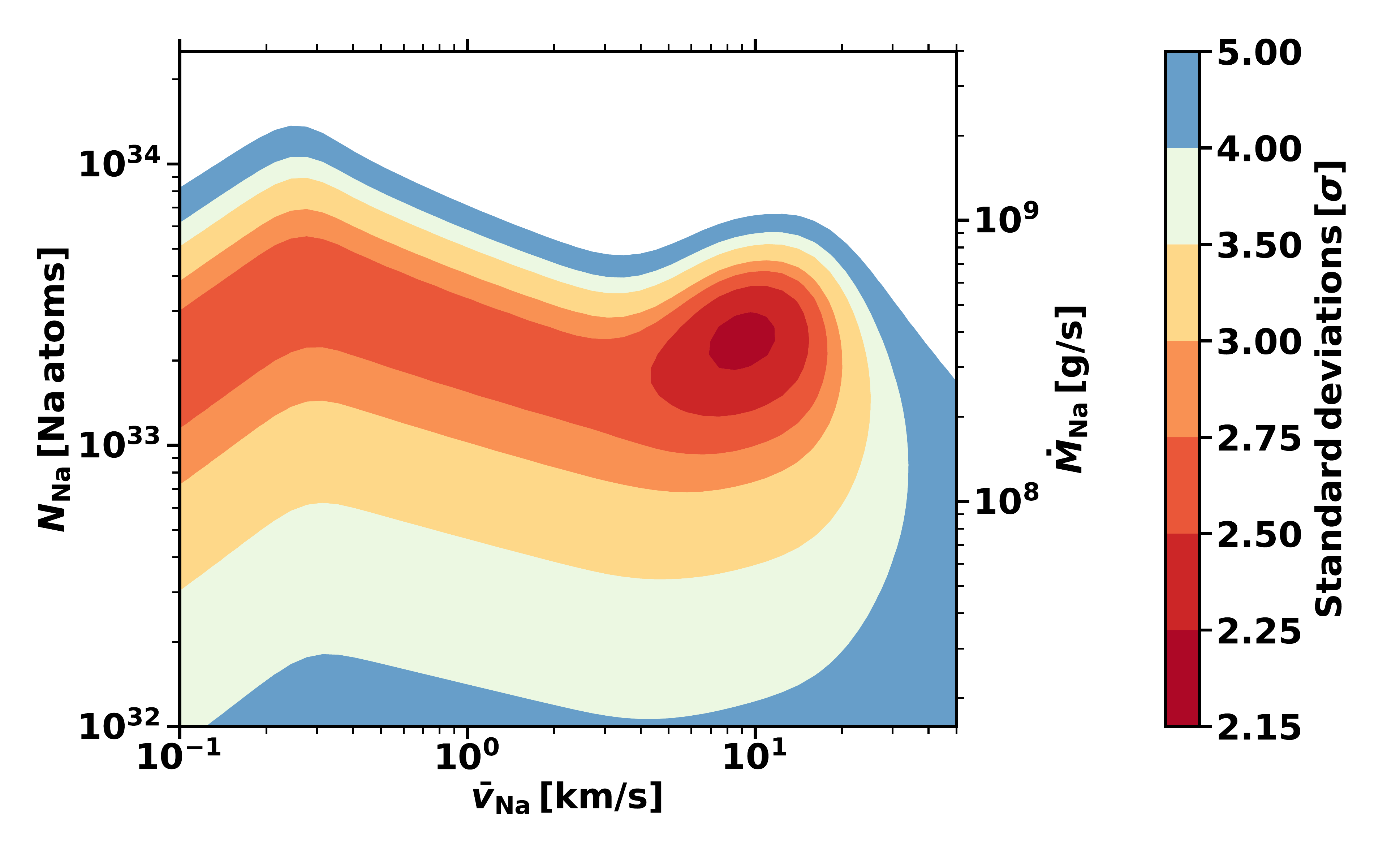}}\quad
    \subfigure[Torus]{\includegraphics[width=\columnwidth]{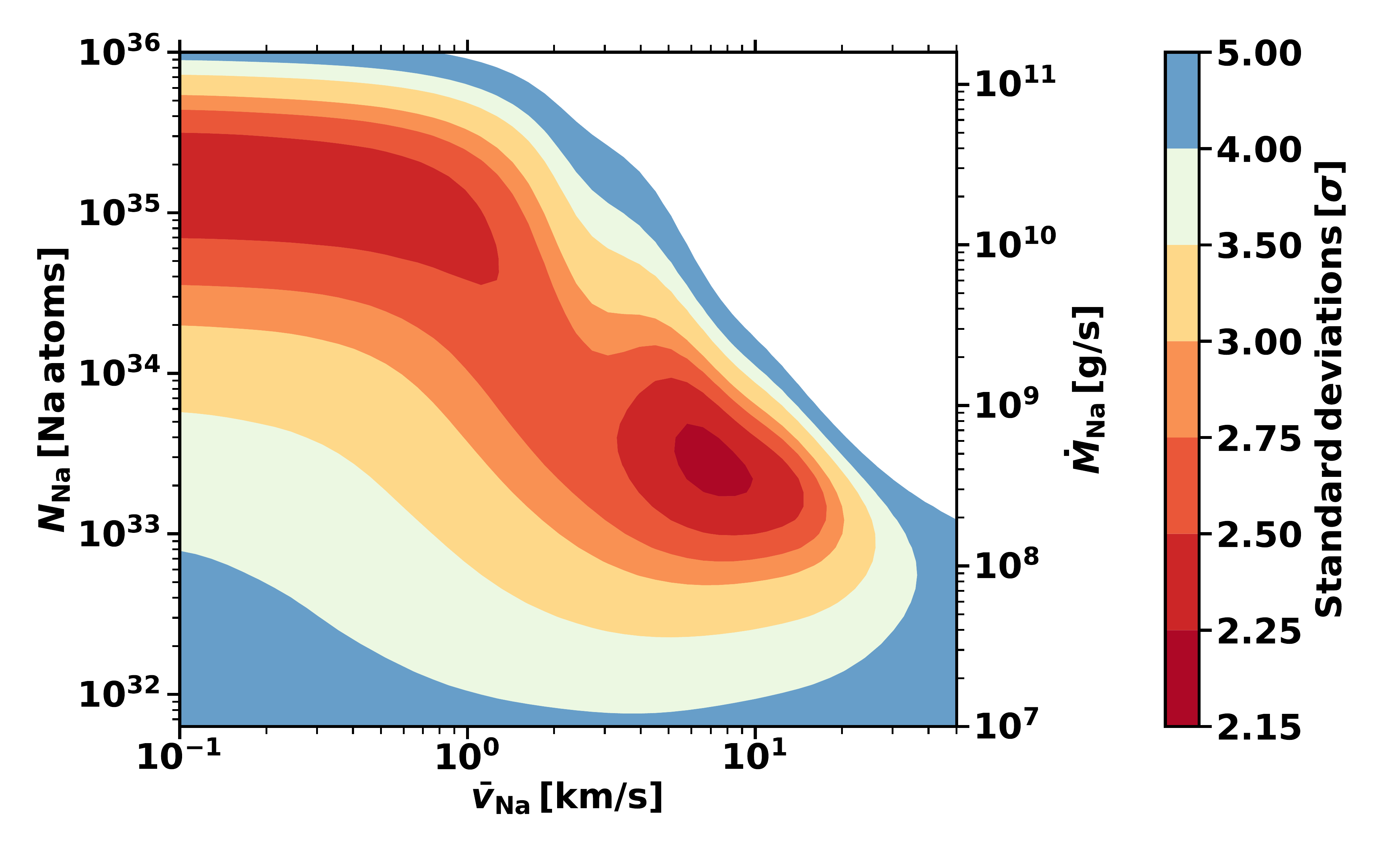}}
    \caption{Parameter retrieval for WASP-49b. The color-coding represents the deviation from $\chi_r^2=1$, in standard deviations of the $\chi_r^2$-distribution ($\sigma=\sqrt{2/\nu}$, where $\nu$ corresponds to the degrees of freedom of the model, see Section \ref{observation}). The contours are calculated on a $50\times50$ parameter grid and interpolated linearly using \texttt{pyplot.contourf}. The auxiliary parameter $P_0$ for the hydrostatic and escaping scenarios is computed using $\chi_{\mathrm{Na}}=1.7\,$ppm and (only for the escaping scenario) $T=T_{\mathrm{eq}}$ at the base of the wind.}
    \label{RetrievalPlotsW49}
\end{figure*}

We observe in the $\chi_r^2$-maps for all four scenarios that the shapes of the best-fitting regions are irregular and extend very far in a particular direction (Figure \ref{RetrievalPlotsW49}). Hence, the retrieved parameters in Table \ref{RetrievalResults} can contain error bars in only one direction for the case of WASP-49b. In the hydrostatic scenario, the parameters are driven to values such that the resulting atmosphere is more optically thin. We retrieve a large temperature of $T=6100_{-3400}\,$K, significantly above the planetary equilibrium temperature of WASP-49b of 1400\,K (\citealt{Wyttenbach2017}). We retrieve similar velocities and source rates in the three evaporative scenarios, with the most prominent difference being the line ratio in the torus scenario of $f_{\mathrm{D2/D1}}=1.24^{+0.07}_{-0.04}$ (indicating absorption mostly in an optically thick regime). Although the torus scenario is evaporative, we see in Figure \ref{tau_profiles} that the optical depth profile is constant over a large range of radii and in the transition region between optically thick and optically thin chords. Since our best-fitting torus scenario requires a sodium mass loss rate comparable to the desorbing torus forward model of HD189733b, the optical depth profile of the best-fitting torus model for WASP-49b resembles the dashed red line in Figure \ref{tau_profiles}, indicating rather optically thick absorption.

The comparison of the retrieved mass loss rates from \texttt{Prometheus} with the calculated rates within \texttt{DISHOOM} is shown in Table \ref{Mdot comparison}. Since \texttt{Prometheus} doesn't directly retrieve $\dot{M}_{\mathrm{Na}}$ but rather $\mathcal{N}_{\mathrm{Na}}$, we use Equation \ref{Mdot} to obtain an upper limit to the mass loss rate from the retrieved $\mathcal{N}_{\mathrm{Na}}$. The retrieved source rate in the escaping scenario is nearly three orders of magnitude larger than the one we calculate using \texttt{DISHOOM}, indicating that an escaping wind can probably not provide enough sodium to generate the observed transit depth. The retrieved source rate in the exomoon scenario is also larger than the one we calculate from \texttt{DISHOOM}, but still within the error bars. For the torus scenarios, direct outgassing (Eqn. \ref{mdotsub}) within \texttt{DISHOOM} significantly underestimates the retrieved source rate, while a desorbing torus (Eqn. \ref{Enhancement torus}) overestimates it (but still within the error bars).

\begin{table}
    \centering
	\caption{Mass loss rate comparison between \texttt{Prometheus} and \texttt{DISHOOM}. We emphasize that the \texttt{Prometheus} mass loss rates are upper limits, due to our usage of Equation \ref{Mdot} and a minimum lifetime of neutral sodium. The sodium source rates within the escaping scenarios fall short of the required rates. While the torus source rates due to direct outgassing (Torus 1, Eqn. \ref{mdotsub}) are also significantly lower than the retrieved ones, we remark that the enhanced rates due to a desorbing torus (Torus 2, Eqn. \ref{Enhancement torus}) are comparable to the retrieved values. Note that the retrieved mass loss rates within \texttt{Prometheus} are larger than the minimal mass loss rates (Table \ref{DISHOOM results}).}
	\label{Mdot comparison}
	\begin{tabular}{llcr} 
\hline
Planet & Scenario & $\dot{M}_{\mathrm{Na}}\,$[kg/s]&$\dot{M}_{\mathrm{Na}}\,$[kg/s] \\
&&\texttt{Prometheus} & \texttt{DISHOOM}\\

\hline

&Escaping&$10^{5.3\pm1}$ &$10^{2.4\pm0.3}$ \\ 

WASP-49b&Exomoon&$10^{5.6\pm0.6}$ &$10^{4.3\pm1.5}$ \\

&Torus 1&$10^{5.7^{+2.3}_{-0.8}}$ &$10^{3.1\pm1.3}$ \\ 

&Torus 2&$10^{5.7^{+2.3}_{-0.8}}$ &$10^{6.4\pm1.3}$ \\ \hline

&Escaping&$10^{4\pm0.2}$&$10^{2.3\pm0.3}$\\ 

HD189733b&Exomoon&$10^{4.3\pm0.2}$&$10^{4\pm1.2}$\\

&Torus 1&$10^{4.2\pm0.2}$&$10^{1.0\pm1.5}$\\
&Torus 2&$10^{4.2\pm0.2}$&$10^{4.1\pm1.5}$\\ \hline
	\end{tabular}
\end{table}

\subsection{Inverse Modeling of HD189733b}

\begin{figure}
    \centering
    \includegraphics[width=\columnwidth]{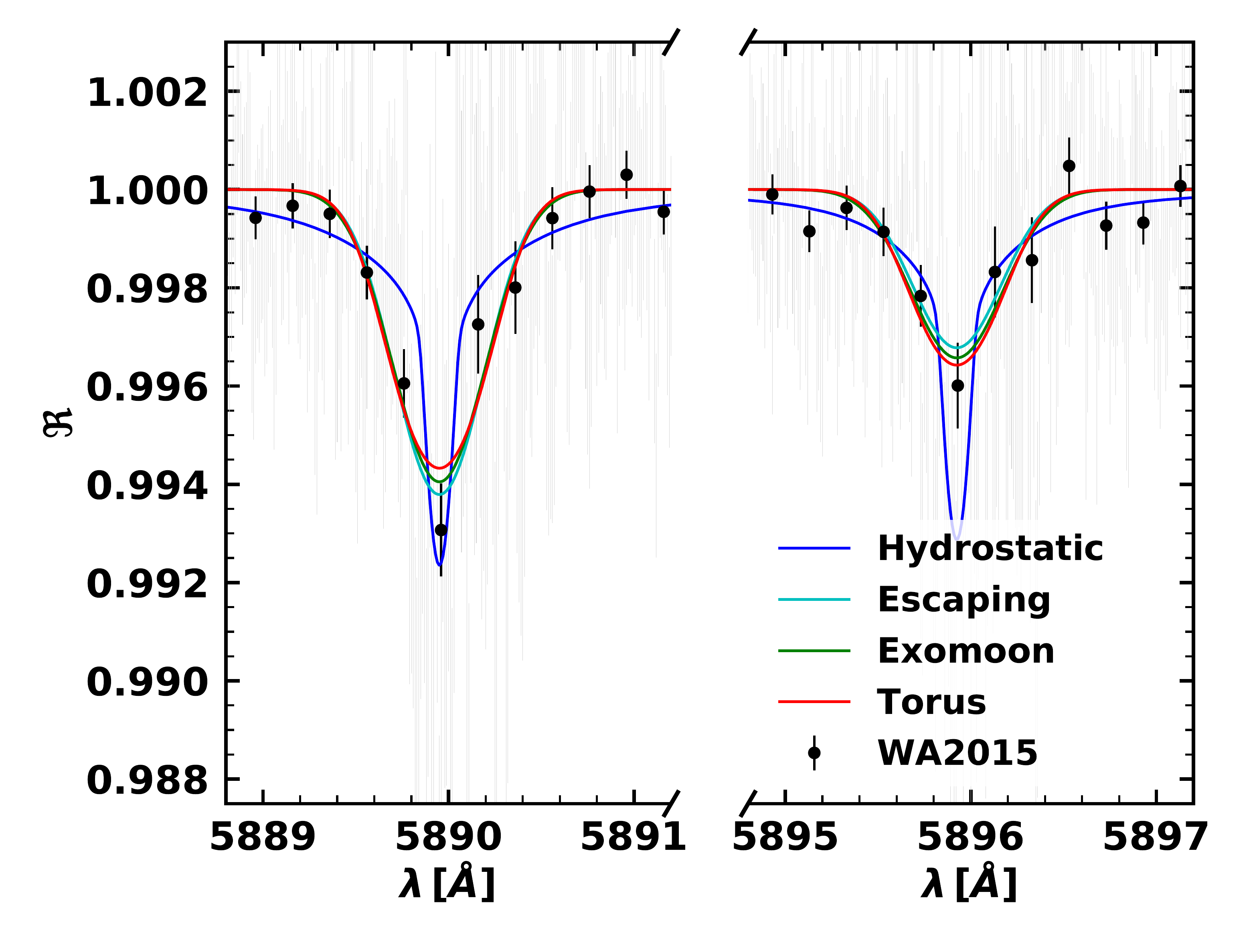}
    \caption{Transit spectrum for HD189733b with our best-fitting models for each scenario. Planetary parameters and data points are taken from \citet{Wyttenbach2015}. Note the break in the x-axis.}
    \label{hd189bestfits}
\end{figure}

\citet{Wyttenbach2015} detected sodium at HD189733b in the Na I doublet. Compared to WASP-49b, this transit spectrum has weaker absorption features (less than one percent absorption at D2 line center), but a larger D2-to-D1 line ratio of $f_{\mathrm{D2/D1}}=1.74\pm0.45$ (again calculated in bands of $0.2\,\angstrom$ centered on the sodium D lines). The line cores for the transit spectrum of HD189733b are broader than the ones of WASP-49b, which leads to more data points lying on the line cores and enabling a more precise retrieval. Therefore, and due to the comparatively smaller error bars for this observation, the best-fitting regions are more regular and significantly smaller for HD189733b (note that we adjusted the color map scale for the parameter retrievals of HD189733b). The observation and the best-fitting models for each scenario are shown in Figure \ref{hd189bestfits}. The retrieved parameters are summarized in Table \ref{RetrievalResults}, we note that the hydrostatic scenario with $f_{\mathrm{D2/D1}}=1.086\pm0.001$ significantly underpredicts the observed line ratio. All scenarios achieve a very similar goodness-of-fit with $1.49<\chi_r^2<1.53$, significantly larger than $\chi_r^2=1$ (see Section \ref{Inverse Modeling W49} for an interpretation of these high $\chi_r^2$-values).

\begin{figure*}
    \centering
    \subfigure[Hydrostatic]{\includegraphics[width=\columnwidth]{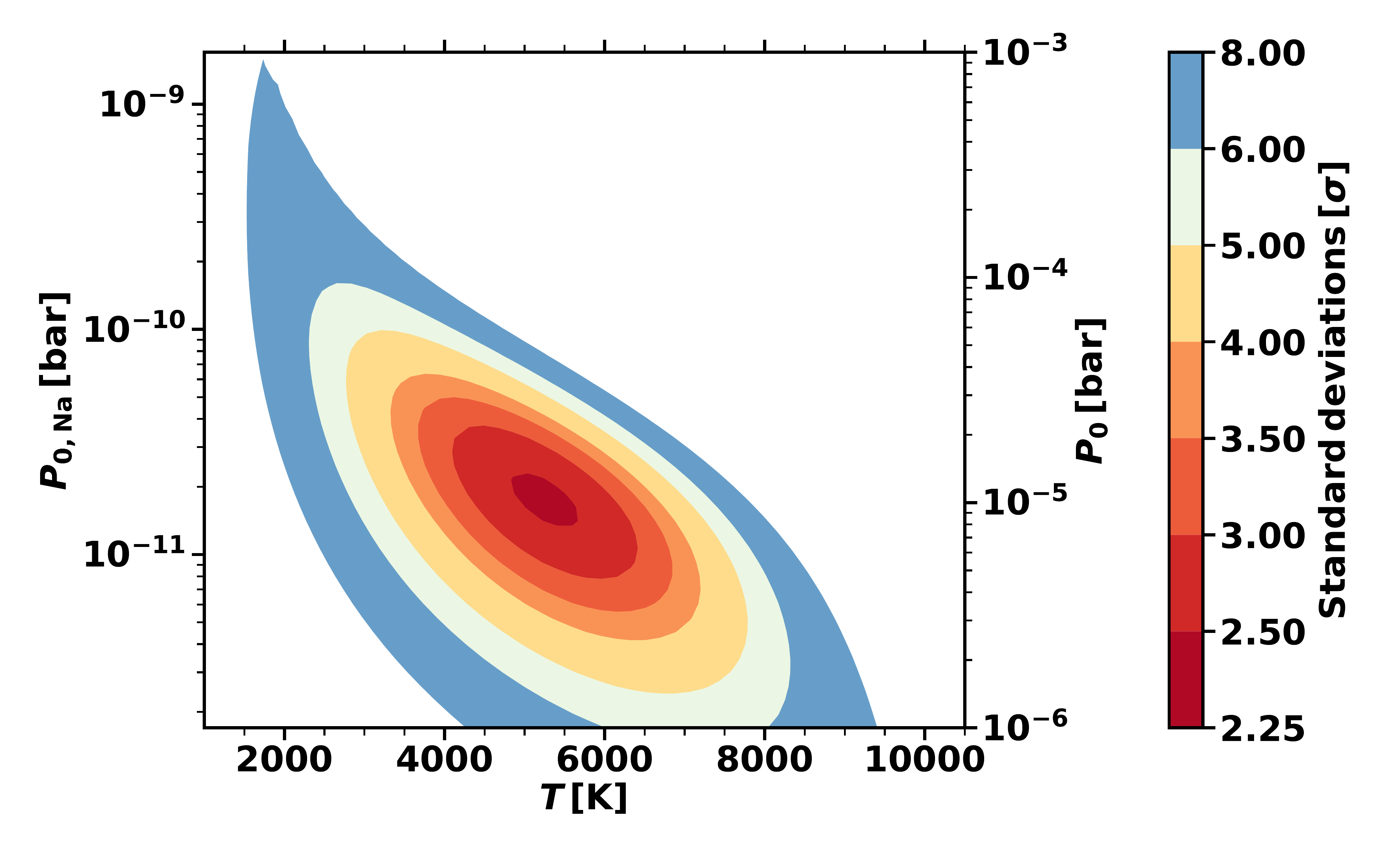}}\quad
    \subfigure[Escaping]{\includegraphics[width=\columnwidth]{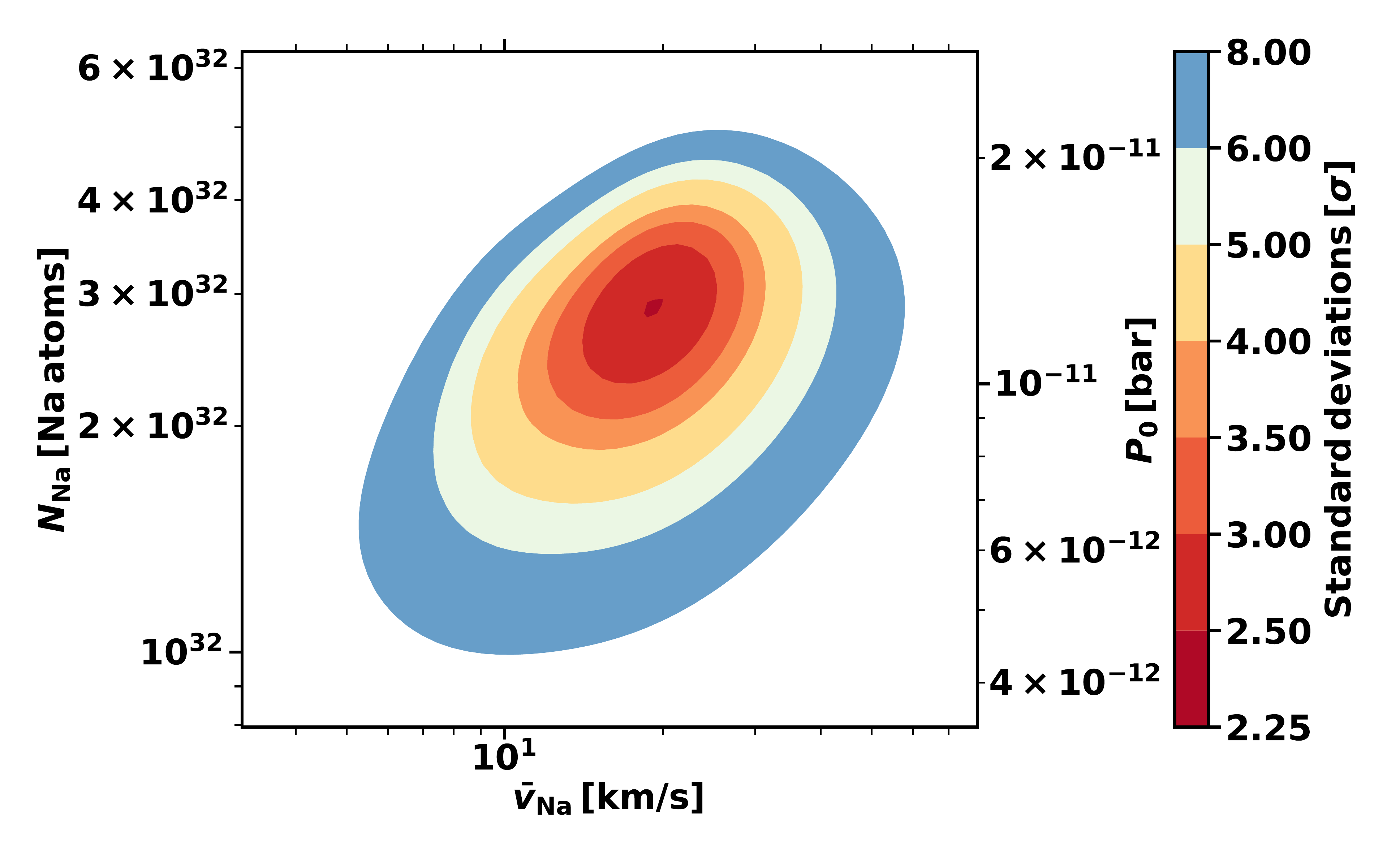}}
    \subfigure[Exomoon]{\includegraphics[width=\columnwidth]{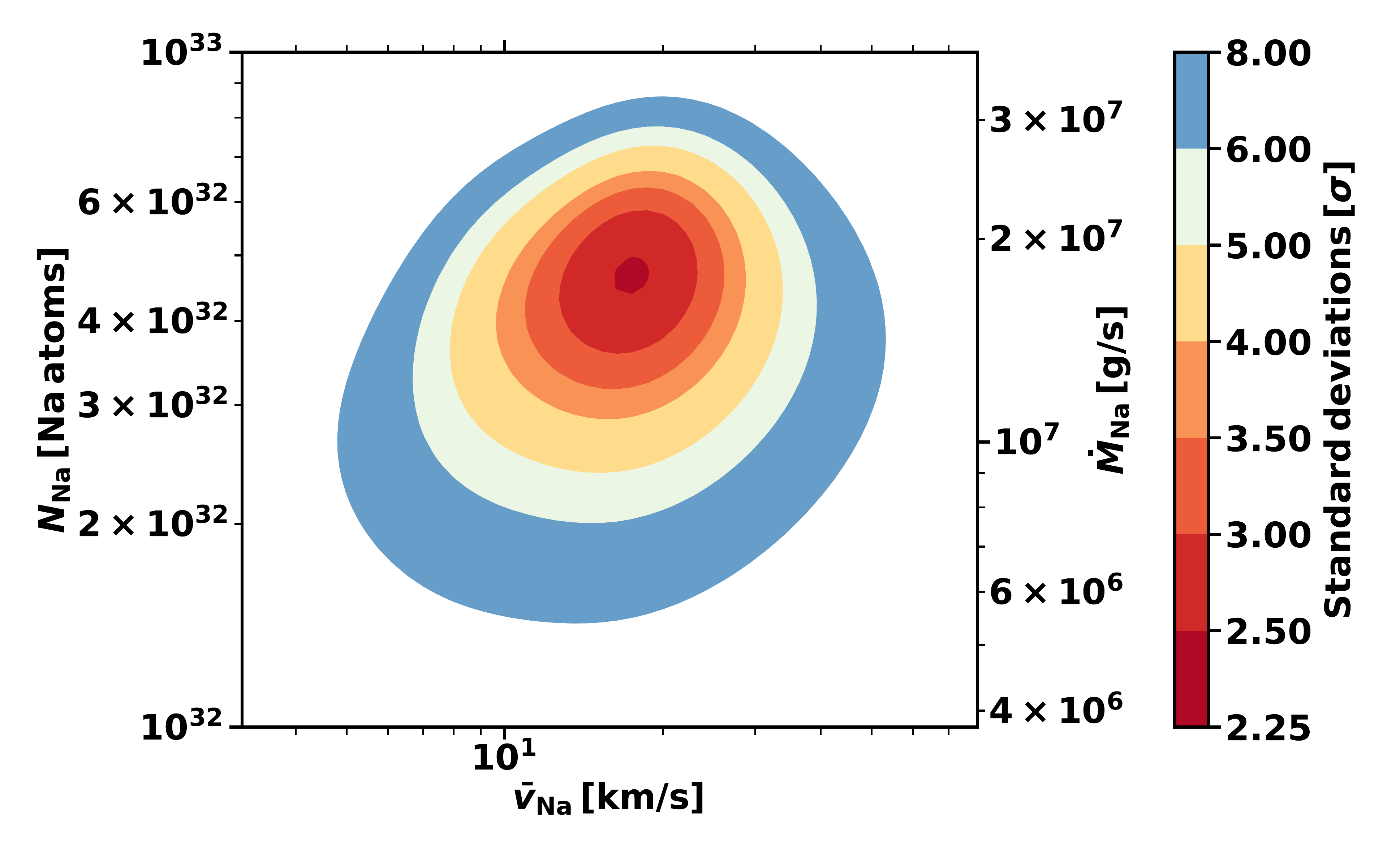}}\quad
    \subfigure[Torus]{\includegraphics[width=\columnwidth]{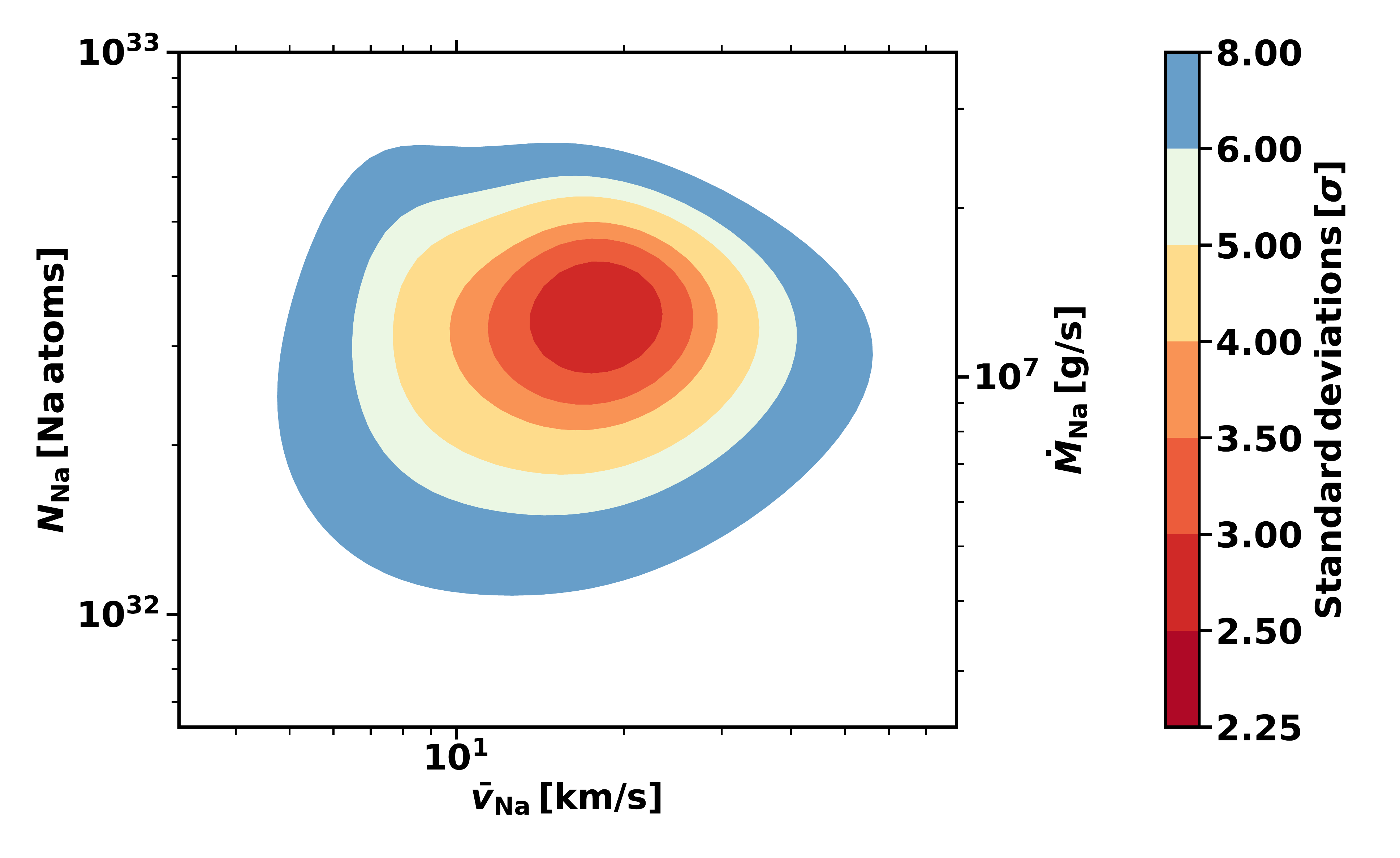}}
    \caption{Parameter retrieval for HD189733b. The color-coding represents the deviation from $\chi_r^2=1$, in standard deviations of the $\chi_r^2$-distribution ($\sigma=\sqrt{2/\nu}$, where $\nu$ corresponds to the degrees of freedom of the model, see Section \ref{observation}). The contours are calculated on a $50\times50$ parameter grid and interpolated linearly using \texttt{pyplot.contourf}. The auxiliary parameter $P_0$ for the hydrostatic and escaping scenarios is computed using $\chi_{\mathrm{Na}}=1.7\,$ppm and (only for the escaping scenario) $T=T_{\mathrm{eq}}$ at the base of the wind.}
    \label{RetrievalPlotsHD189}
\end{figure*}

We show the $\chi_r^2$-maps for all four scenarios in Figure \ref{RetrievalPlotsHD189}. As in the case of WASP-49b, the parameters in the hydrostatic scenario are driven to values such that the resulting atmosphere becomes optically thin. Since there is less absorption compared to WASP-49b, the resulting temperature is also lower: $T=5200\pm1500\,$K, which is clearly above the planetary equilibrium temperature of 1140\,K (\citealt{Wyttenbach2015}). The velocities and mass loss rates are again very similar between the three evaporative scenarios. Compared to WASP-49b, we retrieve larger velocities and smaller source rates for HD189733b, which is due to the broader lines and smaller transit depth in the observation.

We compare the retrieved mass loss rates of \texttt{Prometheus} to the rates computed within \texttt{DISHOOM} in Table \ref{Mdot comparison}. Again, the escaping wind doesn't generate enough sodium (by two orders of magnitude) to reproduce the observed transit depth. While the calculated \texttt{DISHOOM} source rates in the exomoon and desorbing torus (Eqn. \ref{Enhancement torus}) scenarios are very comparable to the retrieved ones, a torus sourced by direct outgassing (Eqn. \ref{mdotsub}) falls short of the retrieved source rate by three orders of magnitude.

\section{Discussion}\label{Discussion}
\subsection{Comparison to other Retrievals} \label{comparisons}
\subsubsection{WASP-49b Studies}
Different authors conducted a parameter retrieval using the same data from \citet{Wyttenbach2017} as we used. In the following we compare our findings to these studies. \citet{Wyttenbach2017} could fit the line cores and the line wings of the spectrum separately with hydrostatic, isothermal and vertically-mixed (constant mixing ratios throughout the atmosphere) models (equivalent to our hydrostatic scenario), but these authors couldn't reproduce the entire spectrum with a hydrostatic model. Their best-fitting model for the line cores has $T=2950^{+400}_{-500}\,$K. \citet{Cubillos2017} attempted to fit the spectrum with a more sophisticated (but still hydrostatic) model (with a $T$-$P$-profile from a hydrodynamic simulation and variable mixing ratios computed with equilibrium chemistry). However, they run into the same difficulty as \citet{Wyttenbach2017}: The observed spectrum at WASP-49b has large absorption on the line cores (approximately two percent on the D2 line), but very little absorption between the line cores. As seen in Section \ref{Forward Modeling}, this property along with $f_{\mathrm{D2/D1}}>1.1$ indicates that absorption occurs mainly in an extended, optically thin region. Since the hydrostatic models of \citet{Wyttenbach2017} and \citet{Pino2018} both retrieve temperatures smaller than $3000\,$K, they model a small and dense atmosphere leading to absorption in the optically thick regime, which isn't able to reproduce the full observed transit spectrum.

\citet{Fisher2019} used yet another hydrostatic model which is isothermal and vertically-mixed, but incorporates NLTE-effects, clouds and less restrictive priors for the temperature range. Since these authors allow for higher temperatures and lower reference pressures (due to cloud decks at high altitudes), their atmosphere is much more extended and optically thin and they were able to reproduce the observed spectrum. These authors retrieve a temperature of $T=8415^{+1020}_{-1526}\,$K (in their LTE scenario $T=7209^{+1763}_{-1892}\,$K), which is in line with our hydrostatic fit which has very high temperature ($T=6100_{-3400}\,$K).

\subsubsection{HD189733b Studies}

As in the case of WASP-49b, different authors conducted a parameter retrieval using the same data from \citet{Wyttenbach2015} as we used. \citet{Wyttenbach2015} could fit different parts of the transit spectrum of HD189733b with isothermal, vertically-mixed, hydrostatic models and interpreted this as a temperature gradient in the atmosphere. Their hottest model to fit the D2 line core has a temperature of $3270\,$K. \citet{Pino2018} combined the high-resolution data with low-resolution transmission spectra for HD189733b (\citealt{Pont2008}; \citealt{Sing2011}; \citealt{Sing2016}) and fit a vertically-mixed ($\chi_{\mathrm{Na}}=1\,\mathrm{ppm}$), hydrostatic model with a customized T-P-profile to the low-resolution data. However, with the parameters retrieved from the low-resolution spectra these authors couldn't fit the high-resolution data from \citet{Wyttenbach2015}, unless the T-P-profile has significantly larger temperatures at low pressures (reaching $\approx8000\,$K at $\sim0.1\,$nbar). Still, this high-temperature model with $f_{\mathrm{D2/D1}}\approx1.1$ doesn't reproduce the large line ratio observed at HD189733b.

\citet{Huang2017} performed a very sophisticated hydrostatic simulation of HD189733b's atmosphere incorporating many different forms of atmospheric chemistry and NLTE-effects. They find that the temperature rises steeply from $2000\,$K at $10\,\mu$bar to $12'000\,$K at $0.1\,$nbar (their Figure 3). These authors furthermore find that at pressures below $1\,\mu$bar most of the sodium atoms are ionized (their Figure 5). \citet{Huang2017} used very different parameters (namely LyC boost and atomic layer base pressure) to fit the full observed spectrum, making a comparison to our scenarios difficult. However, we note that the found atomic layer base pressure of $10\,\mu$bar compares well to our retrieved reference pressure ($P_0=11\,\mu$bar assuming $\chi_{\mathrm{Na}}=1.7\,$ppm) in the hydrostatic scenario, and our temperatures ($T=5200\pm1500\,$K), which are significantly larger than the ones retrieved by \citet{Wyttenbach2015}, are in line with the atmospheric model from \citet{Huang2017}. We also remark that these authors bin the observations in a slightly different way, using 0.05-$\angstrom$-bins. For this particular choice of the bin size, the D2-to-D1 line ratio of HD189733b is approximately 1.1, in line with the hydrostatic models.

\subsection{Potassium in the Atmosphere/Exosphere of Hot Jupiters}\label{potassium}

We have so far focused on the analysis of the Na I doublet. Our codes and the analysis of the D2/D1 line ratio can be readily applied to the K I doublet at $7667\,\angstrom$ and $7701\,\angstrom$. Here, we don't compare our models to observational data and refrain from a normalization and binning routine applied to the spectra as at Na I. We employ a HARPS-like instrumental LSF convolution procedure for our model spectrum. The model serves as a prediction for expected high-resolution K I detections embarked by ESPRESSO (\citealt{Chen2020}) and PEPSI (e.g. \citealt{Keles2019}).

Identical to the forward model presented in Section \ref{Forward Modeling}, we use \texttt{DISHOOM} to calculate K I mass loss rates (Table \ref{K source rates}), which are then converted into $\mathcal{N}_{\mathrm{K}}$ (the number of neutral K I atoms in the system) using Equation \ref{Mdot}. The average lifetime of neutral K I is larger by a factor of 3.75 compared to that of Na I as determined by photoionization (\citealt{Huebner2015}). We set the velocities of the K I atoms in the evaporative scenarios uniformly to that of sputtered atoms $\sim 10\,$km/s. For the hydrostatic scenarios we use the retrieved temperature and reference pressure (Table \ref{RetrievalResults}). For the endogenic, planetary scenarios we use a volumetric sodium-to-potassium ratio of Na/K $ = 15.9$ corresponding to the solar value (\citealt{Asplund2009}). For the exogenic, exomoon \& torus scenarios we use lunar Na/K $ = 6$ (\citealt{PotterMorgan88}) which also corresponds to the maximum Na/K ratios for chondrites as studied by \citet{FegleyZolotov2000}. We note that the evaporative scenarios are strongly dependent on the Na/K ratios due to their optically thin nature. In comparison with the $0.18\%$ K D1 absorption depth detection by \citealt{Keles2019} for HD189733b, we find for the same 0.8\,$\angstrom$ bandpass a K D1 absorption depth of $\approx 0.12\%$ for the hydrostatic scenario, $\approx 0.05\%$ for a the exomoon scenario, and $\approx 0.04\%$ for a desorbing torus (with an enhanced source rate according to Eqn. \ref{Enhancement torus}). At present we believe more observations are needed to make interpretations of the Na/K ratio at exoplanets. 

\begin{table}
    \centering
	\caption{Potassium source rates computed within \texttt{DISHOOM}. $\dot{M}_{\mathrm{tor2,K}}$ corresponds to a desorbing torus (Eqn. \ref{Enhancement torus}).}
	\label{K source rates}
	\begin{tabular}{lcr}
\hline
Source rate & WASP-49b & HD189733b \\ \hline
$\dot{M}_{\mathrm{esc,K}}$ [kg/s] &$10^{1.4\pm0.3}$ &$10^{1.3\pm0.3}$ \\ 
$\dot{M}_{\mathrm{moon,K}}$ [kg/s] &$10^{3.2\pm1.6}$&$10^{2.9\pm1.3}$ \\
$\dot{M}_{\mathrm{tor1,K}}$ [kg/s]&$10^{2.9\pm1.3}$&$10^{-0.6\pm1.5}$ \\
$\dot{M}_{\mathrm{tor2,K}}$ [kg/s] & $10^{6.2\pm1.3}$&$10^{2.6\pm1.5}$ \\ \hline
	\end{tabular}
\end{table}

\begin{figure}
    \centering
    \includegraphics[width=\columnwidth]{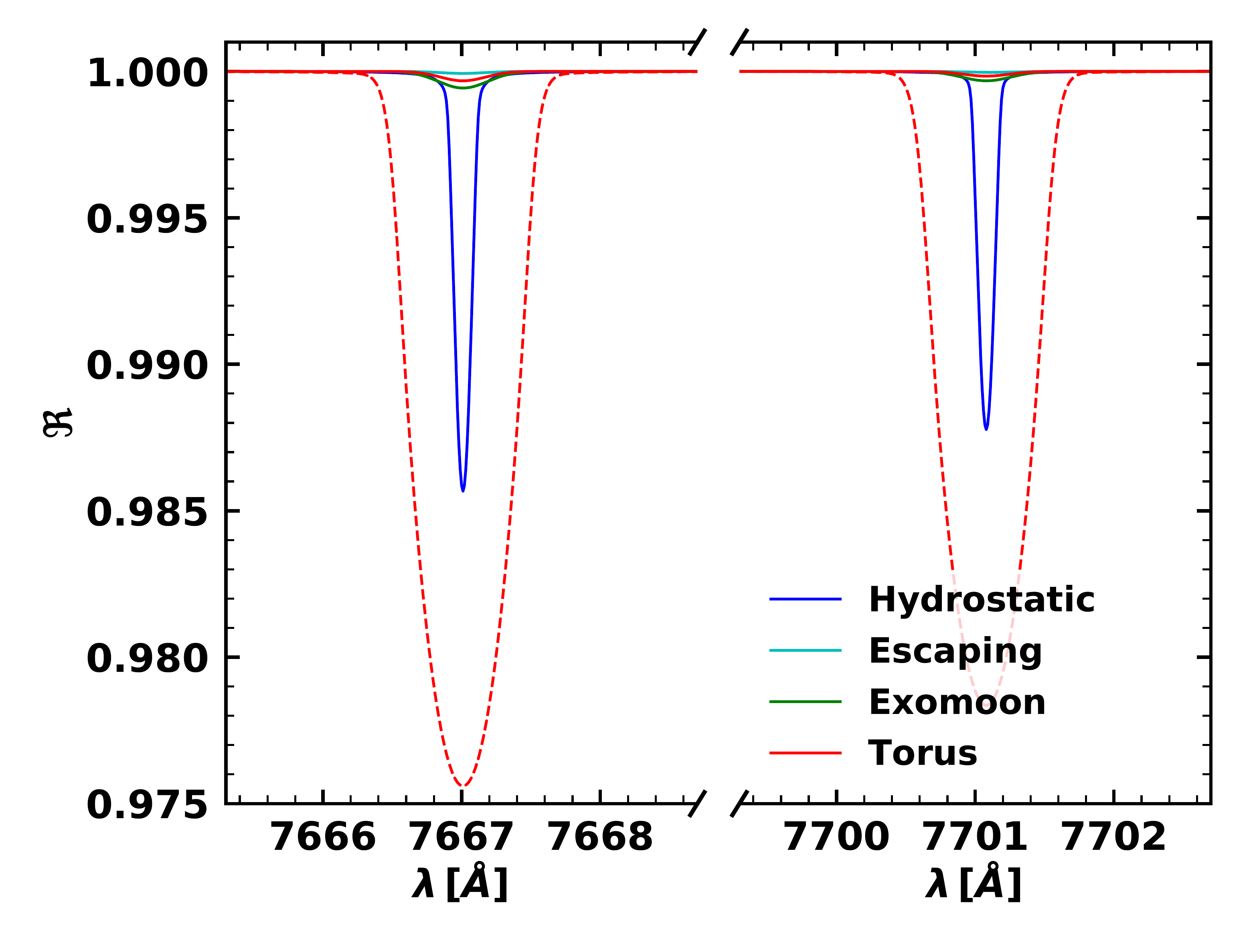}
    \caption{Potassium transmission spectrum for WASP-49b. The dashed red line corresponds to an enhanced potassium source rate for a desorbing torus (Eqn. \ref{Enhancement torus}). Note the break in the x-axis.}
    \label{w49K}
\end{figure}

\begin{figure}
    \centering
    \includegraphics[width=\columnwidth]{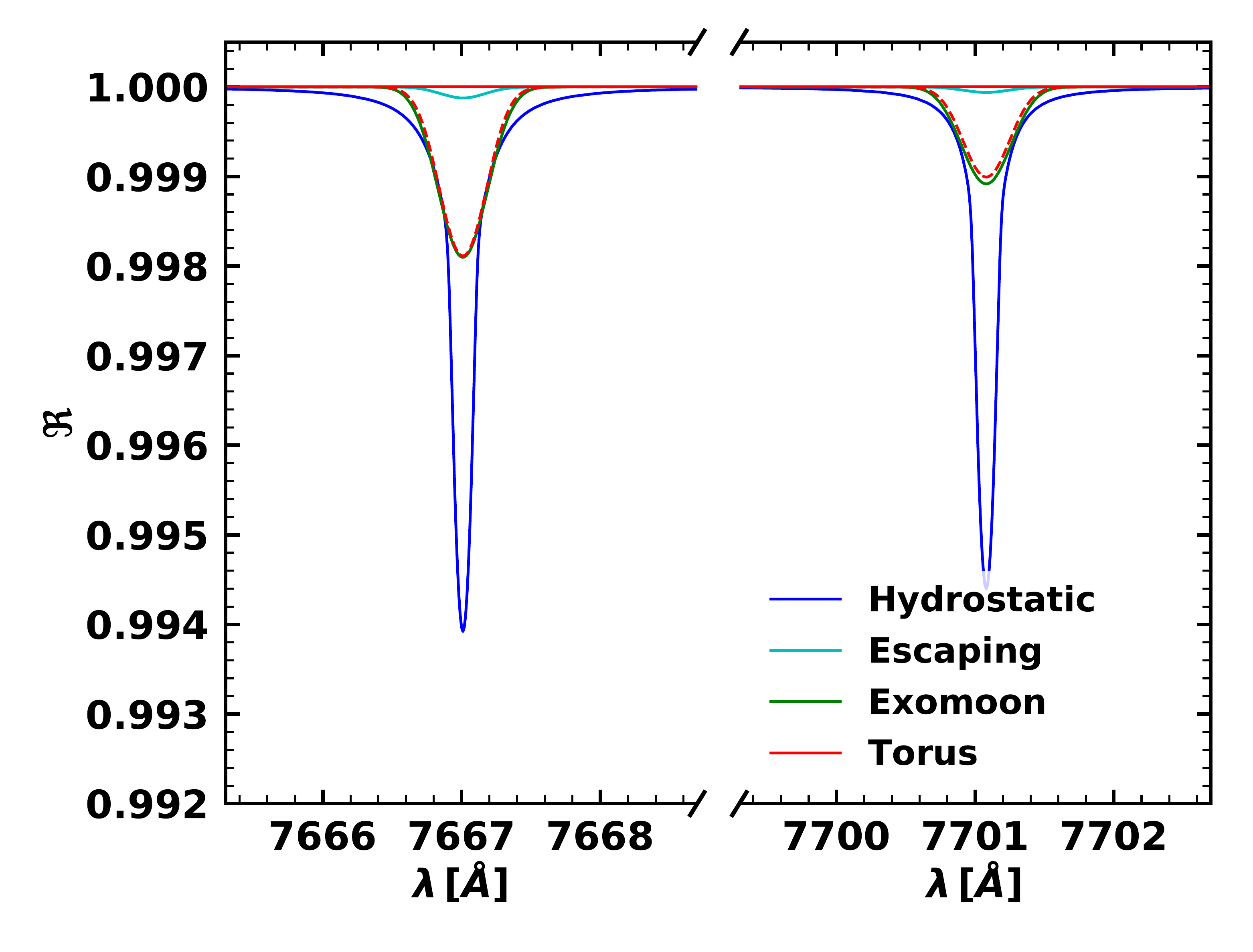}
    \caption{Potassium transmission spectrum for HD189733b. The dashed red line corresponds to an enhanced potassium source rate for a desorbing torus (Eqn. \ref{Enhancement torus}). Note the break in the x-axis.}
    \label{hd189K}
\end{figure}

\subsection{Toroidal Atmospheres/Exospheres at Ultra-Hot Jupiters}\label{ultrahot}
Ultra-hot Jupiters ($T\gtrsim2000\,\mathrm{K}$) have recently been under spectral scrutiny after a remarkable detection of atomic iron at an exoplanet system \citep{Hoeijmakers2018}. Since then, several authors have detected not only Fe I but also Ti, V (\citealt{Hoeijmakers2020}), and other heavy metals (\citealt{Hoeijmakers2018}; \citealt{Hoeijmakers2019}; \citealt{Sing2019}; \citealt{Cabot2020}; \citealt{Gibson2020}). However, the detection diaspora are vastly different at these bodies and the 'ultra-hotness' of the hot Jupiters does not appear to be a sufficient criterion for Fe I detections. For instance, \citet{Wilson-Cauley2020} was unable to detect metals at WASP-189b using PEPSI (R $\sim 50,000$) on the Large Binocular Telescope despite a planetary equilibrium temperature exceeding $2640\,\mathrm{K}$. Therefore atmospheric heating, leading to heavy atom escape as proposed by \citet{Cubillos2020}, should also lead to a Fe I signature at WASP-189b. To better understand the temperature-\textit{independent} metallic detections, we directly compare the latest Na I detections at two ultra-hot Jupiters by the HARPS spectrograph: WASP-76b (\citealt{Seidel2019}) and WASP-121b (\citealt{Cabot2020}; \citealt{Hoeijmakers2020}). In Table \ref{ultrasystemparameters} we provide system parameters of the two systems along with the measured D2-to-D1 ratios of the Na I doublet. Building on our ability to differentiate optically thin and optically thick gases from $f_{\mathrm{D2/D1}}$ in Figure \ref{d2d1ratio} the planets, despite similar equilibrium temperatures (2360 K and 2190 K) and Fe I detections, appear to be in dramatically different gas regimes based on the Na I observations.



\begin{table}
	\centering
	\caption{System parameters for the ultra-hot Jupiters orbiting F stars: WASP-121b (\citealt{Hoeijmakers2020}) and WASP-76b (\citealt{Seidel2019}).  Average sodium lifetimes against photoionization are estimated from \citet{Huebner2015}.}
	\begin{tabular}{lcr} 
\hline
 Parameter& WASP-121b & WASP-76b \\
\hline
$R_{\ast}$ [$R_{\odot}$]& 1.46&1.3\\ 

$R_0$ [$R_J$]& 1.87&1.83\\

$M_p$ [$M_J$]&1.18&0.92\\

$T_{\mathrm{eq}}$ [K]& 2360 &2190\\

$\gamma$ [km/s]&  38 &-1.07\\

$t_{\mathrm{Na}}$ [s] & 50&85 \\

$f_{\mathrm{D2/D1}}$  & 2&1
	\end{tabular}
	\label{ultrasystemparameters}
\end{table}

\subsubsection{Toroidal Atmospheres}\label{toroidalatmospheres}

In light of the stark detection of $f_{\mathrm{D2/D1}} \approx 2$ at WASP-121b in a companion paper by \citet{Hoeijmakers2020}, we decide to model an identical exogenic, toroidal geometry to WASP-76b to better understand the physical mechanism fueling the (presumably temperature-independent) metal detections at ultra-hot Jupiters. 

\begin{figure}
    \centering
    \includegraphics[width=\columnwidth]{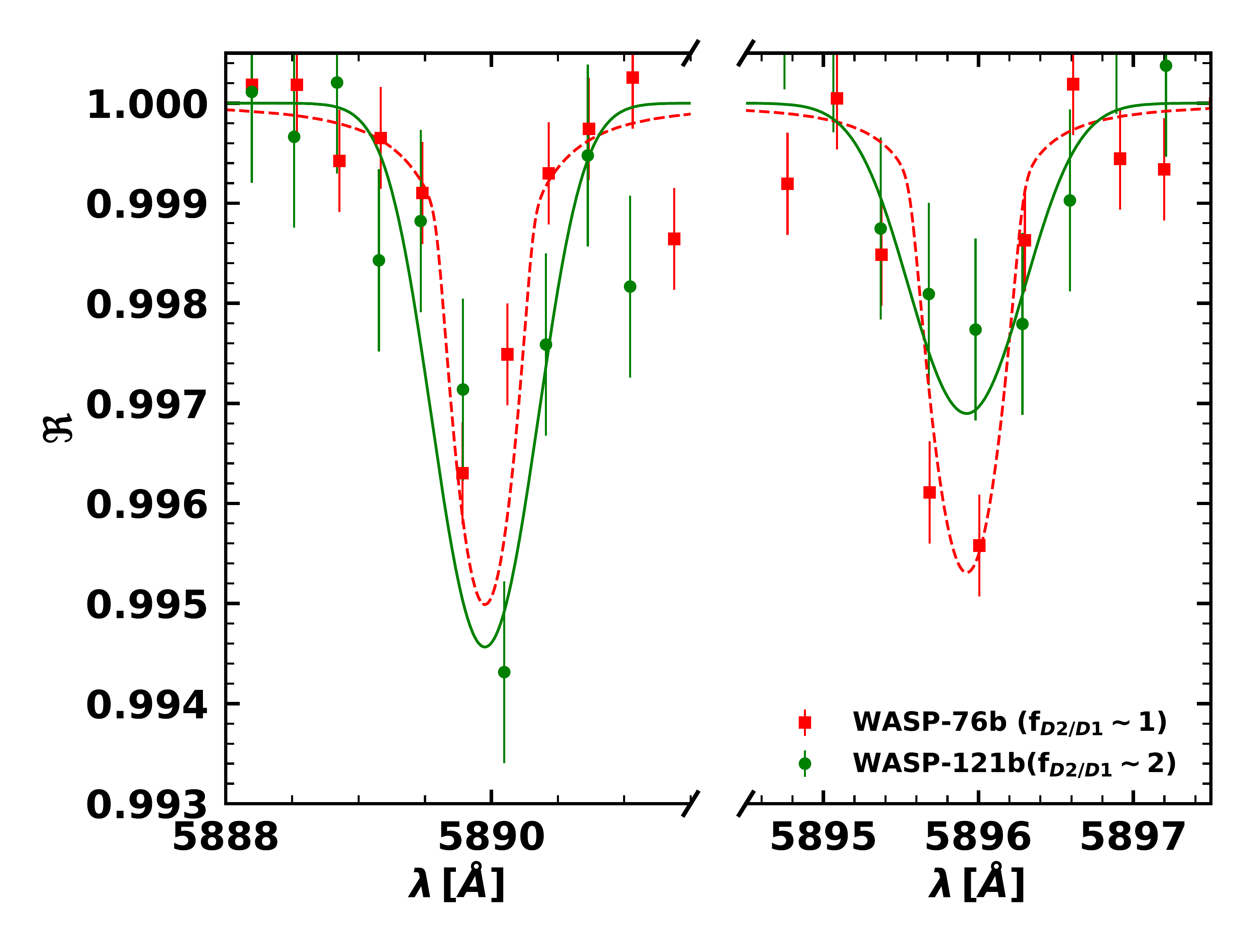}
    \caption{Transit spectrum for WASP 76-b (red squares) and WASP 121-b (green dots) comparing optically thick (red model) and optically thin (green model) toroidal atmospheres and exospheres.    }
    \label{w76121}
\end{figure}

So far, the Na I and Fe I signatures at WASP-76b have been interpreted as endogenic atmospheric winds (\citealt{Seidel2019}; \citealt{Ehrenreich2020}). The scenario of day-night migration coupled with rotation leading to an evening/morning or dusk/dawn asymmetry has surprisingly also been observed in an exospheric regime on other tidally-locked bodies (Ganymede: \citealt{Leblanc2017}; Europa: \citealt{Oza2019A}). In \citet{Oza2018} it was shown that the phenomena of asymmetric 'atmospheric bulges' is degenerate with a collisionless gas on a tidally-locked body. Therefore, while we find an asymmetric Fe I atmospheric wind quite reasonable, we test the degeneracy by simulating Na I rotating in a toroidal atmosphere (red line: WASP-76b $f_{\mathrm{D2/D1}} \approx$ 1) or exosphere (green line: WASP-121b $f_{\mathrm{D2/D1}} \approx$ 2) in Figure \ref{w76121}. 

We find the WASP-121b observations (green) are roughly reproduced by fixing a Na source orbiting at $\sim$ 28 km/s ejecting $\sim 10^9$ g/s extending to $\sim 1.9\, R_0$, equivalent to the planet's Hill radius. This corresponds to a toroidal scale height of $H_{t, W121} \sim R_J/4$ as described in Section \ref{torus} (Case 1) and Eqn. \ref{mdotsub} appropriate for an exo-Enceladus sourcing a toroidal exosphere. In stark contrast, WASP-76b (red) is twice as thick with $H_{t, W76} \sim R_J/2$. At WASP-76b, the torus is confined close to the planetary atmosphere with a much slower orbital speed of $ v_{\mathrm{orb}} \sim$ 7 km/s at $1.125\,R_0$. For grain densities $\sim 3\,\mathrm{g}\,\mathrm{ cm} ^{-3}$, the Roche radius for WASP-76b is roughly $\sim 0.97\, R_0$. On the other hand, the required Na rate here for a silicate exoring is alarmingly large $\sim 10^{9}\,$kg/s of pure atomic Na. This rate is 20 $\times$ as large as the \textit{total} energy-limited escape rate of the H/He envelope, assuming an XUV-efficiency of $\eta_{XUV} = 0.3$ (c.f. Section \ref{escaping}). To our knowledge, the only physical process capable of generating such a high quantity of Na I in a toroidal distribution is the thermal desorption of silicate grains as shown in Table 5 of \citet{Oza2019} and described in Section \ref{torus} (Case 2) and Eqn. \ref{Enhancement torus}. As indicated, an exo-Io at WASP-76b would be catastrophically destroyed due to the mass loss as also implied by the radiative hydrodynamic simulations of \citet{Perez-Becker2013}. The evaporation of such a body would imply a large quantity of metallic gas, as observed.

\section{Conclusions}\label{Conclusions}

The advent of high-resolution transmission spectroscopy of transiting exoplanets has enabled a closer look at the \textit{environment} immediately surrounding the exoplanet. By modeling the sodium D doublet, whose resonance lines are exceptionally bright given even sparse column densities $\gtrsim 10^{10}$ cm$^{-2}$ (c.f. Figure \ref{ecog2}), we investigate three non-hydrostatic, evaporative scenarios in addition to the canonically used hydrostatic model atmosphere. To this end we use a custom-built radiative transfer code (\texttt{Prometheus}), coupled via sodium mass loss rates from a metal evaporation model (\texttt{DISHOOM}) to perform both forward and inverse modeling of high-resolution transmission spectra in the sodium doublet.

The three additional scenarios describe (i) an endogenically escaping medium (Equation \ref{n_esc}) , (ii) an exogenic outgassed cloud sourced by an exomoon (Equation \ref{n_out}), and an (iii) exogenic torus representing circumplanetary material (Equation \ref{n_tor}). Profile (i) of an escaping atmosphere has been simulated in detail and observed across several species over the past decade. Should the exoplanet be close-in, it is highly likely that it is experiencing extreme atmospheric escape on the order of $\sim 10^{7} - 10^{10} $ kg/s \citep{Wyttenbach2020} corresponding to $\sim 10 - 10^{4} $ kg/s of pure sodium escape given solar abundance. The exogenic profiles are based on a recent exomoon study by \citet{Oza2019} as well as comparative solar system studies of Jupiter and Saturn's (cryo-)volcanically active moons Io and Enceladus (\citealt{Johnson2006_OH}, \citealt{Johnson2006b}). We find that both hydrostatic and non-hydrostatic scenarios can fit HARPS transit spectra of WASP-49b and HD189733b. 

In mitigating the apparent sodium degeneracy, we find that first determining whether the absorption occurs in a primarily optically thick or optically thin regime is critical (Section \ref{physics}). A diagnostic based on the ratio of transit depths at the D2 and D1 line centers, $f_{\mathrm{D2/D1}}$, is shown to be indicative of an optically thin or optically thick regime in Section \ref{d2d1method}. Hydrostatic models we find, despite arbitrary heating at $T \sim 10^4\,\mathrm{K}$, cannot achieve line ratios larger than $ \approx 1.2$. This is a consequence of the exponentially decaying number density profile prescribed by hydrostatic equilibrium in contrast to the tenuous and extended profiles of the non-hydrostatic scenarios primarily leading to $f_{\mathrm{D2/D1}} \gg 1$. Given multiple observations with line ratios greater than one (WASP-121b, HD189733b, WASP-49b) an inclusion of evaporative sources of the sodium absorption seems warranted. We find that the evaporative torus scenario is nevertheless able to fit planets with $f_{\mathrm{D2/D1}} \approx 1$ (e.g. WASP-76b) in Section \ref{toroidalatmospheres}. Upon analyzing the Na I source rates required to fit toroidal atmospheres to ultra-hot Jupiter spectra, we remark that a common source for Fe I and Na I due to the evaporation of a rocky body may be reasonable.

Based on an evaporative curve of growth (Figure \ref{ecog2}), we find that for optically thin exospheres (as produced by our three evaporative scenarios), the number of neutral sodium atoms in the system, $\mathcal{N}_{\mathrm{Na}}$, is directly proportional to the transit depth or the equivalent widths of the Na D lines. Equivalently, observed equivalent widths can be used to constrain sodium source rates. For optically thick atmospheres (as in a hydrostatic setting), however, it is not the total amount of sodium which sets the transit depth but rather the (wavelength-dependent) location at which the atmosphere becomes transparent (the transit radius $R_{\lambda}$). While the transit radius can always be calculated from an observed transit depth (Equation \ref{Transit Radius}), the analytical framework presented in Section \ref{hydrostaticgas} is uniquely applicable to hydrostatic atmospheres and breaks down for non-hydrostatic number density profiles. Hence, the transit radius loses its physical meaning as border between optically thick and optically thin chords for the evaporative scenarios, rendering $R_{\lambda}$ obsolete.

For observed high-resolution transmission spectra of WASP-49b and HD189733b, we find that all four scenarios can be fit to the data, from a statistical point of view. From a physical point of view, however, we remark that the escaping scenario and the torus scenario with direct outgassing (Eqn. \ref{mdotsub}) are not able to supply the retrieved source rates, according to our mass loss calculations. Caution should be used in adjusting the solar abundance here in that an atmospheric metal enrichment enhancement $ \gtrsim 100 \chi_{\odot}$ is unlikely \citep{ThorngrenFortney2019}. Furthermore, we note that while the hydrostatic scenario can fit the high-resolution observations as well as the evaporative scenarios in terms of $\chi_r^2$, a few critical issues remain. (1) The atmospheres require significant thermospheric heating ($T\approx5\times T_{\mathrm{eq}}$) and (2) an observed line ratio of $f_{\mathrm{D2/D1}}\geq1.2$ (significant for HD189733b) whereas our best-fitting models yield $f_{\mathrm{D2/D1}}\leq1.2$ despite any aforementioned heating sources at HD189733b and WASP-49b. 

At present, acknowledging that hydrostatic profiles have been excellent approximations to low-resolution transmission spectra, we suggest that radiative transfer and atmospheric escape models would benefit from the non-hydrostatic framework we have outlined in this paper for high-resolution transmission spectra. Further transmission spectra observations, at higher SNR, could validate non-hydrostatic line ratios of the alkaline resonance lines, in significant excess of $\approx 1.2$. Time-resolved observations such as phase curves could provide insight regarding the spatial distribution of alkali atoms.

Looking forward, the recent identification of heavy metals \textit{beyond} the Hill sphere of gas giant exoplanets by high-resolution transmission spectroscopy (\citealt{Hoeijmakers2019}; \citealt{Cubillos2020}) is reminiscent of the remarkable identification of exocomets or falling evaporating bodies at the $\beta$ Pictoris system (e.g. \citealt{Roberge2000}; \citealt{Lecavelier2001}). In this light, based on the evaporative transmission spectra modeling for gas giant exoplanets carried out here, we specifically suggest further modeling of escaping metals subject to ambient plasma fields, along with the likelihood of satellites and tori as at Jupiter and Saturn. \\

 We sincerely extend our gratitude to A. Wyttenbach, J.V. Seidel, and H.J. Hoeijmakers for sharing the high-resolution Na I data analysed here. We thank R.E. Johnson for insight on the exomoon and tori models. We appreciate the radiative transfer discussions with E. Lellouch, C. Huang, and D. Kitzmann during the preparation of the manuscript. We also thank R. Ottersberg for computational help within \texttt{Prometheus}, and acknowledge insightful discussions on mass loss with R. Murray-Clay. AO acknowledges support from SNSF grant 'Planets in Time.' We thank our reviewer I. Waldmann for helpful comments, which improved the quality of this work. We have benefited from the public available programming language \texttt{Python}, including the \texttt{numpy} (\citealt{Walt2011}), \texttt{matplotlib} (\citealt{Hunter2007}) and \texttt{scipy} (\citealt{Virtanen2020}) packages.

Data availability: No new data were generated or analysed in support of this research.





\bibliographystyle{mnras}
\bibliography{bibliography}




\bsp	
\label{lastpage}
\end{document}